\documentclass[12pt]{article}
\pdfoutput=1
\usepackage[]{hyperref}
\usepackage{amsmath}
\usepackage{url}
\usepackage{graphicx,color}

\newcommand{\pa}[1]{\left(#1 \right)}
\newcommand{\PA}[1]{\biggl(#1 \biggr)}

\newcommand{\ca}[1]{\mathcal{#1}}

\def\sinh{{\mathrm{sinh}}}
\def\cosh{{\mathrm{cosh}}}

 \def\ep{{\epsilon}}

 \def\frac#1#2{{#1\over #2}}

 \def\s{\sqrt}

\def\be{\begin{equation}}
\def\ee{\end{equation}}
\def\ba{\begin{eqnarray}}
\def\ea{\end{eqnarray}}
\numberwithin{equation}{section}

 \def\de{\partial}

 \def\f {\frac}
 \def\ti{\tilde}
 \def\ap{\alpha}

 \def\ddd{\cdot\cdot\cdot}
 \def\no{\nonumber \\}

 \def\la{\langle}
 \def\lb{\rangle}
 \def\ep{\epsilon}

\begin{document}

\begin{titlepage}
\thispagestyle{empty}

\begin{flushright}
YITP-18-122
\\
IPMU18-0197
\\
\end{flushright}

\bigskip

\begin{center}
\noindent{{ \textbf{Holographic Quantum Circuits from Splitting/Joining Local Quenches}}}\\
\vspace{2cm}
Teppei Shimaji$^a$, Tadashi Takayanagi$^{a,b}$, and Zixia Wei$^a$
 \vspace{1cm}

{\it
 $^{a}$Center for Gravitational Physics, \\
 Yukawa Institute for Theoretical Physics,
 Kyoto University, \\
 Kyoto 606-8502, Japan\\
$^{b}$Kavli Institute for the Physics and Mathematics of the Universe (WPI),\\
University of Tokyo, Kashiwa, Chiba 277-8582, Japan\\
}

\vskip 2em
\end{center}

\begin{abstract}
We study three different types of local quenches (local operator, splitting and joining) in both the free fermion and holographic CFTs in two dimensions. We show that the computation of a quantity called entanglement density, provides a systematic method to capture essential properties of local quenches. This allows us to clearly understand the differences between the free and holographic CFTs as well as the distinctions between three local quenches. We also analyze holographic geometries of splitting/joining local quenches using the AdS/BCFT prescription. We show that they are essentially
described by time evolutions of boundary surfaces in the bulk AdS. We find that the logarithmic time evolution of entanglement entropy arises from the region behind the Poincar\'{e} horizon as well as the evolutions of boundary surfaces. In the CFT side, our analysis of entanglement density suggests such a logarithmic growth is due to initial non-local quantum entanglement just after the quench.
Finally, by combining our results, we propose a new class of gravity duals, which are analogous to quantum circuits or tensor networks such as MERA, based on the AdS/BCFT construction.
\end{abstract}

\end{titlepage}
\tableofcontents
\newpage

\section{Introduction}

Conformal field theories (CFTs) in two dimensions are very special among quantum field theories (QFTs)
 in that infinite dimensional conformal symmetries constrain their properties.
 Owing to this, we can analytically
 calculate much more physical quantities than those in ordinary QFTs. Nevertheless, we can observe a variety of qualitative difference in the dynamics of two dimensional CFTs (2d CFTs). We can easily come up with two extreme examples of 2d CFTs. One is free CFTs such as massless free fermion or scalar theories. Another one is the strong coupling limits of 2d CFTs with large central charges $c$, namely the holographic CFTs, which have dual descriptions via the AdS/CFT correspondence \cite{Ma,GKPW}.

\begin{figure}
  \centering
  \includegraphics[width=7cm]{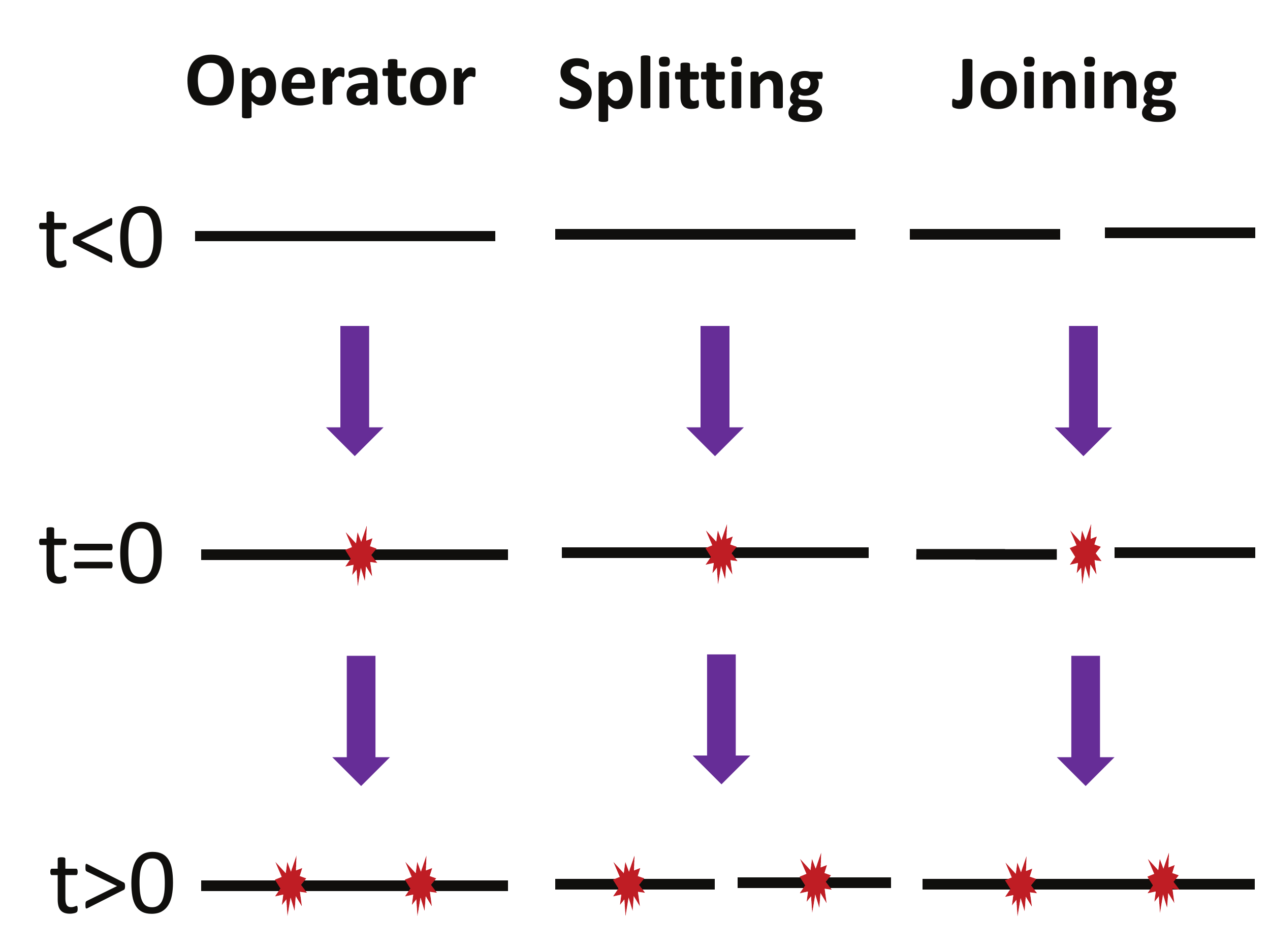}
 \caption{The three local quenches are sketched: the
 local operator quench (left), the splitting local quench (middle), and the joining local quench (right) in two dimensional CFTs. The red points are locations where the energy density is very large.}
\label{fig:SPPS}
\end{figure}

One of the most interesting quantities to characterize the dynamical property of a given quantum state is
the entanglement entropy (EE) \cite{BKLS,Sr,HLW,CC,CH}.  In this paper we would like to explore how differences of 2d CFTs appear in the time evolution of certain excited states. One useful class of excited states is
called local quenches, where we exert a local excitation on a vacuum state. Since it initially modifies the state only locally, they provide clean examples where we can interpret the time evolution easier.
We consider three different types of local quenches: (i) local operator quenches, (ii)  splitting local quenches, and (iii) joining local quenches. We sketched them in Fig.\ref{fig:SPPS}.

The first one (i) is simply defined by acting a local operator on a vacuum and we analyze its time evolution. This was first introduced in \cite{NNT,Nozaki:2014uaa}
and there have already been many related results
both in field theory analysis and holographic analysis. The third one (iii) is defined by joining two semi-infinite lines, as first introduced in \cite{CCL}. The second one (ii) is triggered by splitting a connected line into two disconnected ones. This is a new setup which the present paper will discuss in detail.

In this paper we will point out that it is very helpful and systematic to analyze not the entanglement entropy itself but its second derivatives, called the entanglement density (ED) \cite{HLQ}. Indeed, this quantity extracts the essential behaviors of entanglement entropy under local quenches in term of clear peaks in its graph. By studying the behavior of entanglement density, we can clearly see both similarities and differences among the above three types of local quenches in free CFTs and holographic CFTs, as we will explain later. In particular, this observation resolves an apparent puzzle on the known logarithmic growing entanglement entropy under holographic operator local quenches.

Another purpose of this paper is to explore the spacetime geometries of holographic local quenches. Since the gravity dual geometry of (i) local operator
quenches was already given in \cite{HLQ}, we will focus on (ii) and (iii).  For (iii) joining local quenches, the construction of gravity dual by employing the AdS/BCFT formulation \cite{AdSBCFT} was given in \cite{Ugajin:2013xxa} and we will study more details of the spacetime geometry by using this description in this paper. In addition we will provide the gravity dual geometry for (ii) splitting local quenches. Refer to \cite{NSTW} for other classes of topology changing quantum operations in CFTs such as projections and partial identifications.
Our details geometric analysis clearly will explain the two different sources of logarithmic time evolutions of entanglement entropy observed in the local quenches.

Our analysis of holographic geometry of local quenches is also motivated by the conjectured connection between the AdS/CFT and tensor networks \cite{TNa} such as the MERA \cite{MERA}. In tensor network descriptions, we normally consider discretized lattice theories whose continuum limits correspond to CFTs. Therefore, it is not directly related to the continuous AdS spacetimes, but to their discretized versions as in \cite{TNa,TNc,TNd}. One way to resolve this problem is to consider a continuous tensor networks as in the continuous MERA \cite{cMERA,NRT} or path-integral approaches \cite{TNR,MTW,Caputa:2017urj,Path,Tak}.

However, it is also intriguing to try to realize a discretized version of AdS in a way  that it is naturally derived from the conventional AdS/CFT. The gravity duals of local quenches are useful for this purpose because we can model a class of quantum gates in tensor networks by arranging the splitting and joining procedures in CFTs. As we will discuss in the final part of this paper, we presents a sketch of gravity duals of MERA tensor networks, by combining the holographic local quenches. This indeed qualitatively supports the conjecture.

This paper is organized as follows: In section \ref{sec:EE}, we review basic methods of calculating entanglement entropy by conformal map and also by AdS/(B)CFT. In section \ref{sec:ED}, after we review the definition of
entanglement density, we study its behaviors for global and local operator quenches.
In section \ref{sec:splitting},
we analyze the splitting local quenches and its holographic dual. In section \ref{sec:joining}, we analyze the joining local quenches and its holographic dual. In section \ref{sec:log}, we explain how the logarithmic growth of entanglement entropy arises from a geodesic length in each gravity dual.
In section \ref{sec:tnlq}, we provide qualitative tensor network description of local quenches.
In section \ref{sec:circuits}, we combine the results in this paper to provide a new gravity dual of
a MERA-like (discretized) tensor network, in the framework of AdS/BCFT.
In section \ref{sec:conclusion},  we summarize our conclusions
and discuss future problems. In appendix \ref{dfee}, we show the entanglement density can reproduce the correct entanglement entropy even when the subsystem consists of multiple disconnected intervals in two dimensional massless Dirac fermion CFT. In appendix \ref{sec:HLQ}, we will present the detailed computations of time evolutions of holographic entanglement entropy under splitting/joining local quenches.

\section{Entanglement Entropy and AdS/(B)CFT}\label{sec:EE}

Here we briefly review calculations of entanglement entropy (EE) in two dimensional CFTs, based on conformal mappings and computations of holographic entanglement entropy (HEE) based on the AdS/CFT and
AdS/BCFT. The entanglement entropy is defined by the von-Neumann entropy $S_A=-\mbox{Tr}[\rho_A\log\rho_A]$, where $\rho_A$ is the reduced density matrix defined by
tracing out the original quantum state over all parts of Hilbert space other than $A$.

\subsection{EE for CFT Vacuum in Flat Space}

Consider a two dimensional CFT (2d CFT) on a plane $R^2$, which is described by the complex
coordinate $(w,\bar{w})$. We define the time and space coordinate $(\tau,x)$ as
\be
w=x+i\tau,\ \ \ \bar{w}=x-i\tau.
\ee
The time $\tau$ is Euclidean time and is analytically continued to the real time by
\be
\tau=it.
\ee

A two point function of a primary operator $O$ in a 2d CFT behaves as
\be
\la O(w_1,\bar{w}_1)O(w_2,\bar{w}_2)\lb=\frac{1}{|w_1-w_2|^{2(h+\bar{h})}},
\label{twopp}
\ee
where the primary operator $O$ has the chiral/anti-chiral conformal dimension given by $(h,\bar{h})$.
For the twist operator $\sigma_n$, we have $h=\bar{h}=\frac{c}{24}(n-1/n)$. When we compute the entanglement entropy
(EE) for the subsystem $A$ defined by the interval $0\leq x \leq l$ at a fixed time $\tau=0$, the relevant two point function is
\be
\la \sigma_n(l,l)\bar{\sigma}_n(0,0)\lb =\frac{1}{l^{\frac{c}{6}(n-1/n)}}.
\ee
Therefore the EE $S_A$ for the CFT vacuum reads
\be
S_A=-\frac{\de}{\de n}\log \la \sigma_n(l,l)\bar{\sigma}_n(0,0)\lb\Bigr|_{n=1}=\frac{c}{3}\log \frac{l}{\ep},
\label{gree}
\ee
where $\ep$ is the UV cut off (lattice spacing).

The R\'{e}nyi entanglement entropy, defined by
\be
S^{(n)}_A=\frac{1}{1-n}\log \mbox{Tr}[(\rho_A)^n],  \label{REE}
\ee
can also be computed for the vacuum as
\be
S^{(n)}_A=\frac{c}{6}\left(1+\frac{1}{n}\right)\log \frac{l}{\ep}.
\ee

\subsection{EE for Excited States from Conformal Transformation}

To calculate the EE for a special class of excited states in 2d CFTs,
we can employ the conformal transformation.
We transform the original coordinate system $(w,\bar{w})$ into a new one $(\xi,\bar{\xi})$:
\be
\xi=f(w). \label{CF}
\ee
Also we need to note that the original UV cut off $\ep$, which is a lattice spacing in the $w$ coordinate, is mapped to the one, called $\ti{\ep}_{a,b}$, in the new coordinate $(\xi,\bar{\xi})$ as
\be
\ep=\frac{\ti{\ep}_a}{|f'(w_a)|}=\frac{\ti{\ep}_b}{|f'(w_b)|}. \label{cutr}
\ee
Finally the EE for the excited state is found to be
\ba
S_A=\frac{c}{6}\log\left[\frac{|f(w_a)-f(w_b)|^2}{\ep^2|f'(w_a)||f'(w_b)|}\right].
\label{EEC}
\ea

In our analysis of joining/splitting local quenches, the EE is computed from 
a two point function of twist operators in the presence of a conformal boundary, called
boundary conformal field theory (BCFT).  Since two point functions in BCFT are similar to 
four point functions in CFTs without boundaries, we do not have any universal expression 
such as (\ref{twopp}). However, we can obtain a definite analytical expression for special 
CFTs such as  the Dirac fermion CFT and the holographic CFTs, whose details will be discussed 
later.

\subsection{Conformal Map in AdS$_3/$CFT$_2$}

The AdS/CFT correspondence argues that gravitational theories on AdS$_3$ is equivalent to 2d holographic CFTs, which live on the AdS boundary \cite{Ma}. The physical equivalence is formulated such that
the partition function in each side agrees with the other one \cite{GKPW}, so called the bulk-boundary correspondence.

Consider the Poincar\'{e} metric of AdS$_3$, which is dual to the vacuum of 2d CFT (CFT$_2$):
\ba
ds^2=\frac{d\eta^2+d\xi d\bar{\xi}}{\eta^2},  \label{pol}
\ea
where we set the AdS radius to $1$ for simplicity.
The conformal transformation (\ref{CF}) is equivalent to the following coordinate transformation in AdS$_3$
(see e.g.\cite{Ro}):
\ba
&& \xi=f(w)-\frac{2z^2(f')^2(\bar{f}'')}{4|f'|^2+z^2|f''|^2},\no
&& \bar{\xi}=\bar{f}(\bar{w})-\frac{2z^2(\bar{f}')^2(f'')}{4|f'|^2+z^2|f''|^2},\no
&& \eta=\frac{4z(f'\bar{f}')^{3/2}}{4|f'|^2+z^2|f''|^2}. \label{corads}
\ea
The metric in the coordinate $(w,\bar{w},z)$ reads
\ba
ds^2=\frac{dz^2}{z^2}+T(w)(dw)^2+\bar{T}(\bar{w})(d\bar{w})^2+\left(\frac{1}{z^2}
+z^2T(w)\bar{T}(\bar{w})\right)dwd\bar{w}, \label{metads}
\ea
where
\ba
T(w)=\frac{3(f'')^2-2f'f'''}{4f'^2},\ \ \bar{T}(\bar{w})=\frac{3(\bar{f}'')^2-2\bar{f}'\bar{f}'''}{4\bar{f}'^2},
\ea
are the chiral and anti-chiral energy stress tensor.

\subsection{Holographic Entanglement Entropy in AdS$_3/$CFT$_2$}

In Euclidean setups, the holographic entanglement entropy (HEE) \cite{RT,HRT} is given by
\be
S_A=\frac{L}{4G_N},
\ee
in terms of the length $L$ of shortest geodesic which connects the two boundary points of the subsystem $A$. We also need to impose the homology constraint that this geodesic is homologous to the subsystem $A$ in the AdS geometry.

If we consider a connected geodesic which connects $w_a$ and $w_b$, then its length is given by
\be
L_{ab}=\log \frac{|\xi_a-\xi_b|^2}{\ti{\ep}_a\ti{\ep}_b}=\log\left[\frac{|f(w_a)-f(w_b)|^2}{\ep^2|f'(w_a)||f'(w_b)|}\right],
\label{adseec}
\ee
where note the relation (\ref{cutr}), following from the above coordinate transformation around $\eta= 0$:
$\eta\simeq |f'|z$. Then the HEE $\frac{L_{ab}}{4G_N}=\frac{c}{6}L_{ab}$ reproduces the formula (\ref{EEC}).

\subsection{EE from AdS/BCFT}\label{adsbcfts}

\begin{figure}
  \centering
  \includegraphics[width=7cm]{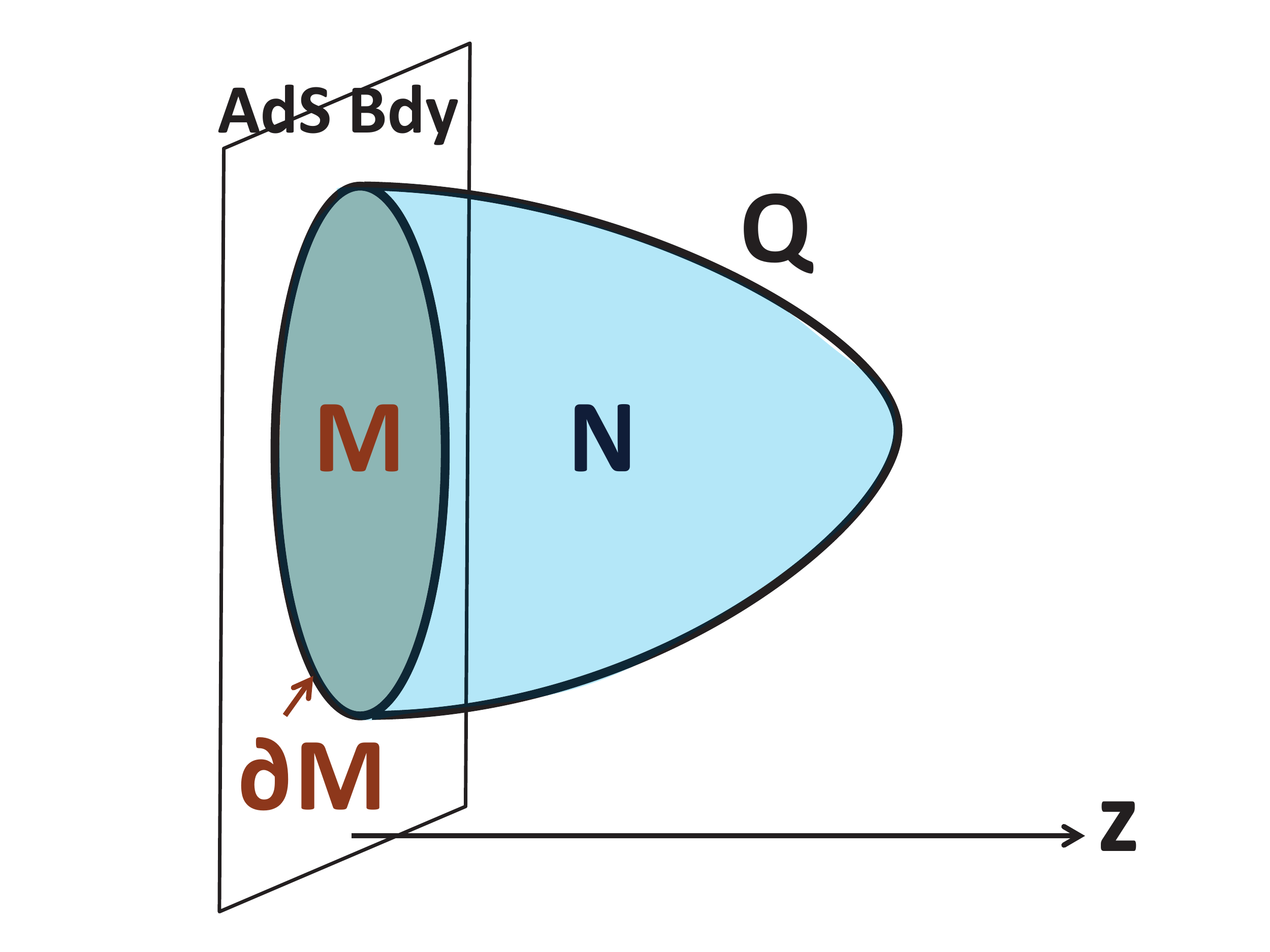}
   \includegraphics[width=7cm]{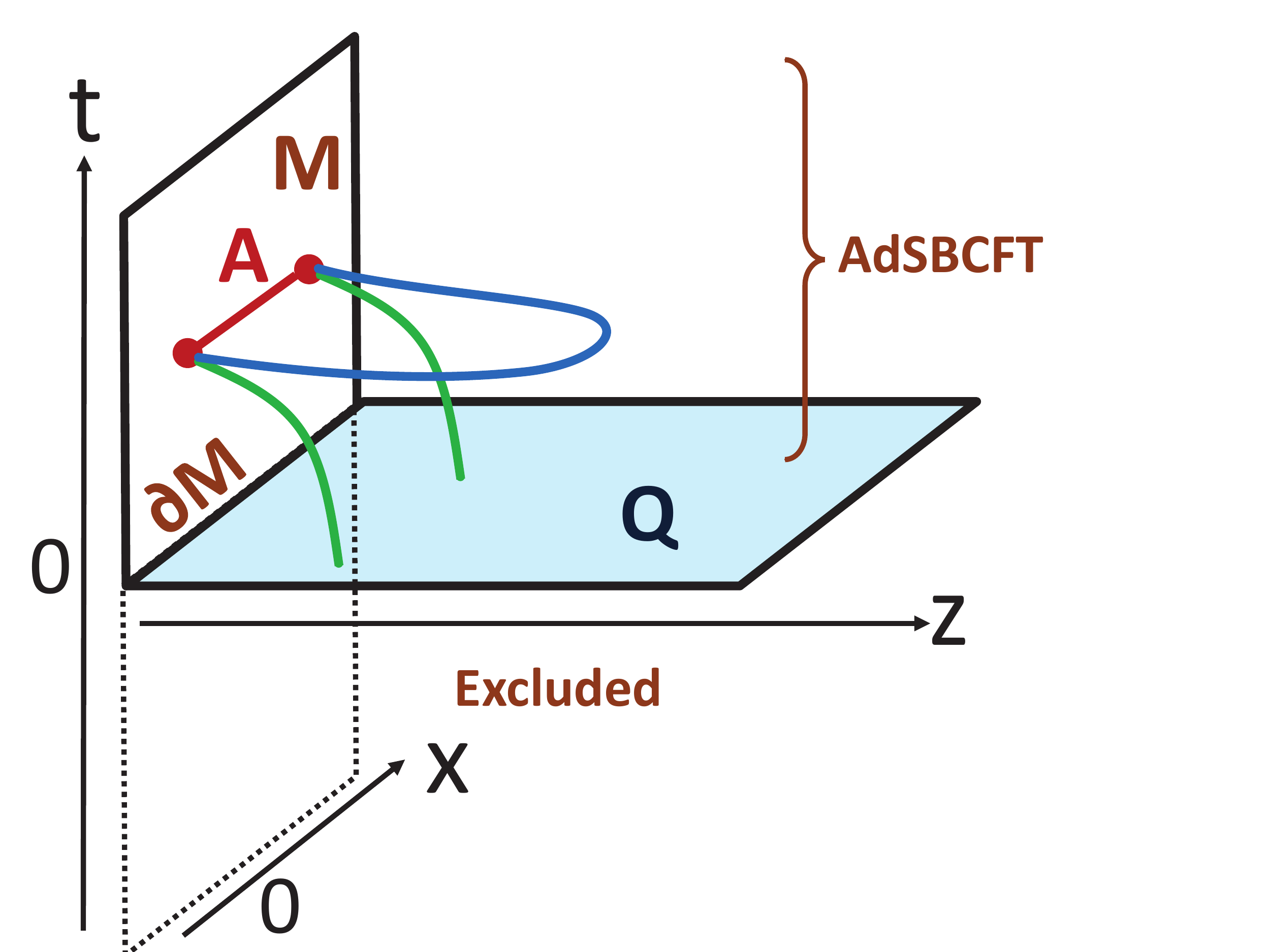}
 \caption{A sketch of AdS/BCFT analysis for AdS$_3$. A holographic CFT on $M$ (with the boundary $\de M$) is dual to gravity on $N$. The boundary of $N$ consists of the surface $Q$ and $M$. The right picture shows the calculation of holographic entanglement entropy. The blue curve gives the connected geodesic contribution and the green ones are the disconnected geodesics which end on the boundary surface $Q$. The correct holographic entanglement entropy is given by the one with a smaller length.}
\label{fig:AdSBCFT}
\end{figure}

Since our coming analysis of joining/splitting local quenches require us to consider a CFT, let us extend the AdS/CFT to a setup where a holographic CFT is defined on a manifold $M$ with boundaries $\de M$. In particular, when a linear combination of conformal symmetry is preserved on $\de M$, we call it a boundary conformal field theory (BCFT). The holographic dual of a BCFT, called AdS/BCFT, can be constructed in the following way \cite{AdSBCFT} (see also \cite{KR} for an earlier
argument). For a typical AdS/BCFT setup, refer to the left picture of Fig.\ref{fig:AdSBCFT}.
Consider a surface $Q$ which ends on
$\de M$ and extends into the bulk. We impose the following condition on $Q$
\be
K_{\mu\nu}-Kh_{\mu\nu}=-T_{BCFT}\cdot h_{\mu\nu}, \label{KT}
\ee
where $h_{\mu\nu}$ is the induced metric on $Q$ and $K_{\mu\nu}$ is the extrinsic curvature on $Q$;
$K$ is the trace $h^{\mu\nu}K_{\mu\nu}$. The constant $T_{BCFT}$ describes the tension of the `brane' $Q$
and can take both positive and negative values in general. This boundary condition (\ref{KT}) arises naturally in the AdS/BCFT setup as follows (for more detail refer to \cite{AdSBCFT}). If we consider the standard gravity action given by the Einstein-Hilbert action plus the Gibbons-Hawking boundary term, it is well-known that there are two boundaries conditions: Dirichlet and Neumann. For AdS/BCFT, we choose the Neumann 
boundary condition as we want to keep the boundary $Q$ dynamical as is so in the BCFT boundary. 
The Neumann boundary condition is given by  $K_{\mu\nu}-Kh_{\mu\nu}=0$. To generalize this boundary condition, we add the tension term of the surface $Q$ to the bulk action, given by $T_{BCFT}\int_Q \s{h}$.
This modified the Neumann boundary condition into the form (\ref{KT}). It is also useful to note that in explicit examples, we can confirm that this boundary condition (\ref{KT})  preserves the boundary conformal symmetry. 

The gravity dual of a CFT on $M$ is given by the AdS gravity solution restricted on the space $N$, defined by the bulk region surrounded by $M$ and $Q$. To find such a solution, we need to solve the Einstein equation with the boundary condition (\ref{KT}), where the presence of $Q$ gives back-reactions and modifies the bulk metric \cite{AdSBCFT,NTU,Chu}. In our examples which we will discuss later, we can analytically find gravity duals of two dimensional BCFTs by using the bulk extension of the conformal map (\ref{corads}).

The region $N$ gets larger as the tension $T_{BCFT}$ increases and this suggests that $T_{BCFT}$ estimates the degrees of freedom on the boundary $\de M$.
Indeed, as shown in \cite{AdSBCFT} in the AdS$_3$ case, the tension is monotonically related to the boundary entropy $S_{bdy}$ introduced in \cite{AFL} as follows:
\be
S_{bdy}=\frac{c}{6}~\mbox{arctanh}(T_{BCFT}). \label{tenrs}
\ee
For notational simplicity, we introduce a positive parameter $k(>0)$ by
\be
S_{bdy}\equiv\frac{c}{12}\log k.  \label{kdef}
\ee
A larger $k$ means a larger boundary entropy or tension $T_{BCFT}$. In particular, we have $k=1$ for $T_{BCFT}=S_{bdy}=0$. Refer to e.g. \cite{Tonni,KNSW} for studies of HEE in higher dimensional setups.

Later we will employ the AdS/BCFT to calculate the holographic entanglement entropy (HEE). Consider the holographic entanglement entropy $S_A$ for an interval $A$ in the AdS$_3$/BCFT$_2$ setup. As in the right picture, there are two possibility: connected one $S^{con}_A$ and the disconnected one $S^{dis}_{A}$.
The latter arises because the geodesic which connects the two end points of $A$ can end on the boundary surface $Q$ in the middle as depicted in the right picture of Fig.\ref{fig:AdSBCFT}.
 The correct holographic entanglement entropy is given by the one with a smaller length.

The connected HEE $S^{con}_A$ is not affected by the boundary surface $Q$. Therefore, this simply 
agrees with the estimation (\ref{EEC}) of the CFT without boundary. On the other hand, the disconnected 
HEE  $S^{dis}_{A}$ is highly affected by the boundary $Q$ as the geodesic ends on $Q$. In the Poincare AdS
setup,   $S^{dis}_{A}$ in Fig.\ref{fig:AdSBCFT}  is computed as (refer to \cite{AdSBCFT} for the derivation):
\ba
S^{dis}_{A}=\frac{c}{6}\log\left(\frac{2s_a}{\ti{\ep}_a}\right)+\frac{c}{6}\log\left(\frac{2s_b}{\ti{\ep}_b}\right)
+2S_{bdy}, \label{disAdS}
\ea
where $s_{a,b}$ are the distance between the surface $Q$ and the two end points of the interval 
$A$; $\ti{\ep}_{a,b}$ are the UV cut off in the Poincare AdS at the two boundary points. 
Intuitively, as the surface $Q$ gets further from the subsystem $A$, then geodesic length (i.e. HEE) gets larger. Thus as the tensor $T_{BCFT}$ gets larger, the HEE gets larger.  This effect is universally 
described by the constant term proportional to $S_{bdy}$ in  (\ref{disAdS}) as follows form the 
result in \cite{AdSBCFT}. Note that this is the sum of two disconnected geodesics and therefore has the doubled contribution of $S_{bdy}$. 

We can also understand this from the CFT viewpoints. In the holographic CFTs (or large central charge 
CFTs with sparse spectrum). As in the standard large $c$ arguments \cite{Asplund:2014coa}, we can approximate a correlation function by a semiclassical saddle point. In our setup with a boundary, there 
are two saddles. One of them is obtained by contracting two twist operators, which give the connected 
EE $S^{con}_A$. The other is given by contracting each of them with its mirror operator 
across the boundary, which leads to the disconnected one  $S^{dis}_{A}$. Both of them agree 
with the results from AdS/BCFT.  The $S_{bdy}$ dependence in (\ref{disAdS}) occurs because the branch cut 
which extends from a twist operator ends on the boundary as is already known in the standard calculation 
of EE in BCFT \cite{CC}.

 More generally, including higher dimensional setups, we can calculate the holographic entanglement entropy in the following way. In the standard holographic entanglement entropy without any boundaries,
$S_A$ is given by the area of minimal or extremal surface $\Gamma_A$ which ends on $\de A$ and which is homologous to $A$. In the presence of boundary surfaces $Q$, we impose the homology condition by regarding the surfaces $Q$ as trivial spaces with the zero size.

\section{Entanglement Density}\label{sec:ED}

Here we study a quantity called entanglement density (ED) introduced in \cite{HLQ}.
First we would like to note that this is not a new quantity as it follows from the values of the entanglement 
entropy (EE). However, the ED provides a helpful way to display the behavior of EE in complicated 
systems, because we can capture the essence of time evolution from the 
behavior of ED as we explain later.
 
The ED is defined from the data of EE in a 2d CFT when the subsystem $A$ is an interval.
Since we have in mind generic excited states, we do not require the translational symmetry.
When $A$ is given by an interval $a\leq x\leq b$ at a fixed time $t$, the EE is written as
$S_A(a,b,t)$ (let us assume $a<b$). The entanglement density $n(a,b,t)$ is defined by
\ba
n(a,b,t)=\frac{1}{2}\frac{\de^2 S_A(a,b,t)}{\de a \de b}=\frac{1}{2}\left(\frac{1}{4}\frac{\de^2}{\de\xi^2}-\frac{\de^2}{\de l^2}\right)S_A(\xi,l,t),
\label{delk}
\ea
where we introduced the center of the interval\footnote{In this paper, $\xi$ is also used to denote another quantity, the coordinate in AdS/BCFT given by (\ref{CF}) and (\ref{corads}). However, since the analysis of ED and discussions in AdS/BCFT do not appear at the same time, its meaning can be easily figured out from the context.} $\xi$ and length $l$ such that
\be
a=\xi-\frac{l}{2},\ \ \ b=\xi+\frac{l}{2}.
\ee
For example, the ED for a CFT vacuum, denoted by $n_0(\xi,l,t)$, takes the universal form:
\be
n_0(\xi,l,t)=\frac{c}{6l^2}. \label{ved}
\ee

By definition, the EE is represented as a double integral of ED:
\ba
S_A(a,b,t)=\left(\int^a_{-\infty} dx \int^{b}_a dy+\int^\infty_{b} dy \int^{b}_a dx \right) n(x,y,t).
\ea
Therefore, if we assume all entanglement comes from the bipartite quantum correlation, the ED measures
the number of EPR pairs between $x=a$ and $x=b$. However we should note that this interpretation is too naive as in general we have multi-partite entanglement in QFTs. This issues becomes serious in the later time behavior of holographic local operator quench \cite{NNT}, as pointed out in \cite{AB}. On the other hand, for the massless Dirac fermion CFT, this interpretation works so well that we can reproduce correct results from the universal ED (\ref{ved}) even when $A$ consists of multiple disconnected intervals as we will explain in the appendix \ref{dfee}.
One of the purposes of the present paper is to emphasize that the entanglement density at least provides a useful and simple way to extract the essential behaviors of the entanglement entropy in a systematical way. We will see this in many examples.

We are especially interested in the difference
\be
\Delta n(\xi,l,t)=n(\xi,l,t)-n_0(\xi,l,t),
\ee
where $n_0(\xi,l,t)$ is given by (\ref{ved}).
In other words, $\Delta n(\xi,l,t)$ is the entanglement density for the entanglement entropy growth
\be
\Delta S_A=S_A-S_{0A},
\ee
where $S_{0A}$ is the entanglement entropy for the ground state.

\subsection{Properties}

As already noted in  \cite{HLQ},  the growth of ED $\Delta n(\xi,l,t)$ enjoys several interesting properties. First of all, owing to the first law of entanglement entropy \cite{Bhattacharya:2012mi,Blanco:2013joa,Wong:2013gua}, in the small size limit
$l\to 0$, we have
\be
\Delta n(\xi,0,t)=-\frac{\pi}{3}T_{tt}(\xi,t), \label{firstq}
\ee
where $T_{tt}$ is the energy density.

Moreover, the growth of ED satisfies a sum rule: $\int da \int db \Delta n(a,b,t)=0$ if
the total state is pure. In \cite{HLQ}, this was proved when we impose the periodic boundary condition, which identifies $x=a$ with $x=b$. More generally, we can prove this even when the total space is an interval $0\leq x\leq L$ as we will explain below. The precise statement of the sum rule is
\ba
\int^L_0 db \int^b_0 da\ \Delta n(a,b,t)=0,  \label{sum}
\ea
at any time $t$. To prove this, we can rewrite it as follows
\ba
\int^L_0 db \int^b_0 da\ \frac{1}{2} \de_a\de_b \Delta S_A(a,b,t)=\int^L_0 db\ \frac{1}{2}\Big[\de_b \Delta S_A(a,b,t)|_{a=b}-\de_b \Delta S_A(a,b,t)|_{a=0} \Big].
\ea
First of all, the quantity $\de_b \Delta S_A|_{a=b}$ is vanishing because the first law (\ref{firstq}) tells us the behavior $\Delta S_A\propto (b-a)^2$ when $|b-a|$ is very small. The second term in the right hand side also follows because
\be
\int^L_0 db\ \de_b \Delta S_A(a,b,t)|_{a=0}=\Delta S_A(a,b,t)|_{b=L,\ a=0}-\Delta S_A(a,b,t)|_{b=0,\ a=0}=0,
\ee
where we employed the first law behavior and the pure state property $S_A=S_{A^C}$ ($A^C$ is the complement of $A$).

It is also intriguing to note that we can also define the entanglement density for R\'{e}nyi entanglement entropy
(\ref{REE}). Since this quantity also has the property $\Delta S^{(n)}_A\propto (b-a)^2$ when $|b-a|$ is very small,
the R\'{e}nyi entanglement density $\Delta n^{(n)}(a,b,t)$ also satisfies the sum rule (\ref{sum}).

\subsection{Example1: Global Quenches}

One of the simplest but non-trivial examples of homogeneous excited states in CFTs  is the global quenches.
This is triggered by a sudden change of the Hamiltonian from a gapped one to a critical one $H$ at a specific time $t=0$. As argued in \cite{CCG}, we can model this process by approximating the state just after the quench by a regularized boundary state $|\Psi(t=0)\lb=e^{-\ap H}|B\lb$, where $\ap$ is an infinitesimally small parameter of the  regularization and $|B\lb$ is a boundary state \cite{Ca,Is}.
In general, The EE of a subsystem with length $l$ shows the linear growth and the saturation \cite{CCG}
\ba
&& \Delta S_A=\frac{\pi c}{6\ap}t \ \ \ (0<t\leq l/2), \no
&& \Delta S_A=\frac{\pi c}{12\ap}l \ \ \ (t> l/2).
\ea
This linear growth can be explained by the entangled pair creations and their relativistic propagations \cite{CCG}.

In this case the ED looks like
\ba
\Delta n(\xi,l,t)=\frac{\pi c}{24\ap} \delta(l-2t)+\Delta n(\xi,l,t)_{UV}. \label{bfv}
\ea
The UV contribution $\Delta n(\xi,l,t)_{UV}\sim O(T_{tt})$
is localized at short distances $l\leq \ap$ such that it obeys the sum rule (\ref{sum}).

\begin{figure}
  \centering
  \includegraphics[width=7cm]{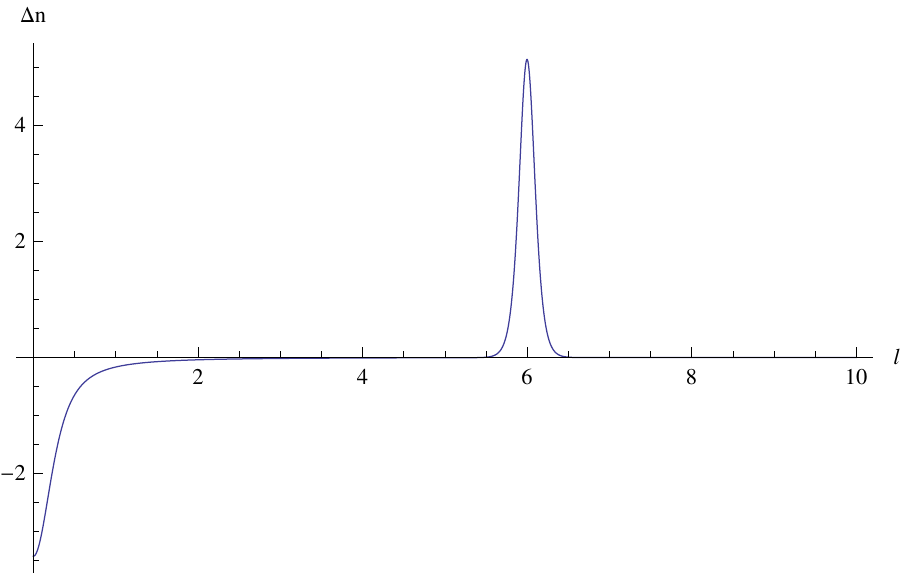}
 \includegraphics[width=7cm]{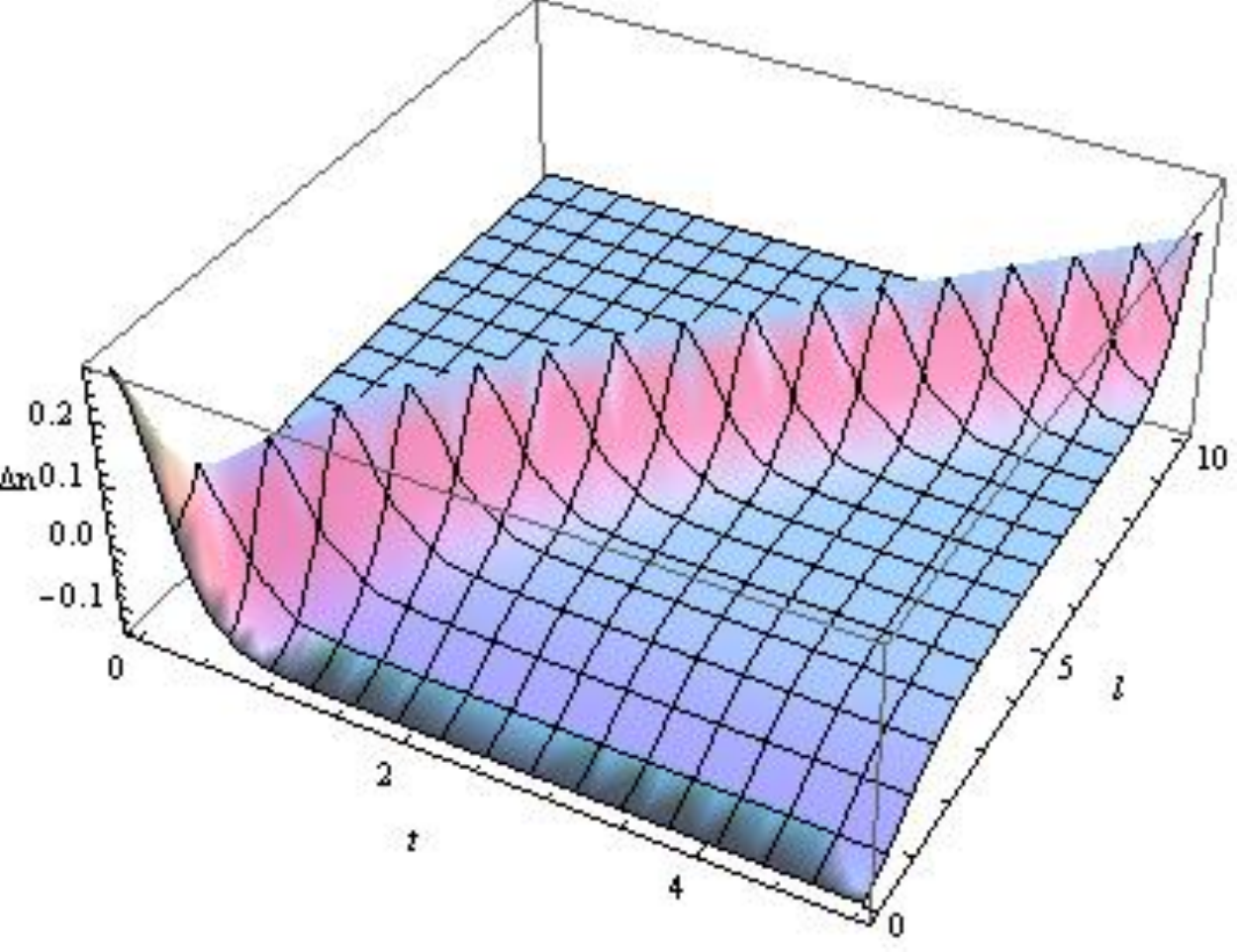}
 \caption{The behavior of entanglement density (ED) under the global quenches.
 The left graph describes the ED $\Delta n(l,t)$ as a function of $l$ at
 $t=3$ (we chose $\ap=0.1$). The right 3D plot shows the behavior of ED $\Delta n(l,t)$
 as a function of the time $t$ (horizontal axis) and the subsystem size $l$ (depth axis), where we set
 $\ap=0.5$.}
\label{fig:GLplot}
\end{figure}

If we consider the massless free Dirac fermion CFT in two dimension ($c=1$) as a solvable example, we explicitly find the following expression of the EE\footnote{
This is found by employing the conformal map $\xi=\exp\Big(\frac{\pi w}{2\ep}\Big)$ and the expression (\ref{saf}) and (\ref{diracf}). The same expression was obtained in \cite{CCG}. For the boundary state we can choose either Dirichlet or Neumann boundary condition for the scalar field obtained from the bosonization,
both of which lead to the same entanglement entropy by an appropriate choice of twist operators
as shown in \cite{UT}.}
\ba
S_A=\frac{1}{6}\log \left[\frac{4\ap}{\pi\ep}\cdot \frac{\cosh^2\left(\frac{\pi t}{2\ap}\right)
\sinh^2\left(\frac{\pi l}{4\ap}\right)}{\cosh\left(\frac{\pi}{2\ap}(t+l/2)\right)\cosh\left(\frac{\pi}{2\ap}(t-l/2)\right)}\right].
\ea
Thus, the ED reads
\ba
\Delta n(l,t)=-\frac{1}{6l^2}+\frac{\pi^2}{192\ap^2}\left(\frac{2}{\sinh^2\left(\frac{\pi l}{4\ap}\right)}+\frac{1}{\cosh^2\left(\frac{\pi}{2\ap}(t-l/2)\right)}
+\frac{1}{\cosh^2\left(\frac{\pi}{2\ap}(t+l/2)\right)}\right).
\ea
Indeed we can confirm  the behavior (\ref{bfv}) in this explicit example. We numerically plotted
the profile of  $\Delta n(l,t)$ in Fig.\ref{fig:GLplot}. We can also confirm that the sum rule
$\int^\infty_0 dl \Delta n(l,t)=0$, explicitly. Similar arguments can also be applied to holographic
calculations of global quenches \cite{AAL,Ba,HaMa,ABGH}.

\subsection{Example2: Local Operator Quenches}\label{sec:loq}

Next we consider the time evolution of an excited state produced by a local insertion of a primary operator $O$ at $x=0$ at the time $t=0$ \cite{NNT,Nozaki:2014uaa}:
\ba
|\Psi(t)\lb={\ca N_O} e^{-iHt}\cdot e^{-\ap H}O(0)|0\lb,  \label{lopqs}
\ea
where $\ap$ is again an infinitesimally small regularization parameter; ${\cal N_O}$ is the normalization factor so that the state has the unit norm.
 We call its time evolution a local operator quench (refer to the left picture of Fig.\ref{fig:SPPS}). We will discuss other types of local quenches (splitting and joining ones) in later sections, which are the main setups we consider in this paper. However, here we briefly discuss the operator quenches because
 they are instructive for our later arguments and the behaviors of their ED have not been studied well before.

Let us focus on two dimensional free CFTs or more generally rational CFTs (RCFTs) such as free scalar fields or minimal models, for simplicity.\footnote{Refer also to the papers \cite{He:2014mwa,Caputa:2014eta,deBoer:2014sna,Guo:2015uwa,Chen:2015usa,Nozaki:2015mca,Caputa:2015tua,
Caputa:2015waa,Rangamani:2015agy,Sivaramakrishnan:2016qdv,Caputa:2016yzn,Numasawa:2016kmo,
Nozaki:2016mcy,David:2016pzn,Caputa:2017tju,Nozaki:2017hby,JaTa,He:2017lrg,KuTa,Ku} for further related calculations.}
We again choose the subsystem $A$ to be the interval $a\leq x \leq b$.
As found in \cite{NNT,Nozaki:2014uaa,He:2014mwa}, the evolution of (R\'{e}nyi) EE looks like (in the limit $\ap\to 0$)
\ba
&& \Delta S^{(n)}_A=0 \ \ \ (0<t<|a|,\ \ t>b), \no
&& \Delta S^{(n)}_A=\log d_O \ \ \ (|a|<t<b), \label{RCFT}
\ea
for any $n$, including the EE $n=1$.
Here we assumed $b>|a|$. The quantity $d_O(\geq 1)$ is called the quantum dimension, which is a finite constant. This behavior is obvious from the picture that the operator insertion creates an entangled pair at $x=0$ and then it splits into the left and right moving quanta, which are entangled with each other as argued in \cite{NNT,Nozaki:2014uaa,He:2014mwa}.

The entanglement density (ED) is found to be\footnote{To see this, note that for any $a$, $b$, we can write
\be
\Delta S_A=\log d_O\cdot \left(\theta(|b|-t)\theta(t-|a|)+\theta(t-|b|)\theta(|a|-t)\right),
\ee
by using the Heaviside step function $\theta(x)$.}
\ba
\Delta n(a,b,t)=-\log d_O\cdot {\rm sgn}(ab)\cdot \delta(t-|a|)\delta(t-|b|)
\label{OLQa}
\ea

\begin{figure}
  \centering
 \includegraphics[width=7cm]{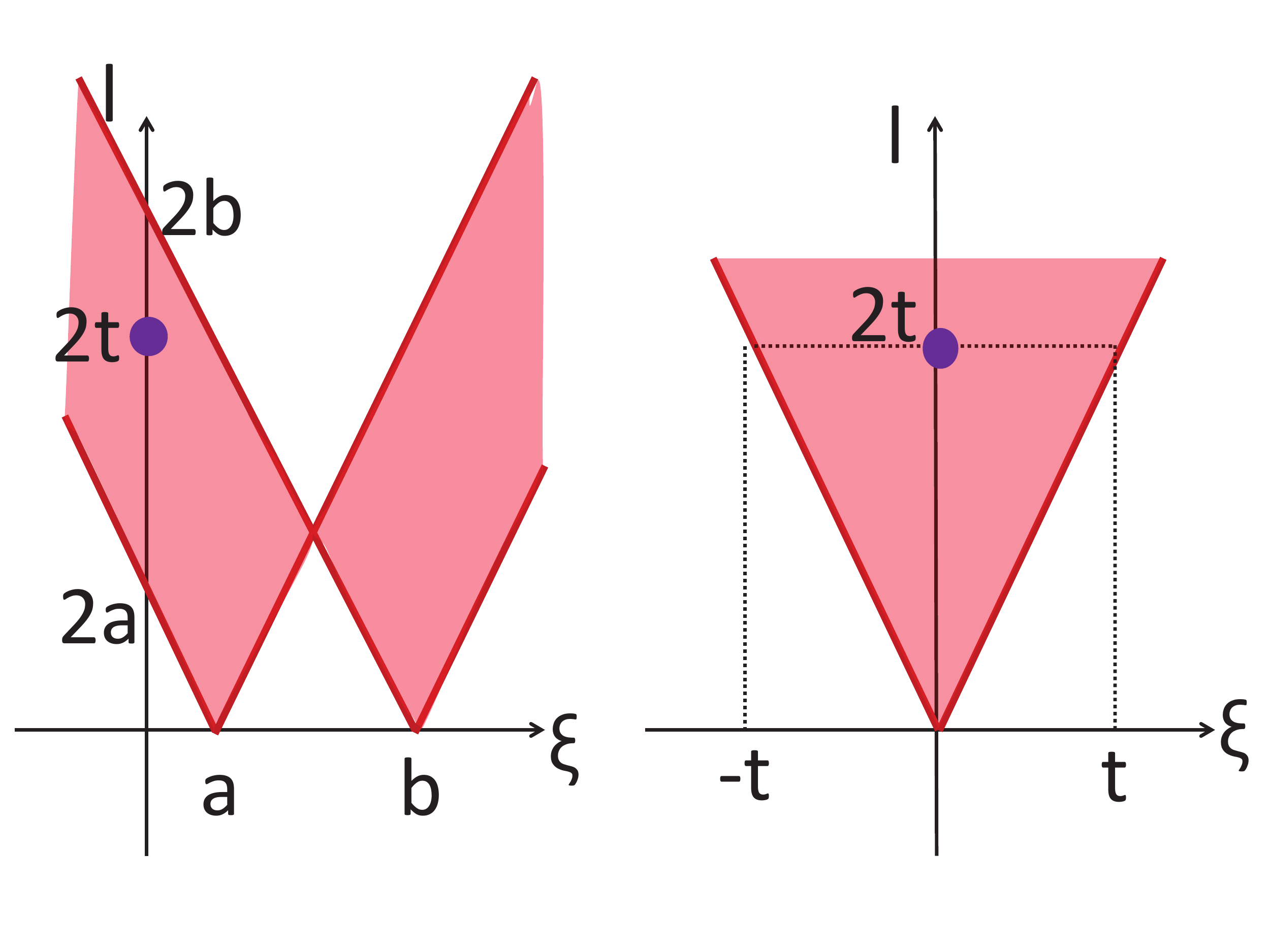}
  \includegraphics[width=7cm]{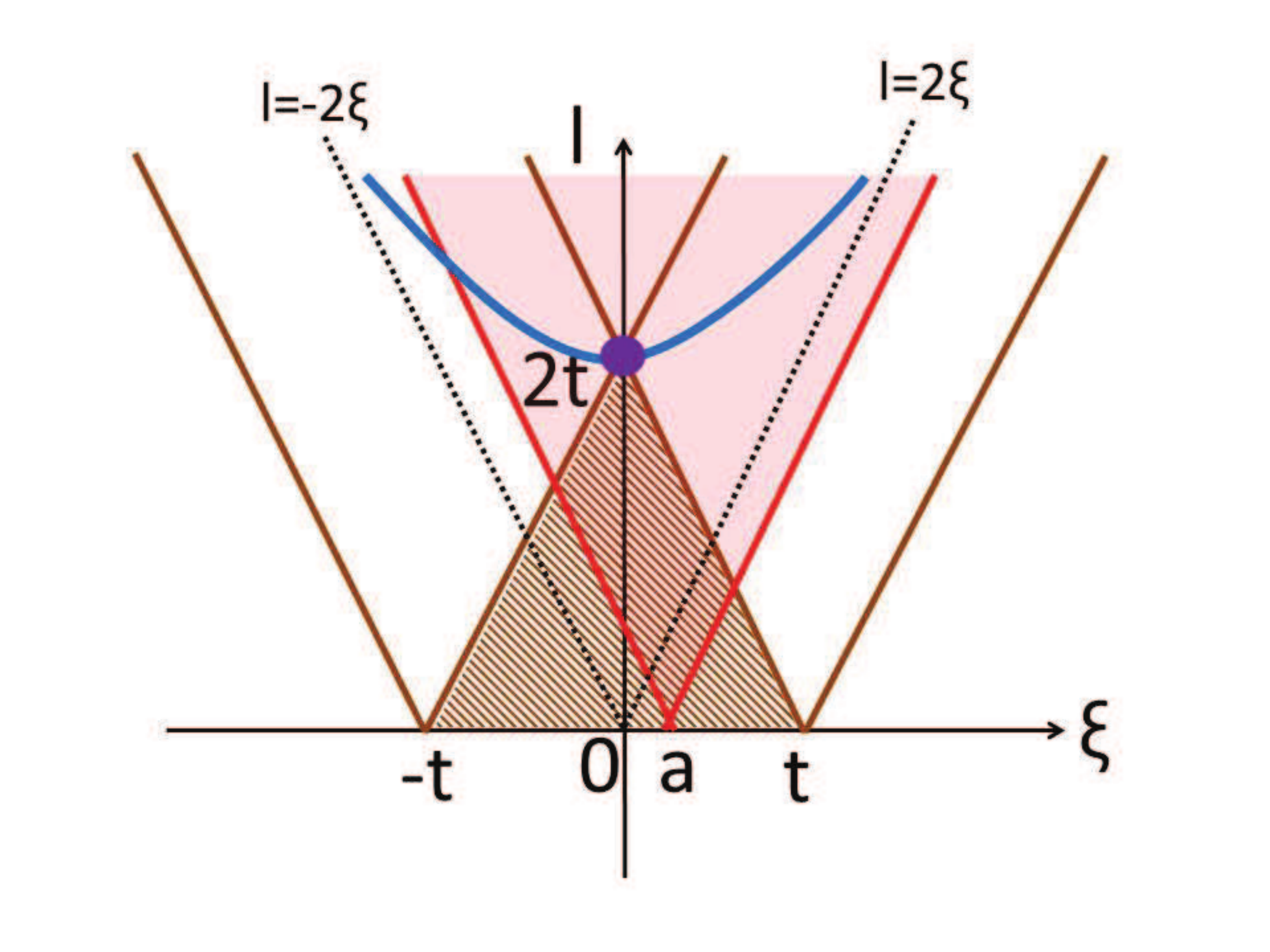}
 \caption{The integration range $\int dl d\xi$ of the ED (red regions) and causality
range (brown shaded region). The left picture shows the region of integration which computes $S_A$ for
 the subsystems $A=[a,b]$. The middle one shows the same one when $A$ is the half line: $a=0$ and $b=\infty$. The brown shaded region in the right picture describe the range of $(\xi,l)$ where the local excitation at $x=t=0$  can make any physical influence assuming causality. In all these three pictures, the purple dot at $(\xi,l)=(0,2t)$ represents the peak of ED due to the local quench. In the right picture, the blue curve near this dot (given by $l=2\s{\xi^2+t^2}$)
 describes the delta functional peaks
 which are peculiar to holographic local quenches. Note that they are outside of the causality region.
 The calculations of $S_A$ for $a\ll t\ll b$ correspond to integrating the red region in the
 right picture. We note the left part of the blue curve gradually gets into the red region, which gives the logarithmic growth of EE.}
\label{fig:Integ}
\end{figure}

It is also straightforward to reproduce the behavior (\ref{RCFT}) by integrating this entanglement density over a suitable region of $(\xi,l)$ as we explain in the left and middle pictures of Fig.\ref{fig:Integ}.\\

Let us confirm this in an explicit example of a massless free scalar field $\phi$.
To obtain an analytical expression, we calculate the growth of 2nd R\'{e}nyi entanglement entropy $\Delta S^{(2)}_A$. We choose the operator $O$ to be $O=e^{i\phi}+e^{-i\phi}$. In this case we have $d_O=2$ and
$\Delta S_A=\log 2$ for $a<t<b$. This is simply because this operator creates the genuine Bell pair
\cite{NNT,Nozaki:2014uaa,He:2014mwa}.  The explicit form of $\Delta S^{(2)}_A$ was obtained in \cite{He:2014mwa} and is given by
\ba
\Delta S^{(2)}_A=\log\left( \frac{2}{1+|z|+|1-z|}\right),
\ea
where $z=\frac{(z_1-z_2)(z_3-z_4)}{(z_1-z_3)(z_2-z_4)}$ is the cross ratio of the four locations:
\be
z_1=-z_3=\s{\frac{a-t-i\ap}{b-t-i\ap}},\ \ \ z_2=-z_4=\s{\frac{a-t+i\ap}{b-t+i\ap}}.
\ee
Indeed, in the limit $\ap\to 0$, we can confirm $(z,\bar{z})\to (0,0)$ for the regions $0<t<|a|$ and $t>b$ , while we have $(z,\bar{z})\to (1,0)$ when $|a|<t<b$. This leads to the result (\ref{RCFT}).

\begin{figure}
  \centering
   \includegraphics[width=7cm]{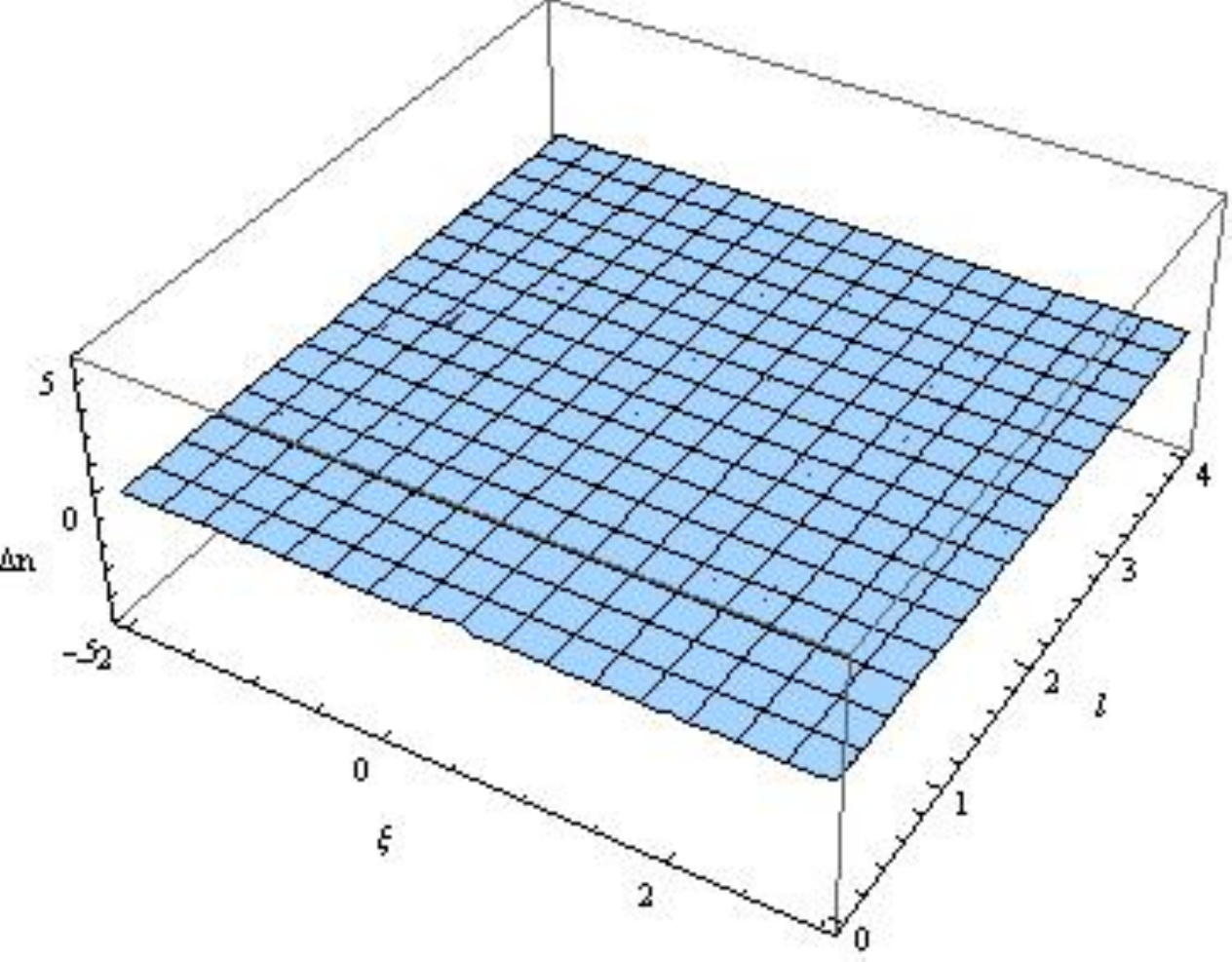}
  \includegraphics[width=7cm]{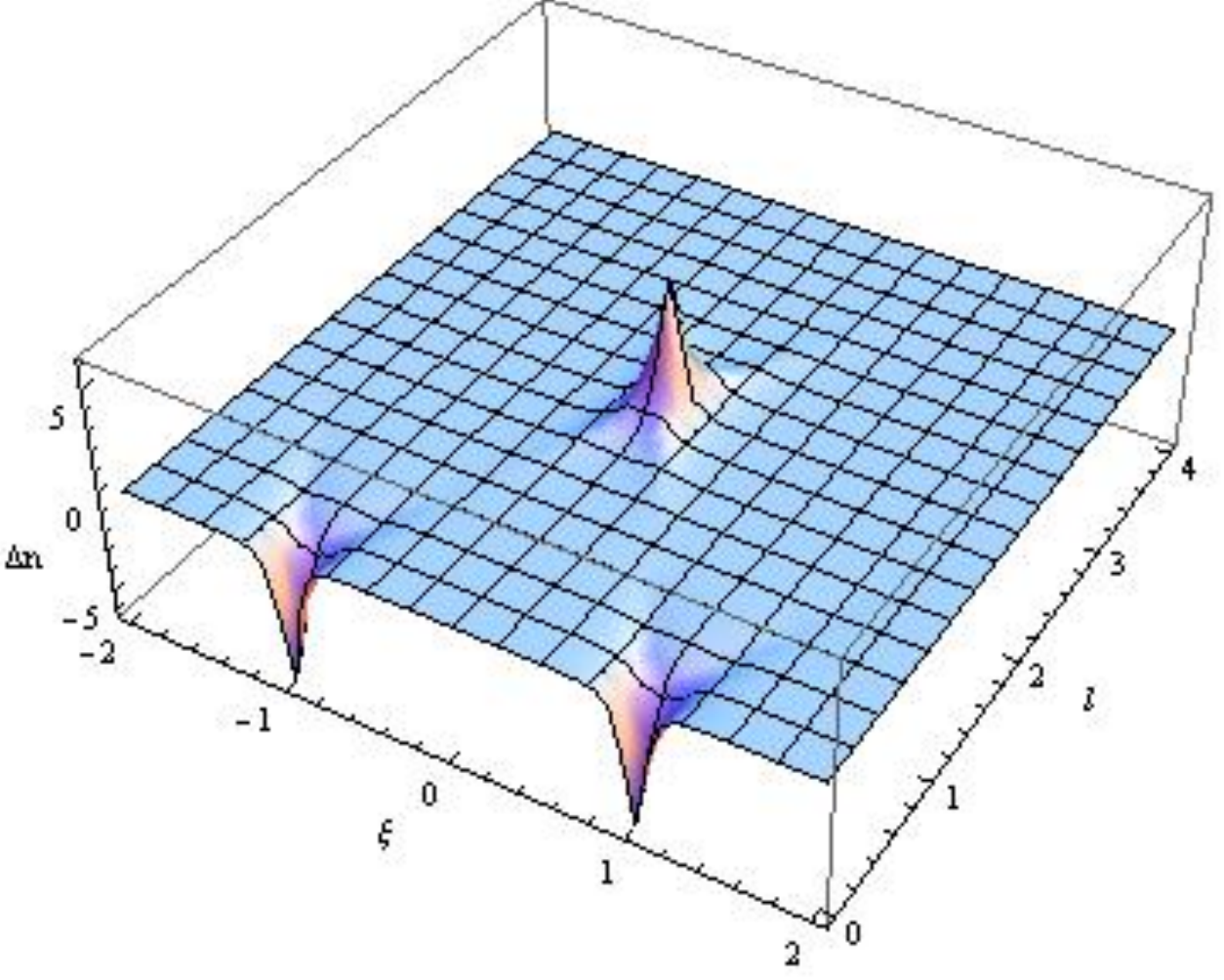}
 \caption{The profile of entanglement density for the 2nd Renyi entropy under the local operator quench in $c=1$ free scalar CFT at the time $t=0$ (left) and $t=1$ (right) with $\ap=0.1$, trigger by the operator $O=e^{i\phi}+e^{-i\phi}$. The horizontal and depth coordinate correspond to $\xi$ and $l$, respectively.}
\label{fig:OpEDEN}
\end{figure}

We plotted the ED obtained from this R\'{e}nyi entropy in Fig.\ref{fig:OpEDEN} for $\ap=0.1$. Indeed, we can numerically confirm the behavior (\ref{OLQa}) as well as the sum rule.
It is also useful to note that the ED is vanishing completely at $t=0$ as the positive and negative delta-functional peaks coincide and cancel each other.

\vspace{5mm}
{\bf Holographic Case}

One may wonder if a similar result is true for local operator quenches in holographic CFTs.
A holographic calculation for the operator local quench was given in \cite{HLQ} (see also \cite{Caputa:2014vaa}) and its entanglement entropy was reproduced from CFT analysis in \cite{Asplund:2014coa}. Especially, the resulting entanglement entropy shows a logarithmic growth
\ba
S_A\simeq \frac{c}{6}\log \frac{t}{\ap} +\frac{c}{3}\log \frac{l}{\ep}, \label{oplog}
\ea
when $|a|\ll t\ll b$. We will explain the geometric explanation in gravity dual in section \ref{sec:log}.

This logarithmic growth is clearly different from the previous RCFT case (\ref{RCFT}). This looks puzzling because the relativistic propagation picture leads to the behavior (\ref{RCFT}) and there seems to be no way to explain the slow logarithmic growth (\ref{oplog}) in relativistic theories.

Interestingly, we can resolve this puzzle by looking at the ED instead of EE.
Our argument here will be very brief because we will study this more closely in the joining local quench model, which has the similarity on this aspect. The plot of ED shows that the peak at $(\xi,l)=(2t,0)$ is not localized as opposed to the Dirac fermion case, but it is continuously distributed on a curve, which is depicted as the blue curve in the right picture of Fig.\ref{fig:Integ}. We observe that
the left part of peak curve (blue) enters into the integration region (red) as time evolves.
Thus, we find that $S_A$ increases gradually under the time evolution. This qualitatively explains the behavior (\ref{oplog}).

Note that the continuous peaks are outside of the region which we expect from the causality propagation of the local excitation inserted at $x=t=0$. The holographic analysis in \cite{HLQ} shows they are on
the curve $l=2\s{\xi^2+t^2}$. In particular, at $t=0$, the peaks are distributed on $l=2|\xi|$.
This is clear different from the RCFT examples, where the ED vanishes everywhere at $t=0$.
From this, we learn that in holographic CFTs, the operator local quench (\ref{lopqs}) initially generates highly non-local entanglement, as opposed to what we expect from the naive causality argument. The relativistic propagation of this non-local entanglement is the reason why we find the logarithmic growth (\ref{oplog}). This property is also expected from the fact that the insertion of local operator is not a local unitary transformation \cite{Sivaramakrishnan:2016qdv}. Note that this initial non-local entanglement can be negligible in the operator local quench in RCFTs, whose ED only has the localized peak at $(\xi,l)=(2t,0)$.

\section{Splitting Local Quench}\label{sec:splitting}

\begin{figure}
  \centering
  \includegraphics[width=7cm]{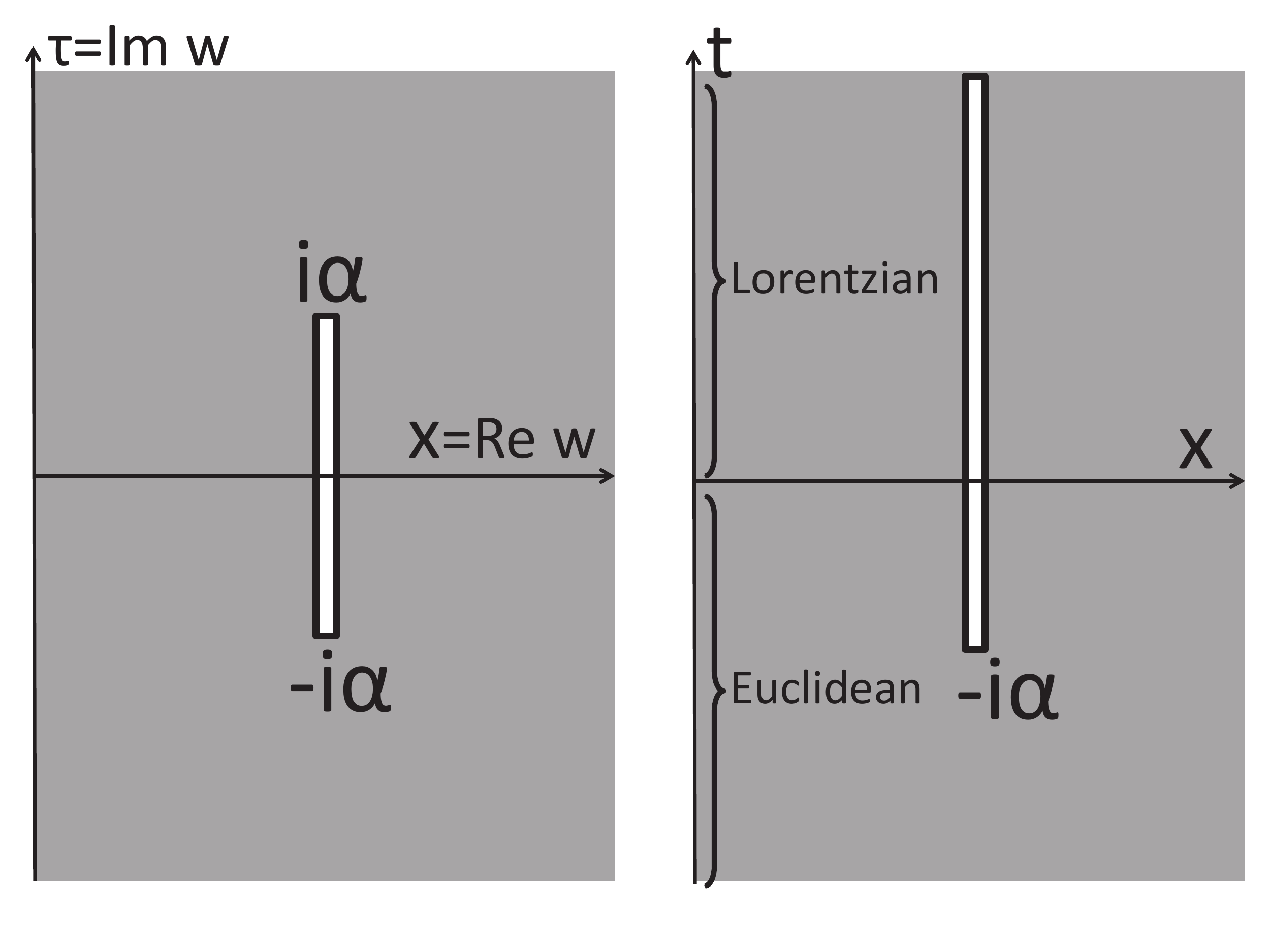}
 \caption{The geometries which realize the splitting local quenches. The left figure describes the space on which we perform the Euclidean path-integral. This is mapped into an upper half plane by the map
 (\ref{fwsp}). The right picture describes the path-integral realization of the time evolution after
 the split process happened at $t=0$, where the Euclidean path-integral for $\tau<0$ creates the state just after the splitting process.}
\label{fig:SPw}
\end{figure}

Now we move on to the main analysis of this paper. Consider the splitting process in a 2d CFT. Namely, we start with a 2d CFT on a connected line $-\infty < x < \infty$ and at $t=0$ we cut it into two halves at $x=0$, as sketched in the middle picture of Fig.\ref{fig:SPPS}.
The preparation of the quantum state at $t=0$ just after the splitting process can be done by considering the Euclidean path-integral as in Fig.\ref{fig:SPw}. The parameter $\ap$ corresponds to the regularization of local quench. This setup is described by the conformal transformation:
\ba
\xi=i\s{\f{w+i\ap}{w-i\ap}}\equiv f(w), \label{fwsp}
\ea
which maps the Euclidean geometry with the cut: $-\ap< \mbox{Im}[w]< \ap$ and Re$[w]=0$
(the left picture of Fig.\ref{fig:SPw}) into an upper half plane Im$[\xi]\geq 0$.

\subsection{Splitting Local Quench in Dirac Free Fermion CFT}

One example where we can calculate the time evolution of EE under the splitting quench is
the massless Dirac fermion CFT in two dimensions ($c=1$).
We choose the subsystem $A$ to be an interval $a\leq x\leq b$, or equally $l=b-a$ and $\xi=(a+b)/2$.
Due to the parity symmetry $x\to -x$, we can assume
\be
|a|<b,\ \ \  b>0,  \label{asu}
 \ee
without losing generality. In the rest of this paper, we always focus on the entanglement entropy $S_A$ for this interval.

In the Dirac free fermion CFT, the twist operator $\sigma_n$ is described by a bosonization
$\sigma_n=e^{\frac{i}{n}X}$, where $X$ is the free massless scalar field dual to the Dirac 
fermion $\psi$ via the standard relation $\psi=e^{iX}$. Since we know the analytical form of 
correlation functions of free scalar on an upper half plane, we can calculate the two point function of 
the twist operators even in the presence of the boundary 
as follows (this is identical to eq.(2.31) of \cite{NSTW} see also \cite{CH,ANT,UT}) :
\ba
S_A=-\frac{1}{6}\log (F\ep^2), \label{saf}
\ea
where $F$ is given by
\ba
F=\left|\frac{d\xi_a}{dw_a}\right|\cdot\left|\frac{d\xi_b}{dw_b}\right|\cdot
\frac{(\xi_a-\bar{\xi}_b)(\bar{\xi}_a-\xi_b)}{|\xi_a-\bar{\xi}_a||\xi_b-\bar{\xi}_b|
(\xi_a-\xi_b)(\bar{\xi}_a-\bar{\xi}_b)}.  \label{diracf}
\ea

The analytical expressions of $S_A$ when $\ap$ is infinitesimally small are given as follows.
In the early time period, $0<t<|a|$, we have
\ba
 S_A&=&\frac{1}{3}\log (b-a)/\ep.  \label{saa}
\ea
When $|a|<t<b$, we have
\ba
S_A=\frac{1}{6}\log \frac{4|a|(b-a)(t-|a|)(b^2-t^2)}{(a+b)(t+|a|)\ap\ep^2}. \label{sab}
\ea
At late time, $t>b$, we have
\ba
S_A=\frac{1}{6}\log \frac{4|a|b(b-|a|)^2}{(b+a)^2\ep^2}. \label{sac}
\ea

\begin{figure}
  \centering
  \includegraphics[width=5cm]{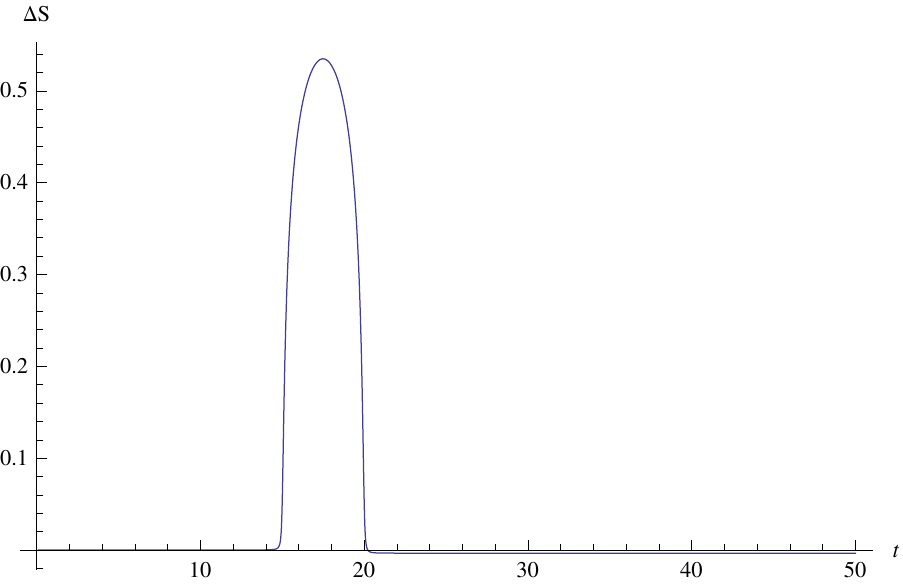}
  \includegraphics[width=5cm]{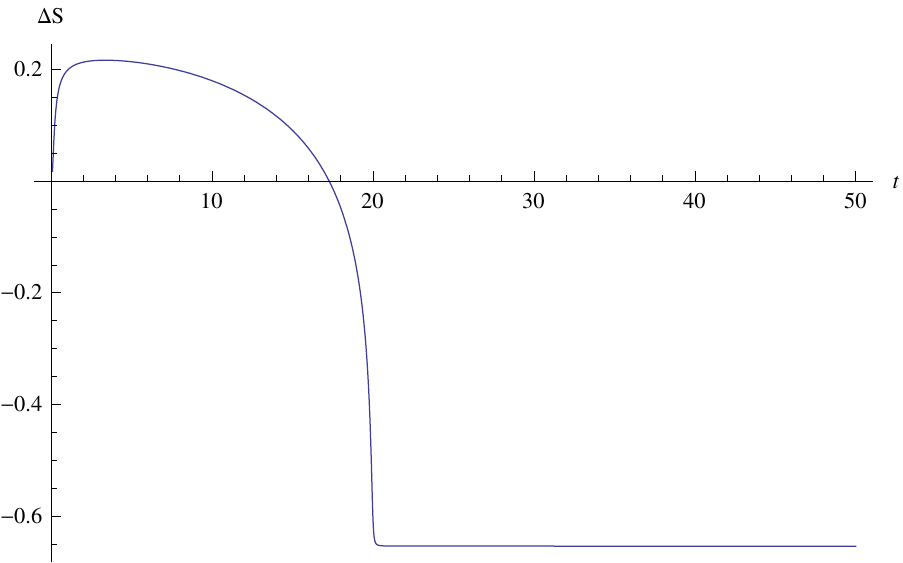}
  \includegraphics[width=5cm]{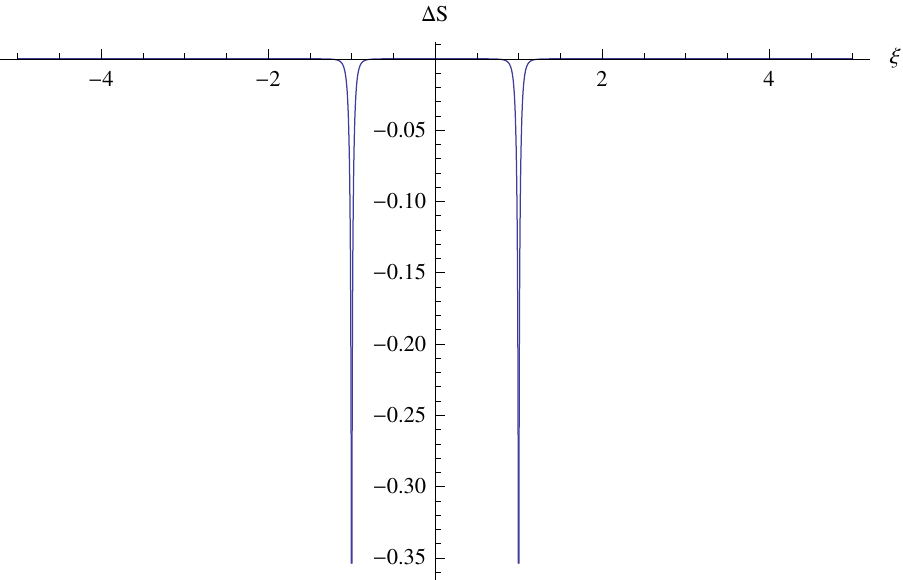}\\
 \caption{The plots of the entanglement entropy growth $S_A-S^{(0)}_A$ for $\ap=0.1$.
 The left graph describes the time evolution when we take $(a,b)=(15,20)$. The middle one
 is the time evolution for  $(a,b)=(0.1,20)$. The right one
 is the plot for various $\xi$ when we fix $l=2$ and $t=0$.}
\label{fig:DLQ}
\end{figure}

First, from (\ref{saa}),(\ref{sab}) and (\ref{sac}), we find that when the subsystem $A$ is symmetric
around the origin i.e. $\xi=0$, the EE coincides with that for the ground state $S^{(0)}_A$:
\be
S_A(0,l,t)=S^{(0)}_A=\frac{1}{3}\log \frac{l}{\ep}.
\ee

We also plotted the numerical values of EE in Fig.\ref{fig:DLQ}. When $\xi\neq 0$, we find $S_A-S^{(0)}_A$ gets non-trivially positive only for the period $|a| < t < b$.
This agrees with the relativistic particle propagation picture of free massless Dirac fermion CFT.
For $0 < t < |a|$, the EE is the same as that for the vacuum.
At the late time region $t > b$, it approaches to the EE for the ground state of the separated system. This is manifest in the left picture in Fig.\ref{fig:DLQ}.
The middle picture in Fig.\ref{fig:DLQ} shows that when we take $|a|\ll b$, the EE is gradually
decreasing just after the sudden initial rise. Indeed, from (\ref{sab}), we find the analytical
profile for $|a|\ll t<b$:
\be
S_A\simeq \frac{1}{6}\log\frac{4|a|(b^2-t^2)}{\ap\ep^2}.  \label{splog}
\ee

The right picture in Fig.\ref{fig:DLQ} shows that the EE is reduced when either end point of $A$
gets closer to $x = 0$, i.e. at the splitting point. This is because in this case, the EE only
comes from that for a half-line.

\begin{figure}
  \centering
   \includegraphics[width=6cm]{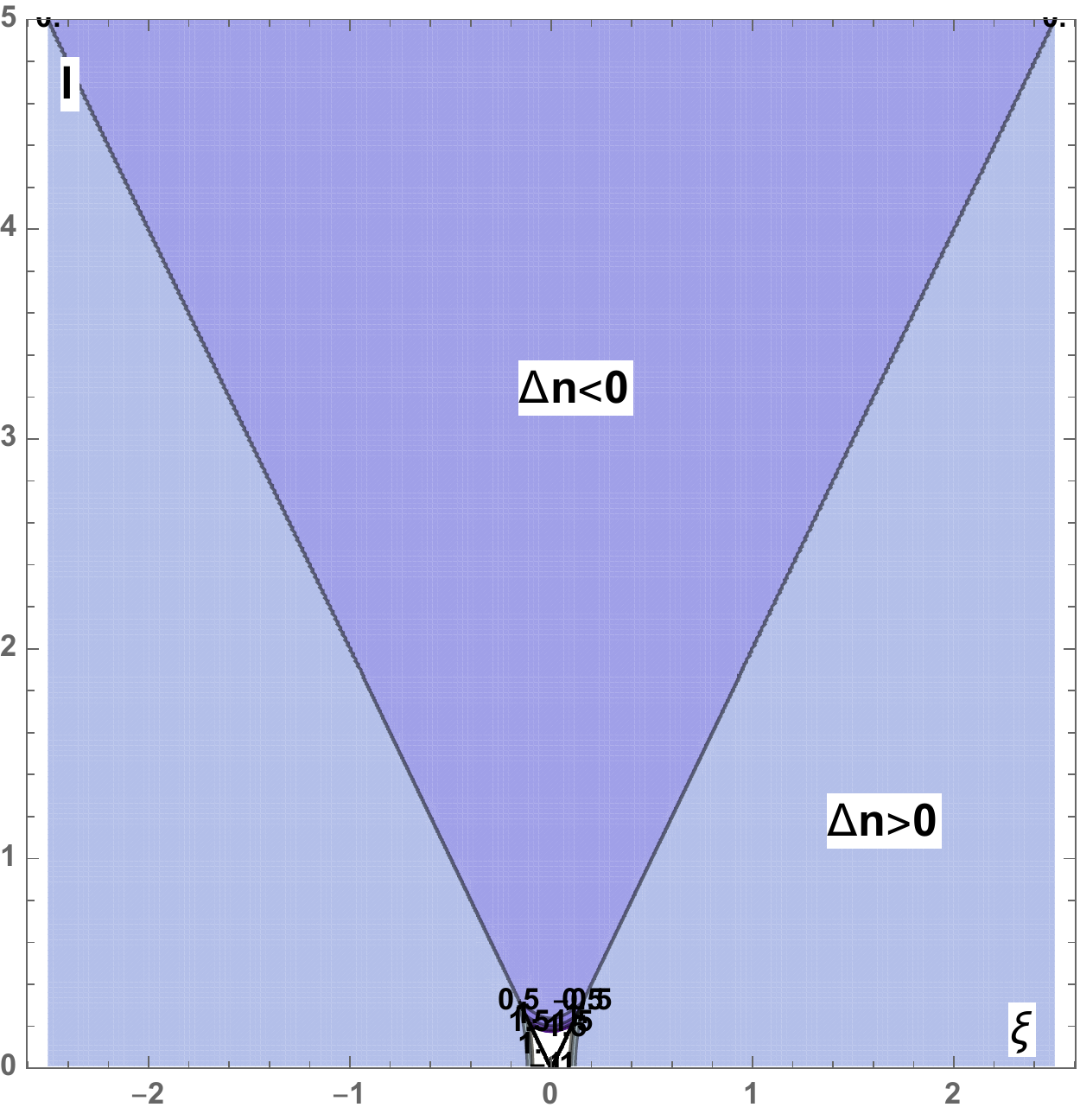}
  \includegraphics[width=6cm]{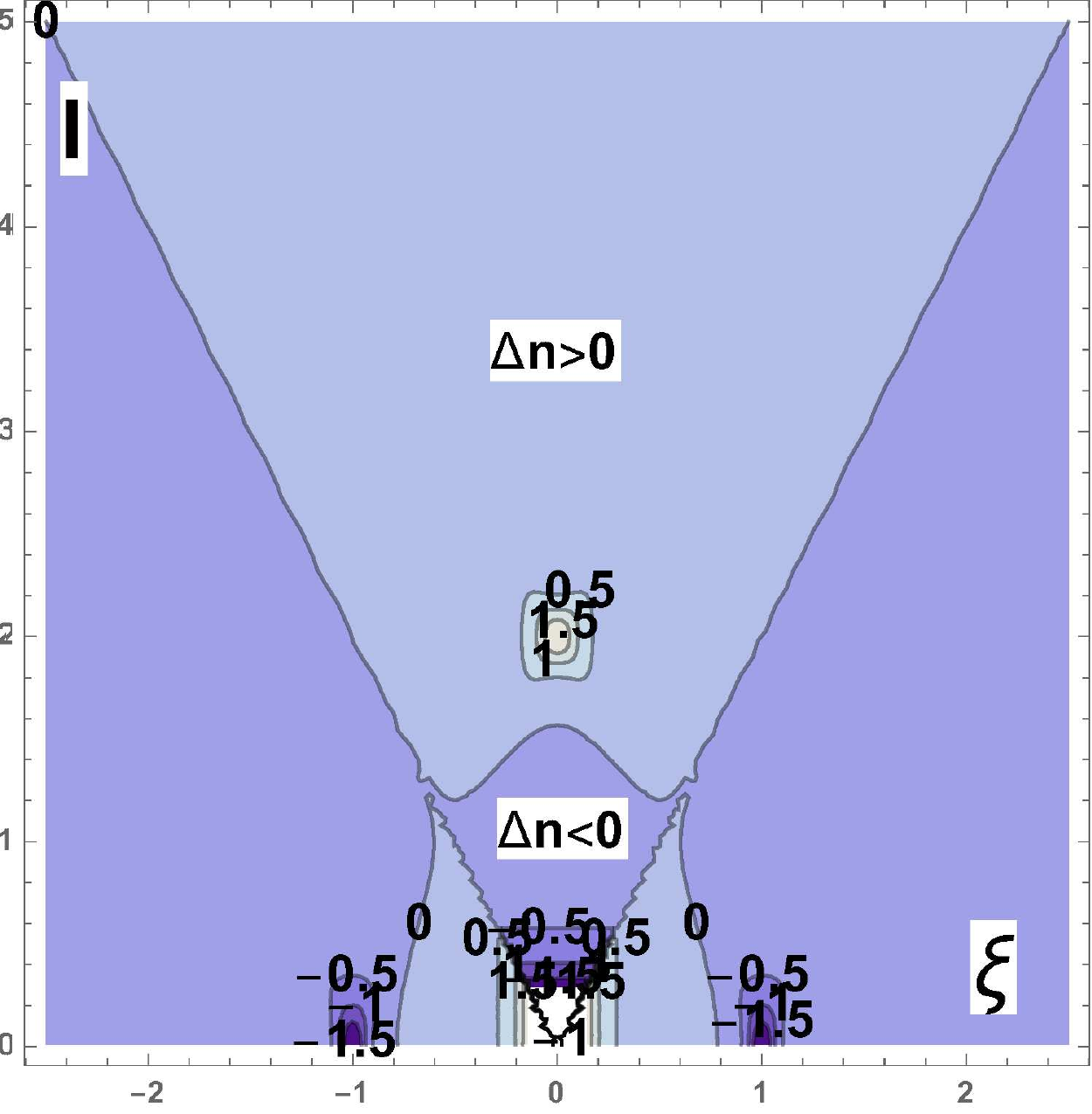}
 \includegraphics[width=6cm]{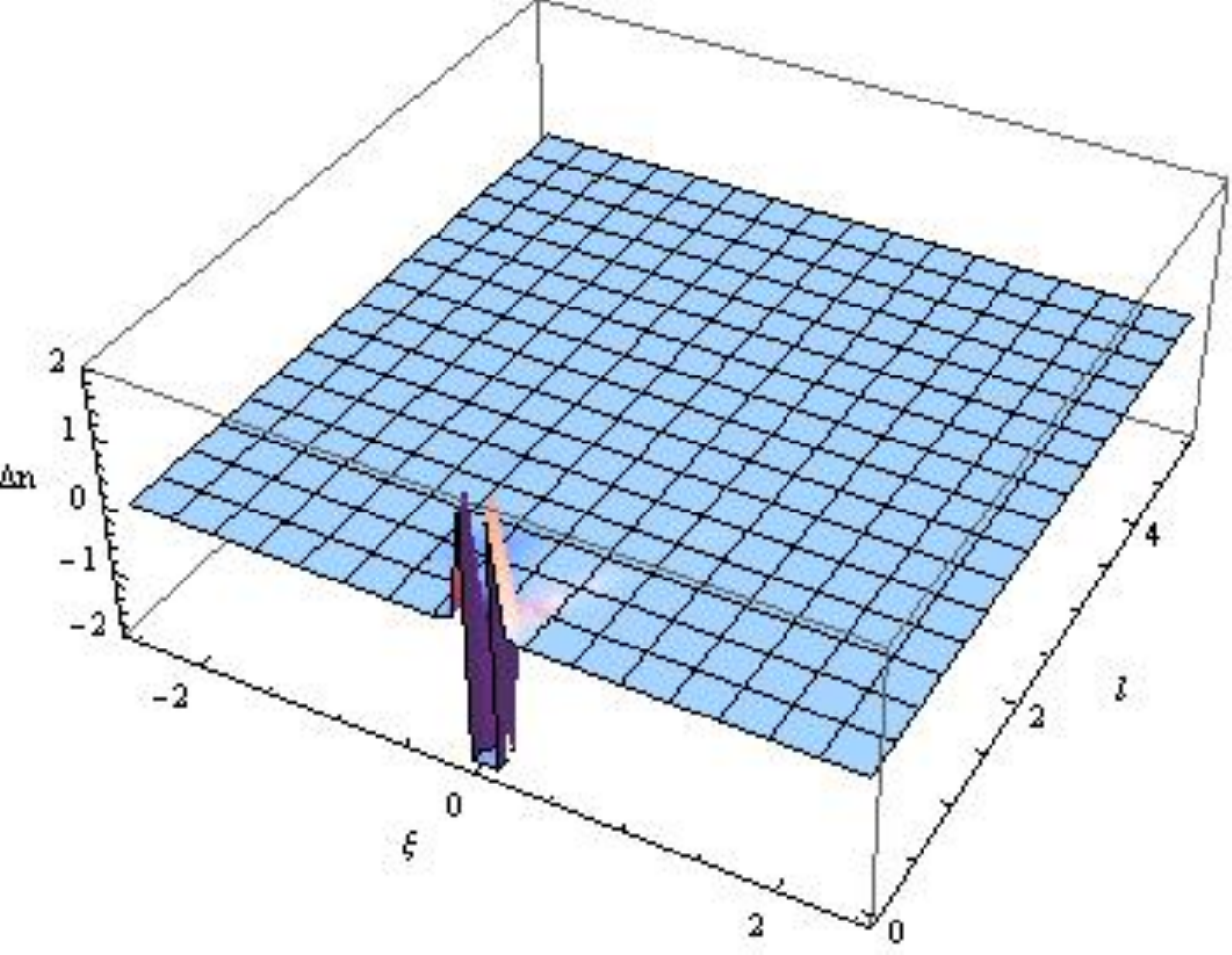}
   \includegraphics[width=6cm]{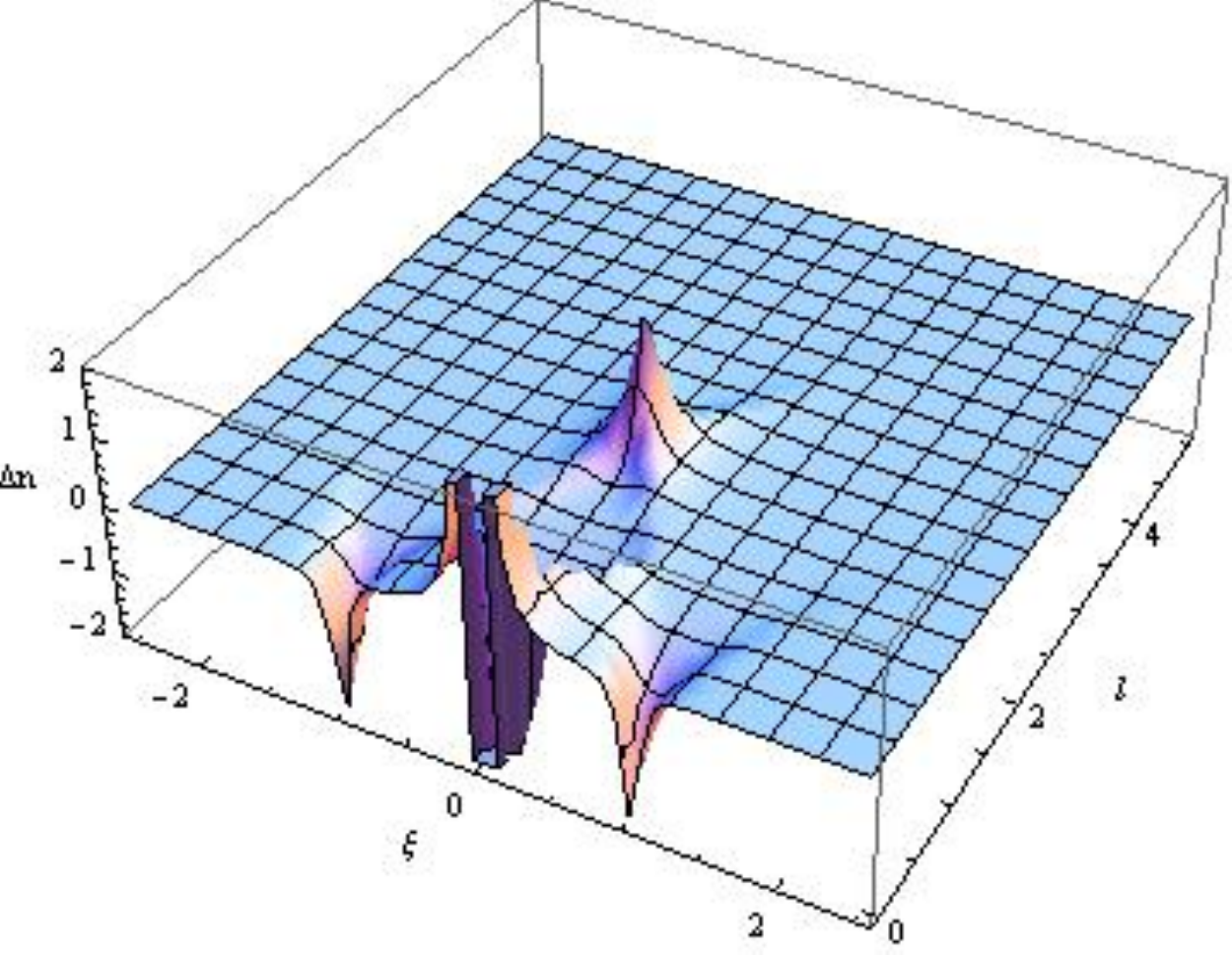}\\
 \caption{The plots of the entanglement density $\Delta n(\xi,l,t)$ for $\ap=0.1$
 at $t=0$ (left two graphs) and $t=1$ (right two graphs). The upper and lower graphs are the contour
 and 3D plots. The horizontal and depth coordinate corresponds to $\xi$ and $l$, respectively.}
\label{fig:DEPD}
\end{figure}

The entanglement density (ED) can also be computed from the second differentiation of the EE (\ref{delk}). The results are plotted in Fig.\ref{fig:DEPD}. We can clearly see the positive peak at $(l,\xi)=(2t,0)$ as well as the negative peaks at $(l,\xi)=(0,\pm t)$, similarly to the local operator quench in RCFTs (\ref{OLQa}). The former agrees with the picture of propagating relativistic particles. The latter coincides the peaks of the energy density predicted from the first law of EE. In addition, there are also peaks near the origin $\xi=l=0$ and this is due to the cut along $x=0$ where the space is divided into the left and right part, which is special to the splitting local quenches.

It is also interesting to note that a negative value region $\Delta n < 0$ is expanding between these three peaks under the time evolution. This is responsible for the decreasing behavior of the EE (\ref{splog}) when we integrate the ED over the region given by the middle picture in Fig.\ref{fig:Integ}.

\subsection{Splitting Local Quench in Holographic CFT}

Another example which allows us to calculate the EE analytically is 2d holographic CFTs.
For the holographic CFT, we can apply the coordinate transformation (\ref{corads}) to calculate the EE. The dual Euclidean geometry is given by Im$[\xi]\geq 0$ in the Poincar\'{e} AdS coordinate (\ref{pol}). The Lorentzian time evolution is obtained by performing the analytic continuation.

The holographic entanglement entropy (HEE) for the connected geodesic $S^{con}_A$ of the subsystem $A(=[w_a,w_b])$
is computed by using (\ref{adseec}) as follows:
\be
S^{con}_A=\frac{c}{6}\log\left[\frac{|f(w_a)-f(w_b)|^2}{\ep^2|f'(w_a)||f'(w_b)|}\right], \label{conhol}
\ee
where $f(w)$ is given by (\ref{fwsp}).

On the other hand, the HEE for the disconnected geodesics is found by applying the AdS/BCFT
as we explained in eq.(\ref{disAdS}). The final result reads
\ba
S^{dis}_A=\frac{c}{6}\log\left(\frac{4(\mbox{Im}f(w_a))(\mbox{Im}f(w_b))}{\ep^2|f'(w_a)||f'(w_b)|}\right)
+2S_{bdy}.
\label{dishol}
\ea
Note that the boundary entropy $S_{bdy}$ contribution arises because the location of boundary surface gets tilted according to the value of the tension $T_{BCFT}$ following (\ref{tenrs}),
so that it solves the boundary condition (\ref{KT}). The boundary entropy $S_{bdy}$ is parameterized by $k$ as in (\ref{kdef}).
Since $S^{dis}$ is the sum of two disconnected geodesic, we have the doubled contribution of $S_{bdy}$ in (\ref{dishol}). Refer to the right picture of Fig.\ref{fig:AdSBCFT}.

The final HEE is given by the smaller of the two:
\be
S_A=\min \{S^{con}_A,S^{dis}_A \}.
\ee

\begin{figure}
  \centering
  \includegraphics[width=5cm]{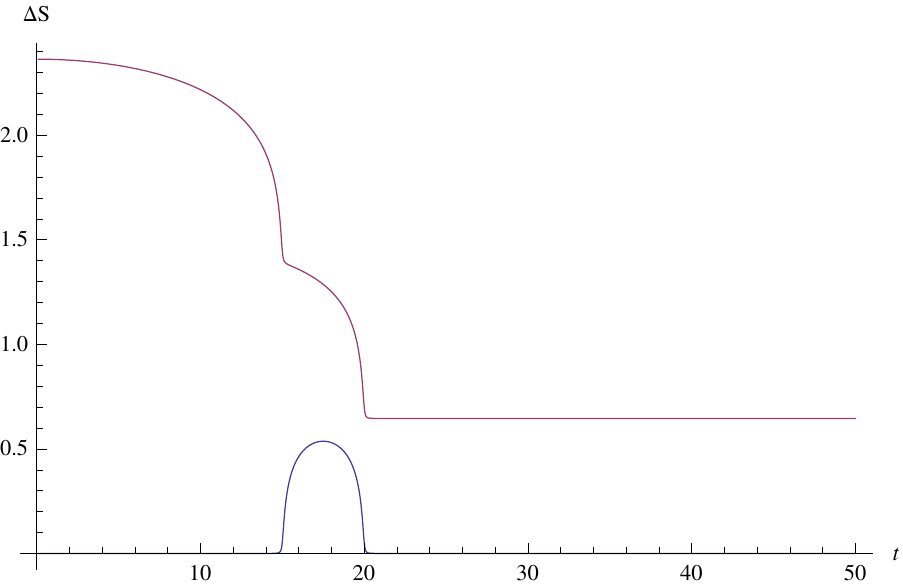}
  \includegraphics[width=5cm]{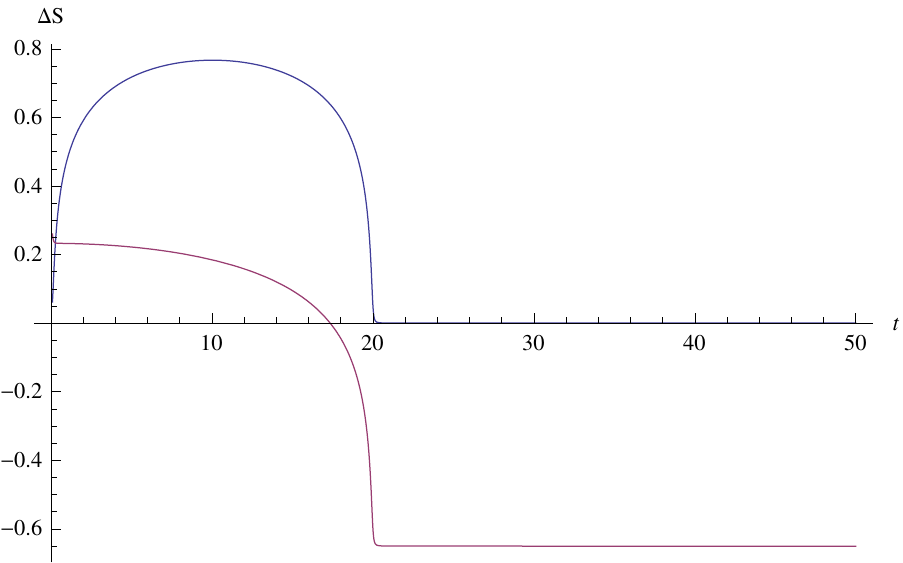}
  \includegraphics[width=5cm]{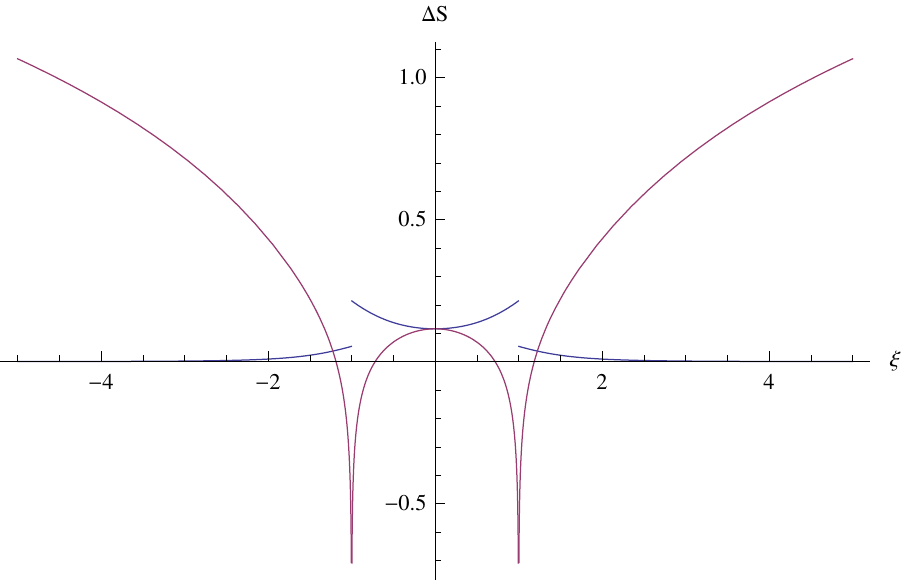}\\
 \caption{The plots of the holographic entanglement entropy growth $S_A-S^{(0)}_A$ at $\ap=0.1$.
 We assume the tension
 of the boundary surface $Q$ is vanishing $T_{BCFT}=0$. The blue/red graph describes the connected/disconnected contribution.
 The left graph describes the time evolution when we take $(a,b)=(15,20)$.
 The middle one is the time evolution for  $(a,b)=(0.1,20)$. The right one
 is the plot for various $\xi$ when we fix $l=2$ and $t=0$.}
\label{fig:HSP}
\end{figure}

The behaviors of EE are plotted in Fig.\ref{fig:HSP} by choose the vanishing tension
$T_{BCFT}=0$ or equally vanishing boundary entropy $S_{bdy}=0$.

When $\ap$ is infinitesimally small, we obtain the following analytical expressions (we take the boundary entropy arbitrary) as explained in the appendix B:

In the early time period, $0<t<|a|$, we have
\ba
 S^{con}_A&=&\frac{c}{3}\log (b-a)/\ep, \ \ \ \no
 S^{dis}_A&=&\frac{c}{6}\log \frac{4(a^2-t^2)(b^2-t^2)}{\ap^2\ep^2}+2S_{bdy}. \label{hpsa}
\ea

When $|a|<t<b$, we have
\ba
S^{con}_A&=&\left\{\begin{aligned}\frac{c}{6}\log \frac{2(b-a)(t-a)(b-t)}{\ap\ep^2},\ \ (a>0)\\
\frac{c}{6}\log \frac{2(b-a)(t+a)(b+t)}{\ap\ep^2},\ \ (a<0)\end{aligned}\right. \no
 S^{dis}_A&=&\frac{c}{6}\log \frac{4|a|(b^2-t^2)}{\ap\ep^2}+2S_{bdy}.  \label{hpsb}
\ea
At late time, $t>b$, we have
\ba
S^{con}_A&=&\left\{\begin{aligned}\frac{c}{3}\log (b-a)/\ep, \quad\qquad\qquad  (a>0)\\
\frac{c}{6}\log \frac{4(t^2-a^2)(t^2-b^2)}{\ap^2\ep^2},\ \ (a<0)\end{aligned}\right. \no
S^{dis}_A&=&\frac{c}{6}\log \frac{4|a|b}{\ep^2}+2S_{bdy}. \label{hpsc}
\ea
It is clear that at the late time limit $t\to \infty$, the EE approaches to a constant value
\be
S_A(t\to\infty)=\mbox{min}\left[\frac{c}{3}\log \f{b-a}{\ep},\ \frac{c}{6}\log \frac{4|a|b}{\ep^2}+2S_{bdy}\right],
\ee
which agrees with the expected result for the separated two half lines.

The behavior of entanglement entropy is also numerically plotted in Fig.\ref{fig:HSP} by choosing
the vanishing tension $T_{BCFT} = 0$ or equally vanishing boundary entropy $S_{bdy} = 0$. The
left graph looks very similar to the one in the Dirac fermion case, which is interpreted
by the propagation of relativistic particles. However, it is intriguing to note that if the
boundary entropy $S_{bdy}$ is positive and very large, then $S^{con}_A$ can dominate in some
region and gives a qualitative discrepancy from the Dirac fermion result.

It is interesting to ask the time evolution of EE when we choose the subsystem $A$ to be almost a half of the total system i.e. $a\ll t\ll b$. If we choose the boundary entropy $S_{bdy}$ is very large
such that $k\gg 1$ (remember the definition of $k$ (\ref{kdef})), then $S^{con}_A$ is favored for the period $a\ll t\ll (2k+1)a$, which results in the logarithmic growth
\be
S^{con}_A\simeq \frac{c}{6}\log\frac{2t}{\ap}+\frac{c}{3}\log \frac{l}{\ep}. \label{finee}
\ee
This is peculiar to the holographic CFTs. However, if $k$ is order one, this logarithmic growth is missing and the EE monotonically decreases as in the previous Dirac fermion example.

The right picture (assuming $S_{bdy} = 0$) in Fig.\ref{fig:HSP} is qualitatively similar to what we obtained for the free Dirac fermion CFT.
For finite $\ap>0$, one may notice that the entanglement entropy for the connected curve $S^{con}_A$ is not continuous as in the third graph in Fig.\ref{fig:HSP}. However, this occurs when one of the endpoints of the subsystem $A$ coincides with $x=0$ and thus this discontinuity should happen because the left and right at $x=0$ are disconnected in the splitting quench. This can be easily understood if we go to the Poincar\'{e} coordinate (\ref{pol}).

\begin{figure}
  \centering
 \includegraphics[width=6cm]{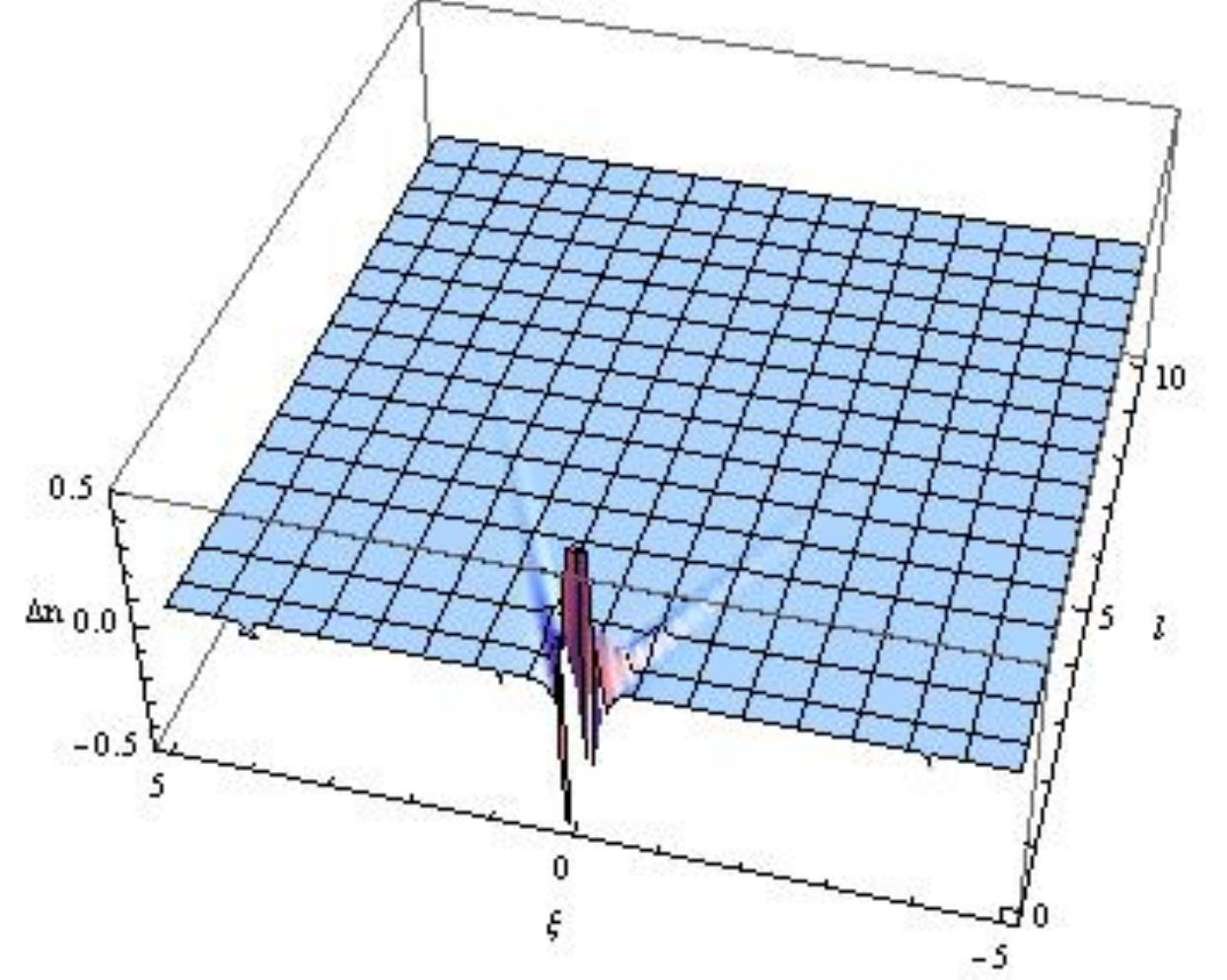}
   \includegraphics[width=6cm]{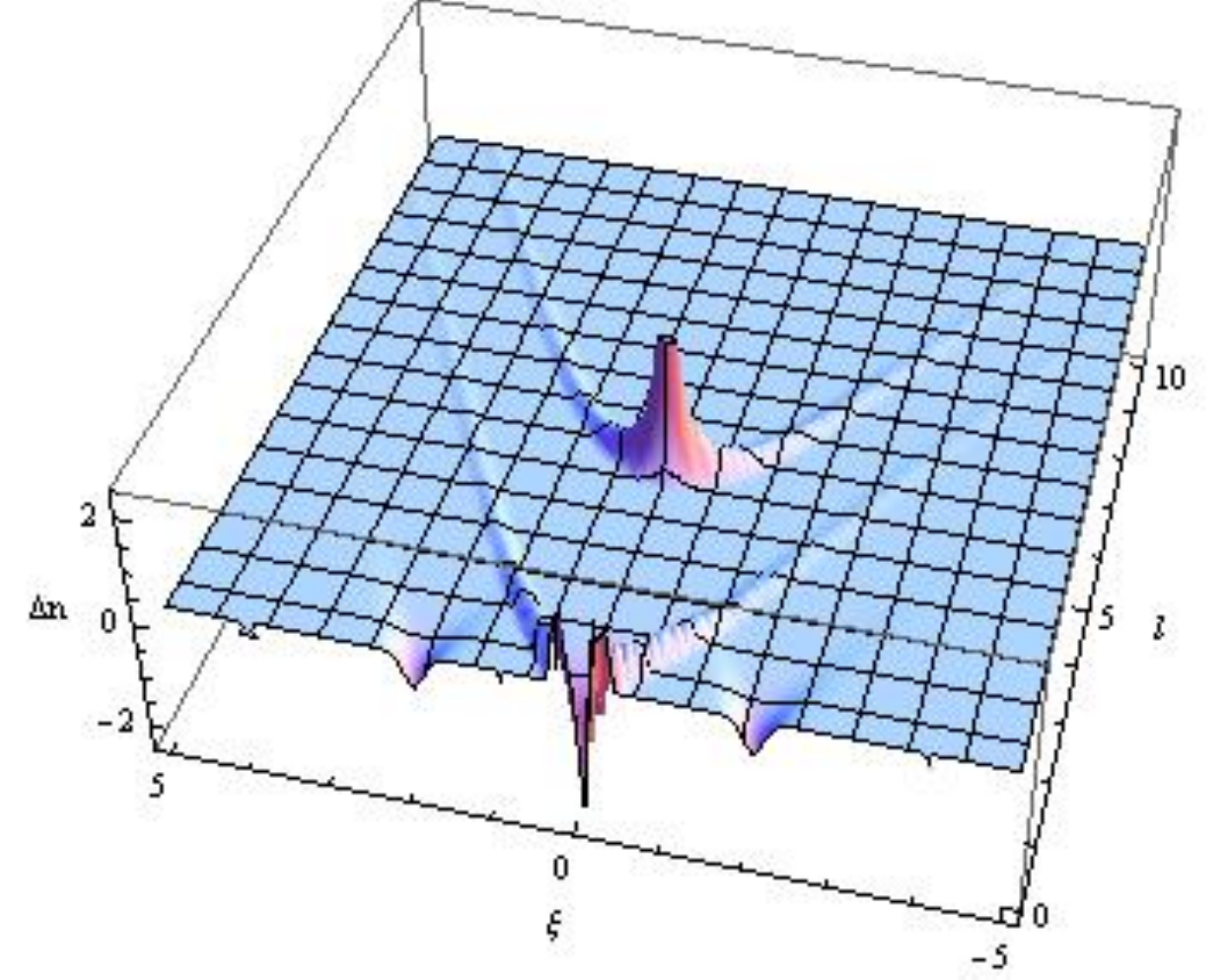}
   \includegraphics[width=6cm]{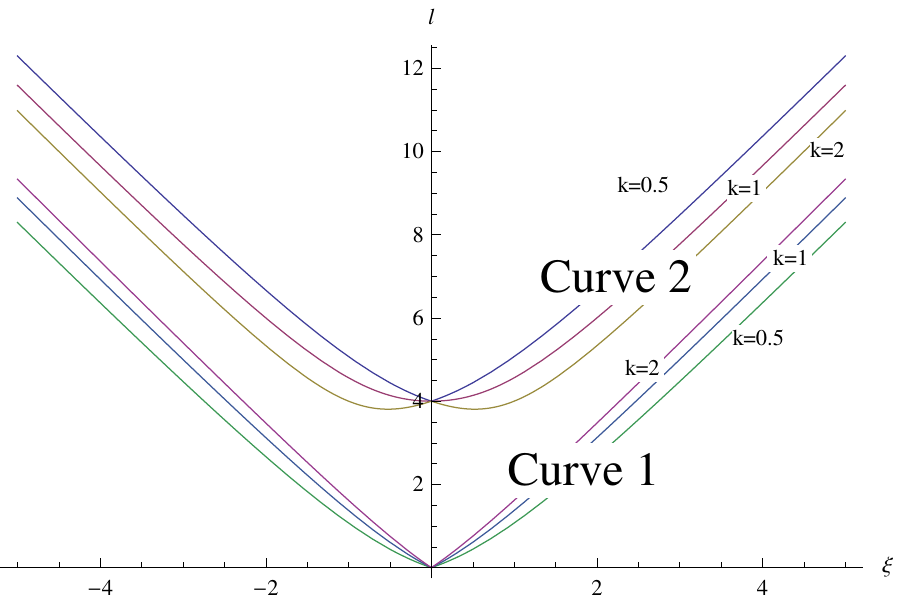}
      \includegraphics[width=6cm]{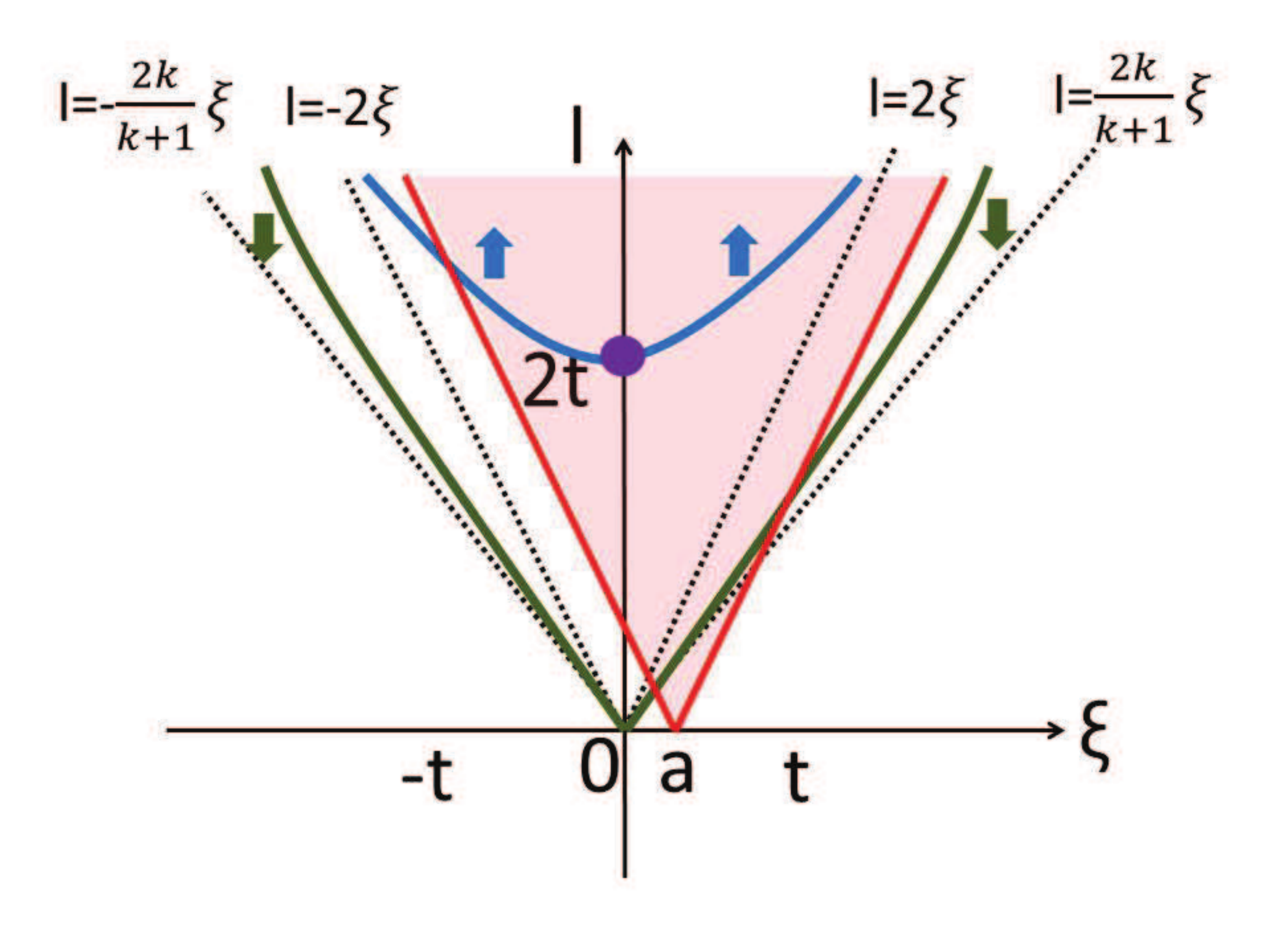}
 \caption{The behaviors of a smeared entanglement density (ED) for the holographic splitting local quench. The upper left and right graph are the plots of $\Delta n(\xi,l,t)$ at $t=0$ and $t=2$, respectively with $\ap=0.1$ (again we set $T_{BCFT}=0$). The horizontal and depth coordinate corresponds to $\xi$ and $l$, respectively. Here we smeared the originally delta functional behavior of the ED by replacing the second derivatives w.r.t $l$ and $\xi$ with finite-differences $\delta=0.2$.
 We observe delta functional behaviors on the two curves (\ref{curveSP0}) and (\ref{curveSP}) at $k=1$.
 The lower left picture shows these two curves at $t=2$ for $k=0.5,1,2$.
 The lower right picture describes the calculation of $S_A$ for $a\ll t\ll b$ by integrating the ED over
the red region. The green and blue curves represent curve 1: (\ref{curveSP0}) and curve 2: (\ref{curveSP}). We find that the blue curve gets into the red region, while the green curve goes out. The cancellations
between them leads to the absence of logarithmic growth.}
\label{fig:HSPED}
\end{figure}

The entanglement density is plotted in Fig.\ref{fig:HSPED}. In the upper graphs, we
smeared the derivatives in (\ref{delk}) by replacing it with the finite difference:
e.g. $\de_a S(a,b,t)\to (S(a+\delta/2,b,t)- S(a-\delta/2,b,t))/\delta$. We did so because
in this holographic case, we expect genuine delta-functional behaviors due to the phase transitions of HEE. In addition to the expected positive peak at $(\xi,l)=(0,2t)$  and the two
negative ones at $(\xi,l)=(\pm t,0)$, we observe continuous peaks along the following two curves
\ba
&& \mbox{Curve 1}: l=-t+\frac{|\xi|}{1+k}+\s{\left(t-\frac{|\xi|}{1+k}\right)^2+\frac{4k}{k+1}(t+|\xi|)|\xi|},
\label{curveSP0}\\
&& \mbox{Curve 2}:  l=t+\frac{|\xi|}{1+k}+\s{\left(t+\frac{|\xi|}{1+k}\right)^2-\frac{4k}{k+1}(t-|\xi|)|\xi|}, \label{curveSP}
\ea
where we used the positive parameter $k$ defined in (\ref{kdef}).
These two curves are located at phase transition points between $S^{con}$ and $S^{dis}$  as we can see from (\ref{hpsb}), where (\ref{curveSP0}) and (\ref{curveSP}) are situated in $a>0$ and $a<0$, respectively.
Their profile is plotted in the lower left picture of Fig.\ref{fig:HSPED}. Both of them coincides with
$l=2|\xi|$ at $t=0$. In the limits $t\to \infty$ and $|\xi|\to \infty$, the curves are approximated by
\ba
\mbox{Curve 1}: && l\simeq \frac{2k}{k+1}|\xi|,\ \ (t\to \infty), \ \  \ \
l \simeq 2|\xi|-\frac{2t}{2k+1},\ \ (|\xi|\to \infty)\no
\mbox{Curve 2}: && l\simeq 2t-\frac{2(k-1)}{k+1}|\xi|,\ \ (t\to \infty), \ \ \ \  l \simeq  2|\xi|+\frac{2t}{2k+1}.\ \ (|\xi|\to \infty) \nonumber
\ea

The presence of these peak curves is peculiar to holographic CFTs and is missing in the free Dirac fermion CFTs or more generally RCFTs. They are outside of the causality zone and therefore the initial state already has highly non-local entanglement. In the lower right picture of Fig.\ref{fig:HSPED}, we sketched
the calculation of EE, which is given by the integration of ED over the red region. For this, we notice that
the blue curve (\ref{curveSP}) gets into the red region, while the green curve (\ref{curveSP0}) goes out. Thus, we can explain the absence of the logarithmic growth of EE for the splitting local quenches due to the cancellations between these two opposite effects.\footnote{However, if we consider
the case $k\gg 1$, then  (\ref{curveSP0})  approaches $l=2\xi$ and thus this curve is fully included in the red region until $t$ gets very large. This explains the logarithmic growth (\ref{finee}), peculiar to
the large $k$ case.} Also we will provide the geometric explanation of the $\log t$ behaviors in section \ref{sec:log}.

\subsection{Holographic Geometry of Splitting Local Quench}

Now let us study the holographic geometry dual to the splitting local quench.
The metric of the gravity dual is given by (\ref{metads}) with
\be
T(w)=\frac{3\ap^2}{4(w^2+\ap^2)^2},\ \ \  \bar{T}(\bar{w})=\frac{3\ap^2}{4(\bar{w}^2+\ap^2)^2},
\label{sttw}
\ee
where $w=x+i\tau$ and $\bar{w}=x-i\tau$. For its Lorentzian extension, we can set $w=x-t$ and $\bar{w}=x+t$. The important ingredient of AdS/BCFT as reviewed in section \ref{adsbcfts} is the boundary surface $Q$ which extends from the AdS boundary $z=0$ to the bulk AdS. At $z=0$, $Q$ coincides with the cut in Fig.\ref{fig:SPw}, which describes the splitting process in the 2d CFT.
 Therefore, we have to be careful in its global geometry i.e. which part of the geometry we should pick up and where the boundary surface $Q$ in the AdS/BCFT prescription is located.

Let us consider the Euclidean geometry dual to the left CFT picture of Fig.\ref{fig:SPw}.
 It is important to note that the gravity dual is precisely defined by mapping an upper half $\mbox{Im}[\xi]\geq 0$ of the Poincar\'{e} AdS$_3$ (\ref{pol}) using the transformation (\ref{corads}) with (\ref{fwsp}), assuming that the tension is vanishing $T_{BCFT}=0$. The boundary $Q$ in the AdS/BCFT, given by $\mbox{Im}[\xi]=0$ in the latter setup, is mapped to a region on $x=0$.

 First we focus on the time slice $\tau=0$ (i.e. $\mbox{Im} [w]=0$) of the gravity dual. The metric on this time slice is given by
 \ba
 ds^2=\frac{dz^2}{z^2}+\left(\frac{1}{z}+\frac{3\ap^2z}{4(\ap^2+x^2)^2}\right)^2 dx^2.
 \ea

\begin{figure}
  \centering
  \includegraphics[width=10cm]{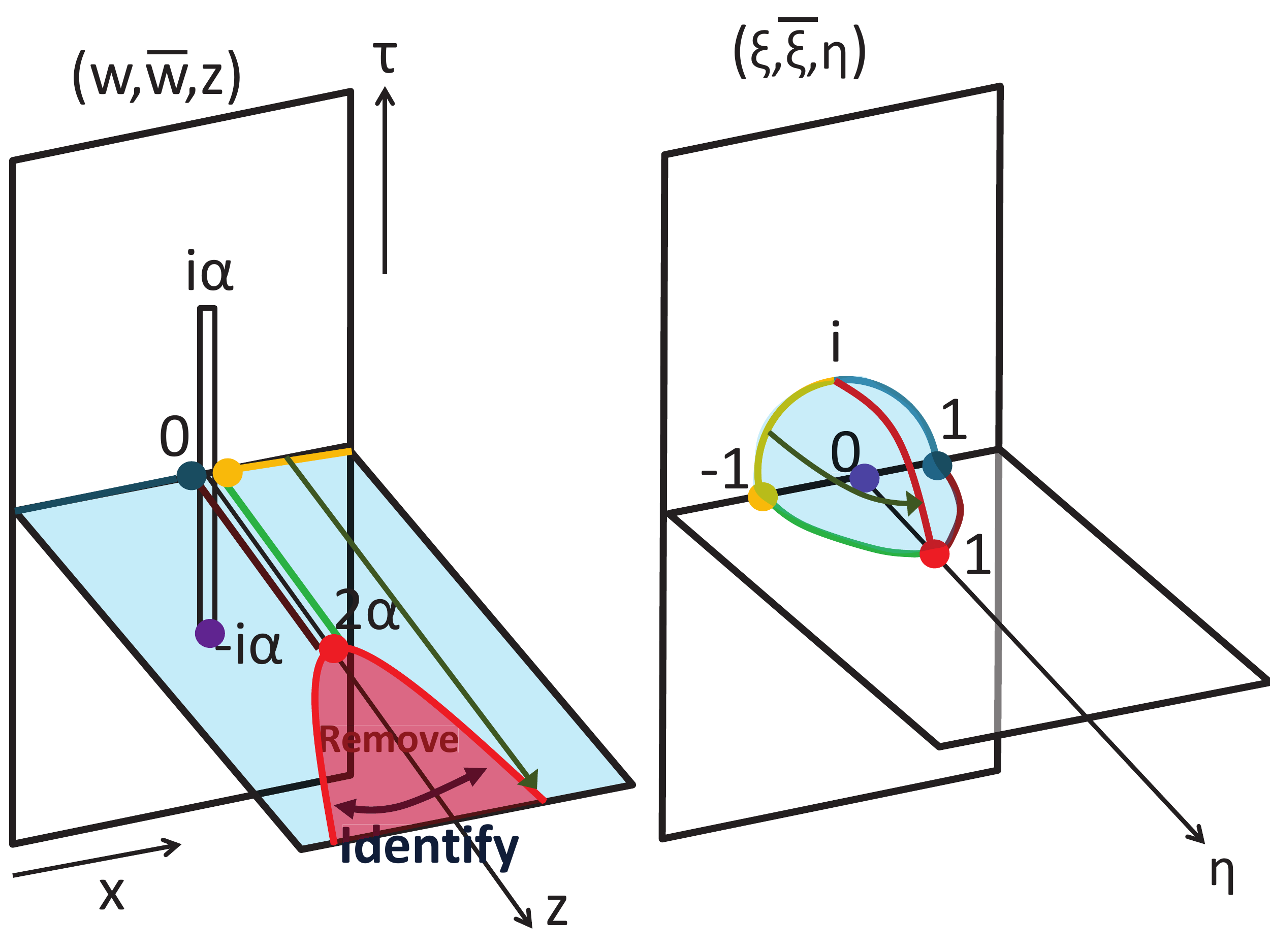}
 \caption{The map (\ref{corads}) between the gravity dual of splitting local quench (left) and
 the upper Poincar\'{e} AdS (right). The red curve in the left is given by $z=\frac{2(x^2+\ap^2)}{\ap}$ and the region inside this curve should be removed with the identification $x$ with $-x$ on the curve.
 The length of red curve gives the HEE between two separated lines. The green and dark brown curve describe the boundary surface $Q$.}
\label{fig:SPmapt0}
\end{figure}

 We can confirm that this slice is mapped into a quarter of the sphere given by $|\xi|^2+\eta^2=1$ with $\eta\geq 0$ and $\mbox{Im}[\xi]\geq 0$.
 However, to realize one to one map, we need to remove the region
 \be
 z>\frac{2(x^2+\ap^2)}{\ap}, \label{identif}
\ee
 and identify $x$ with $-x$ for any $x$ along the curve $z=\frac{2(x^2+\ap^2)}{\ap}$.
 A sketch of the map (\ref{corads}) at $t=0$ is depicted in Fig.\ref{fig:SPmapt0}.
 The boundary surface $Q$ extends from the AdS boundary $z=0$ toward the IR region but it ends at $z=2\ap$ due to the identification. This shows that the two half lines $x>0$ and $x<0$ at the AdS boundary $z=0$, is connected in the bulk through the horizon given by the identified curve $z=\frac{2(x^2+\ap^2)}{\ap}$.
 Therefore the CFTs on these two half lines are entangled and the entanglement entropy is estimated
 as the length of this curve:
 \ba
 S_A=\frac{c}{6}\int^{x_\infty}_0 \frac{2dx}{\s{x^2+\ap^2}}\simeq \frac{c}{6}\log\frac{z_\infty}{\ap}, \label{sax}
 \ea
 where $z_\infty=2x_\infty^2/\ap$ is the IR cut off. This agrees with our expectation because the state before the splitting of the system was the ground state of the CFT on a line. Note that in the Poincar\'e coordinate, this is equal to  a quarter of the largest circle in the sphere $|\xi|^2+\eta^2=1$, depicted as the red curve
in the left picture of Fig.\ref{fig:SPmapt0}.

\begin{figure}
  \centering
  \includegraphics[width=7cm]{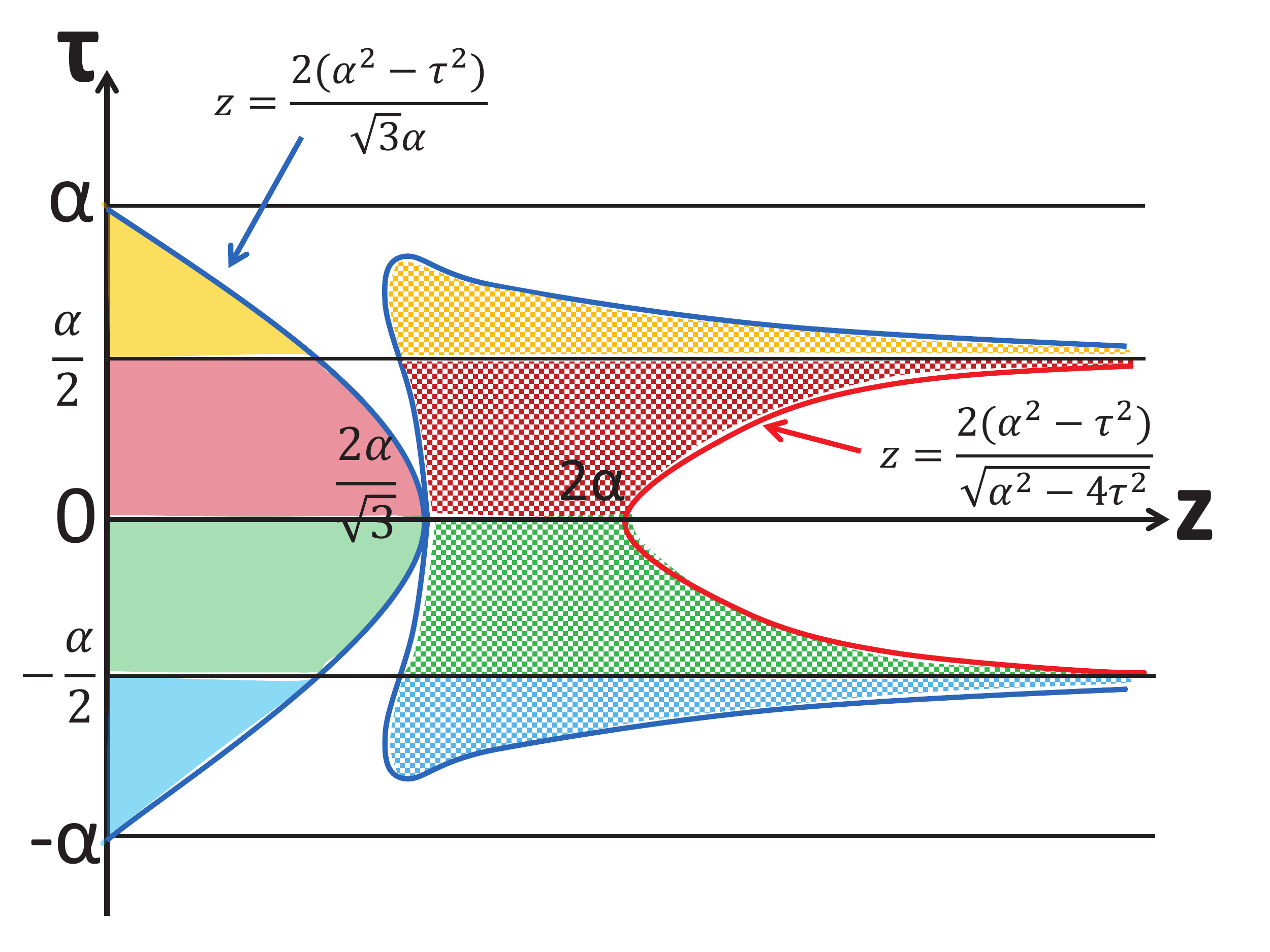}
   \includegraphics[width=7cm]{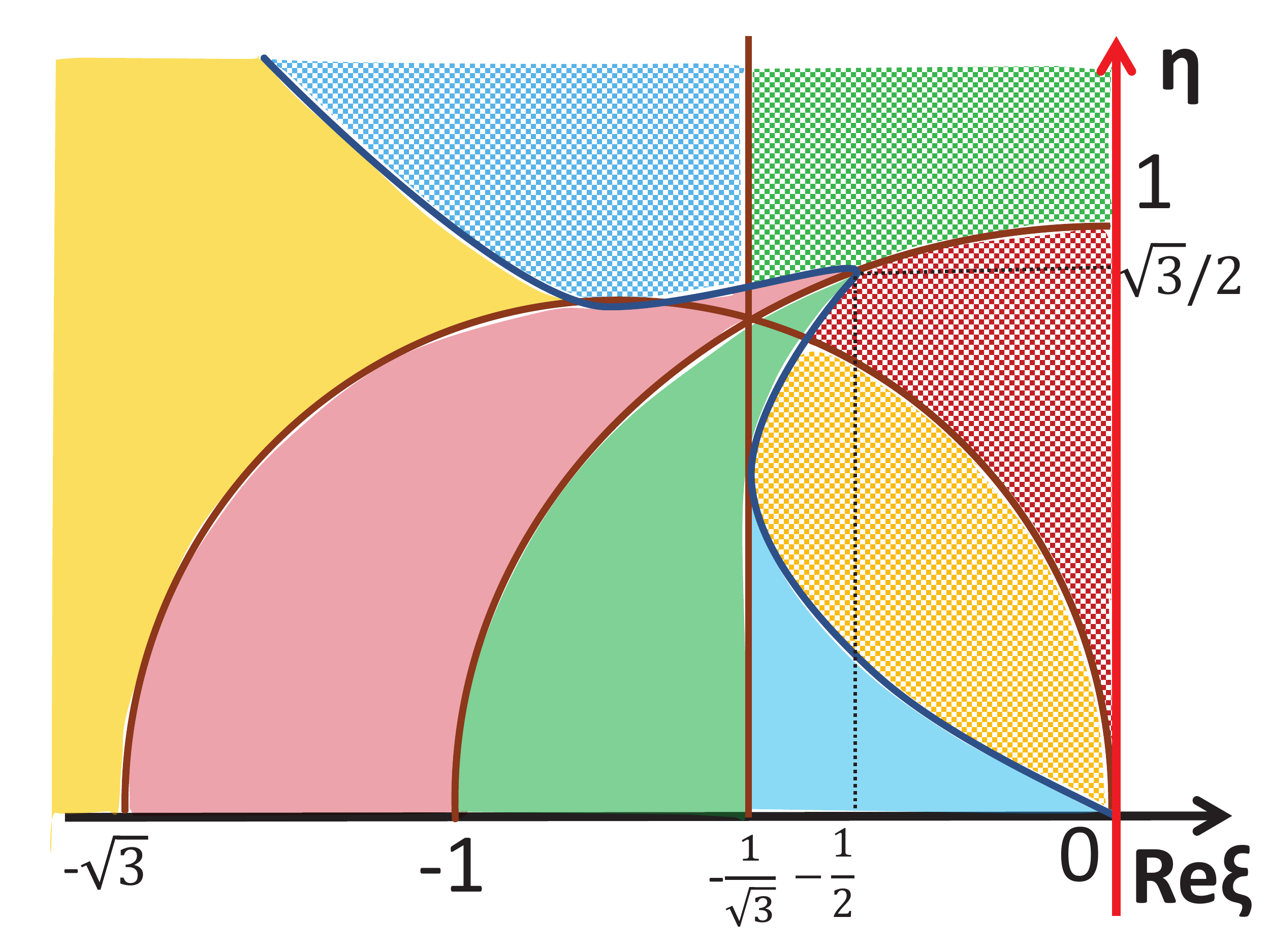}
 \caption{The boundary surface $Q_+$ for $T_{BCFT}=0$ in the gravity dual of splitting local quench.
 The left picture describes in the coordinate $(w,\bar{w},z)$ setting $x=+\delta\to 0$. The right picture does in the Poincar\'{e} coordinate $(\xi,\bar{\xi},\eta)$ on the boundary Im$[\xi]=0$. The red curve in the left picture is mapped into the Re$[\xi]=0$ in the right one. The two blue curves in the left are pasted with each other in a way they are mapped into the blue curve in the right picture. The eight colored regions in the left are mapped into those in the right. Note that there are one more boundary surface corresponding to
 $x=-\delta \to 0$, which is given by the same structure and is mapped to the Re$[\xi]>0$ region.
 The white region in $-\frac{\ap}{2}<\tau<\frac{\ap}{2}$ should also be removed
  to secure the map is one to one.}
\label{fig:SPmapt0a}
\end{figure}

Next we would like to examine how the boundary $Q$ in the AdS/BCFT looks like in our gravity dual.
To see this we analyze which region is mapped into Im$[\xi]=0$. It is straightforward to see that the boundary $Q$ should be on the slice $x=0$ (i.e. Re$[w]=0$). Notice that we need to distinguish the two segments $Q_+$ and
$Q_-$ of the boundary surface $Q$ at $x=+\delta$ and $x=-\delta$ ($\delta>0$ is infinitesimal), which corresponds to Re$[\xi]<0$ and Re$[\xi]>0$ in the Poincar\'{e} AdS. We focus on the time period $-\pi\ap <\tau<\pi\ap$. In Fig.\ref{fig:SPmapt0a}, we sketched how the boundary surface $Q_+$ looks like in the $(w,\bar{w},z)$ coordinate and how it is mapped into the boundary Im$[\xi]=0$ and Re$[\xi]<0$ in the Poincar\'{e} coordinate for $x=+\delta$.

The tip $z=2\alpha$ of the identification region (\ref{identif}) at the time slice $\tau=0$ is extended along
\be
z=\frac{2(\ap^2-\tau^2)}{\s{\ap^2-4\tau^2}}.  \label{idenc}
\ee
for $-\frac{\ap}{2}<\tau<\frac{\ap}{2}$. This is the red curve in Fig.\ref{fig:SPmapt0a}.
Note that this tip is the end point of the boundary surface $Q$ in the case of $T_{BCFT}=0$. Therefore the two sheets ($x=+\delta$ and $x=-\delta$) are identified along this curve (\ref{idenc}) with each other.

There is one more non-trivial issue. To secure one to one mapping into the half of Poincar\'{e} AdS, we need to remove the white region in the left picture in Fig.\ref{fig:SPmapt0a} and join the two blue curves with each other such that it agrees with the right picture i.e. Im$[\xi]=0$ in the Poincar\'{e} coordinate.
Finally we can confirm that the colored regions in the left picture  are mapped into those in the right one in Fig.\ref{fig:SPmapt0a}. The full boundary surface $Q$ is given by joining two copies of such a space.

We will not get into details of the geometry in other parts as we can understand how the Lorentzian
space looks like from the above observations. Nevertheless, it is helpful to point out that
the identification surface (\ref{idenc}) at non-zero $x$ is found to be
\ba
z = \frac{2\sqrt{x^2+(\alpha+\tau)^2}\sqrt{x^2+(\alpha-\tau)^2}}{\sqrt{\alpha^2-\Big(\sqrt{x^2+(\alpha+\tau)^2}-\sqrt{x^2+(\alpha-\tau)^2}\Big)^2}} \equiv G_E(x,\tau).  
\label{idencc}
\ea

\begin{figure}
  \centering
  \includegraphics[width=7cm]{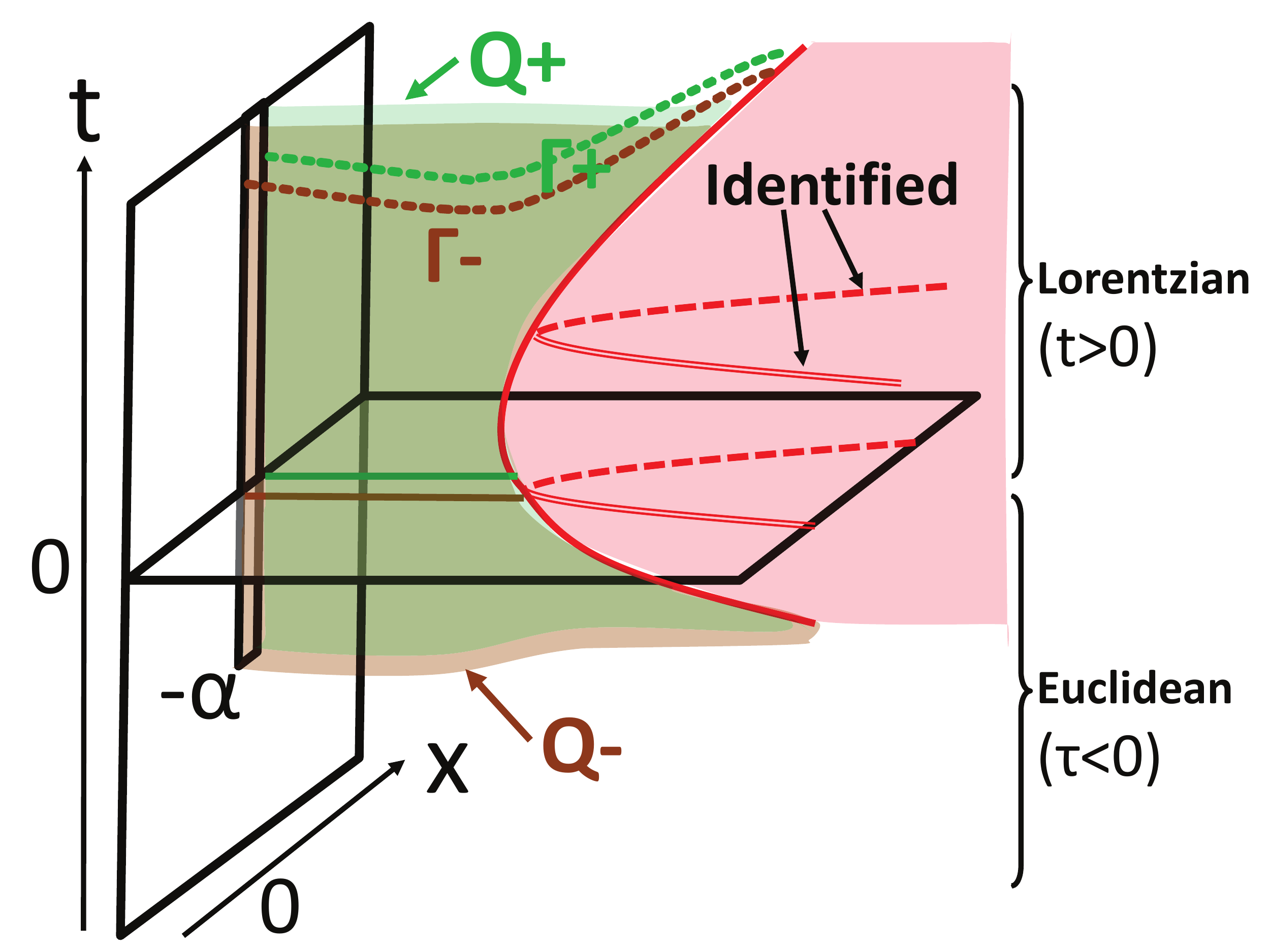}
 \caption{The spacetime geometry of gravity dual of the splitting local quench.
 The green surface ($Q_+$) and brown one ($Q_-$) are the boundary surface $Q$ in the AdS/BCFT.
 The red region defined by  $z>G_L(x,t)$ should be removed
 with the dotted red curve and the doubled red curve identified. Thus this identified red region corresponds to
 the horizon. The region $x>0$ and $x<0$, which are
 separated at the AdS boundary $z=0$ are connected (in a space-like way) through this horizon in the bulk.}
\label{fig:SPHOLgeo}
\end{figure}

In the Lorentzian signature, this surface looks like
\ba
z = \frac{2\sqrt{x^2+(\alpha+it)^2}\sqrt{x^2+(\alpha-it)^2}}{\sqrt{\alpha^2-\Big(\sqrt{x^2+(\alpha+it)^2}-\sqrt{x^2+(\alpha-it)^2}\Big)^2}} \equiv G_L(x,t).  
\label{idenccl}
\ea
Thus the Lorentzian geometry for $t>0$ in the $(w,\bar{w},z)$ is given by removing the
part $z>G_L(x,t)$ and by identifying each two
points on (\ref{idenccl}) as $(\tau,x,z)\sim (\tau,-x,z)$. The boundary surface $Q$ consist of
identical two surfaces $Q_+$ and $Q_-$ which extend in $z<\frac{2(\ap^2+t^2)}{\s{\ap^2+4t^2}}$
and are localized at $x=\delta$ and $x=-\delta$, respectively. The final spacetime of the gravity dual is depicted
in Fig.\ref{fig:SPHOLgeo}.

At late time we can approximate this as
\be
z\simeq t.
\ee
This behavior $z\sim t$ at $x=0$ clearly shows that the two half lines at the boundary are not causally connected in a marginal sense. Therefore the region $x>0$ and $x<0$, which are
 separated at the AdS boundary $z=0$ are connected in a space-like way (i.e. it is a non-traversable wormhole) through this horizon in the bulk. This is consistent with the fact that in the CFT side there is no direct interaction between these two CFTs for $t>0$, though they are entangled.

As in our arguments in this subsection we assumed $T_{BCFT}=0$ for simplicity, we would like to
briefly mention how the gravity dual is modified when $T_{BCFT}\neq 0$. When $T_{BCFT}$ is non-zero, the boundary surface $Q$ in $(\xi,\bar{\xi},\eta)$ will be tilted as
\be
\mbox{Im}[\xi]=-\frac{T_{BCFT}}{\s{1-T_{BCFT}^2}}\eta. \label{tensionb}
\ee
When $T_{BCFT}<0$,
the boundary surfaces $Q_+$ and $Q_-$ tilted in the bulk such that they still coincide at $z=0$
and the angle between them takes a fixed positive value. Therefore, the bulk region gets squeezed and the entanglement entropy $S_A$ (\ref{sax}) is decreased. When $T_{BCFT}>0$, the boundary surfaces are modified in an opposite way and the bulk region expands. The entanglement entropy $S_A$ gets increased.

In this subsection we studied the Lorentzian motion of the boundary surface $Q$.
For the holographic entanglement entropy, we actually need the profile of geodesic in Lorentzian 
geometry, though in the previous subsection we employed the formal Wick rotation arguments to avoid this
analysis. We will explicitly study how the Lorentzian geodesic looks like in section \ref{sec:log}.

\subsection{Splitting Local Quench in A Simple Spin System}

Before we go on, we would like to present numerical results for a splitting local quench in a simple spin model for a reference. We start with a spin-$1/2$ transverse Ising chain with $L$ sites and free boundary condition. Cut the interaction between site $j$ and $j+1$ to separate it into two independent chains at $t=0$. In this case Hamiltonian before quench $H_{before}$ and that after quench $H_{after}$ turns out to be
\begin{align}
H_{before} = -J\sum_{i=1}^{L-1} \sigma_{i}^z\sigma_{i+1}^z + h\sum_{i=1}^L \sigma^x_i
\end{align}
\begin{align}
H_{after} = -J\bigg(\sum_{i=1}^{L-1} \sigma_{i}^z\sigma_{i+1}^z -\sigma_{j}^z\sigma_{j+1}^z\bigg) + h\sum_{i=1}^L \sigma^x_i
\end{align}

\begin{figure}[h]
  \begin{center}
    \begin{tabular}{c}

      \begin{minipage}{0.45\hsize}
        \begin{center}
          \includegraphics[clip, width=6.5cm]{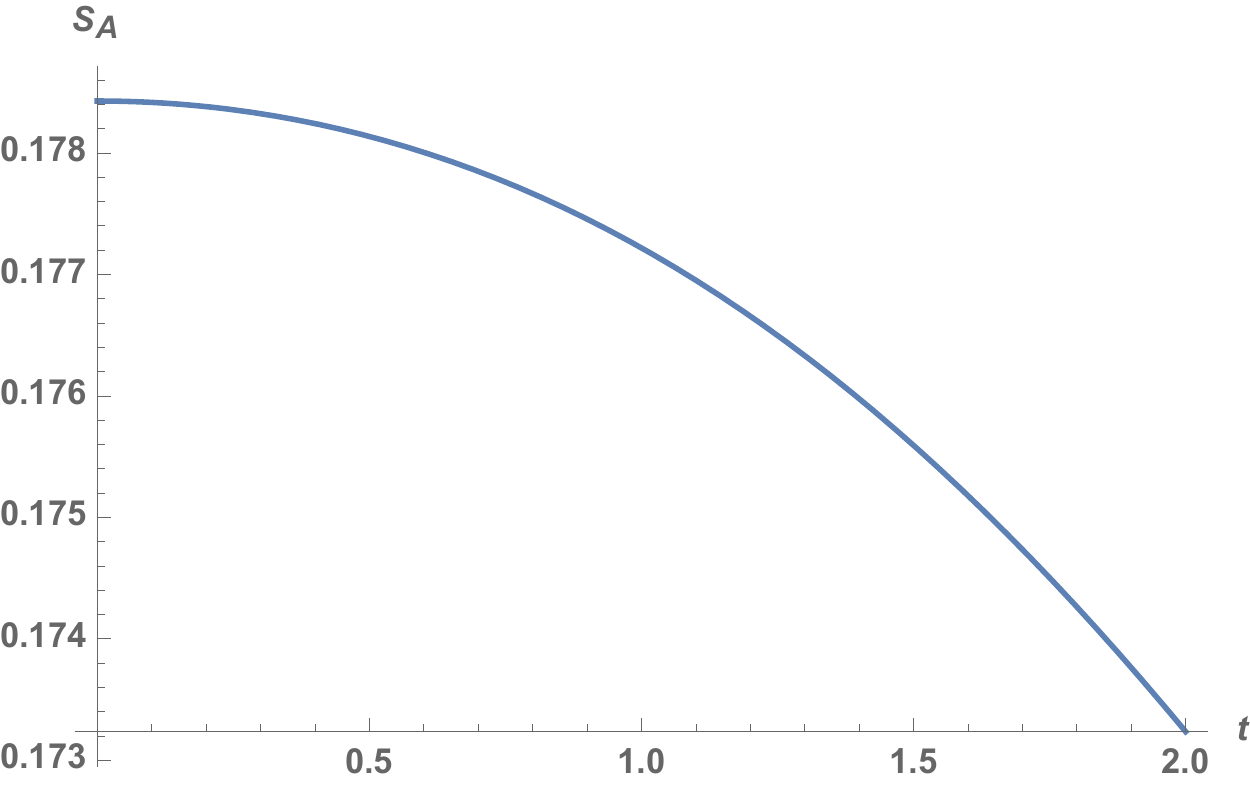}
          \hspace{1.6cm} (a) $0\leq t\leq 2$
        \end{center}
      \end{minipage}

      \begin{minipage}{0.45\hsize}
        \begin{center}
          \includegraphics[clip, width=6.5cm]{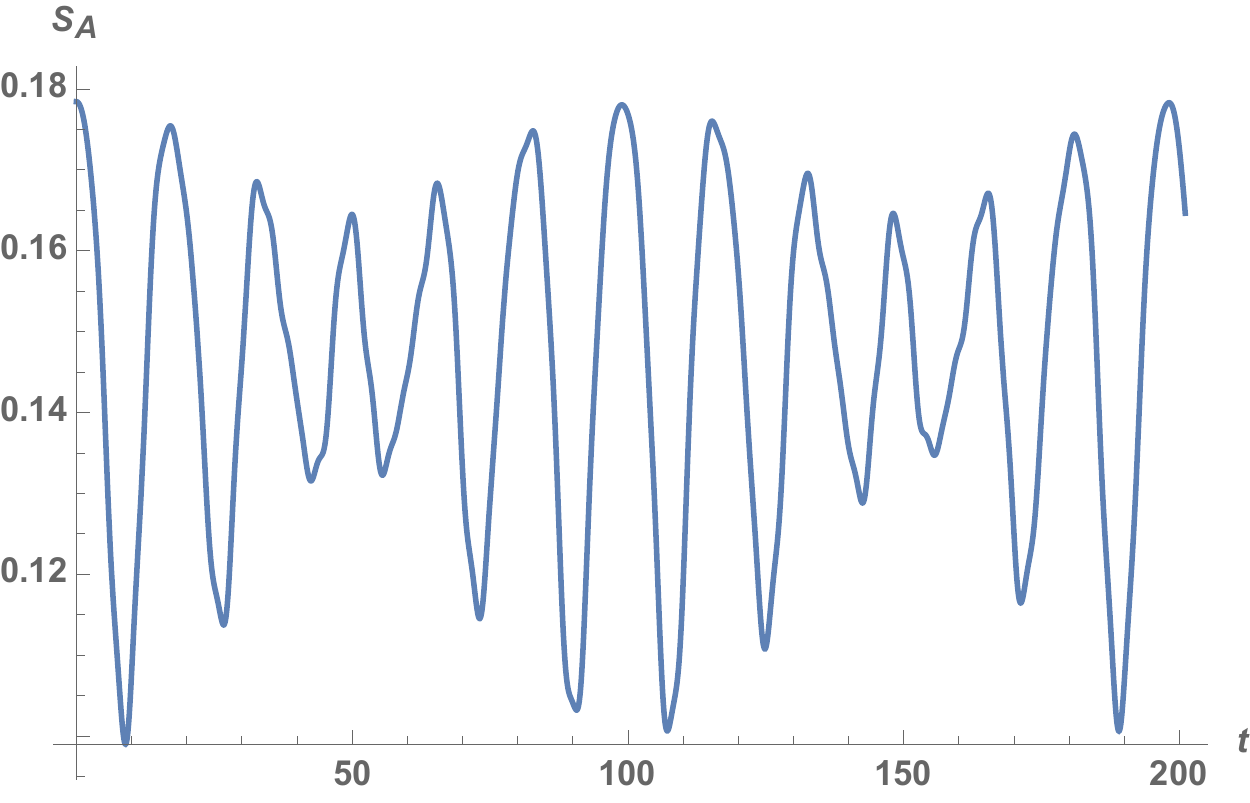}
          \hspace{1.6cm} (b) $0\leq t\leq 64\pi$
        \end{center}
      \end{minipage}
      \end{tabular}
    \caption{Splitting local quench in spin-1/2 transverse Ising model: time evolution of EE $S_A$ for $J=h=1$, $L=6$, and $j=3$, where the subsystem $A$ is chosen to be $A = \{i|i=4,5\}$. We can see oscillation in long time range, while the short-time behavior of $S_A$ is very similar to those in CFT cases.\label{fig:STI}}
  \end{center}
\end{figure}

Then let us start with the ground state of $H_{before}$ and investigate its dynamics under $H_{after}$ at $t\geq 0$. We presented our numerical results in Fig.\ref{fig:STI}.
We can see the behavior of its EE at an early time is very similar to
the result in free fermion  (\ref{sab}) and the disconnected contribution in holographic CFTs (\ref{hpsb}), qualitatively. At late time we observe oscillatory behavior because our spin system has a finite size.

\section{Joining Local Quench}\label{sec:joining}

\begin{figure}
  \centering
  \includegraphics[width=7cm]{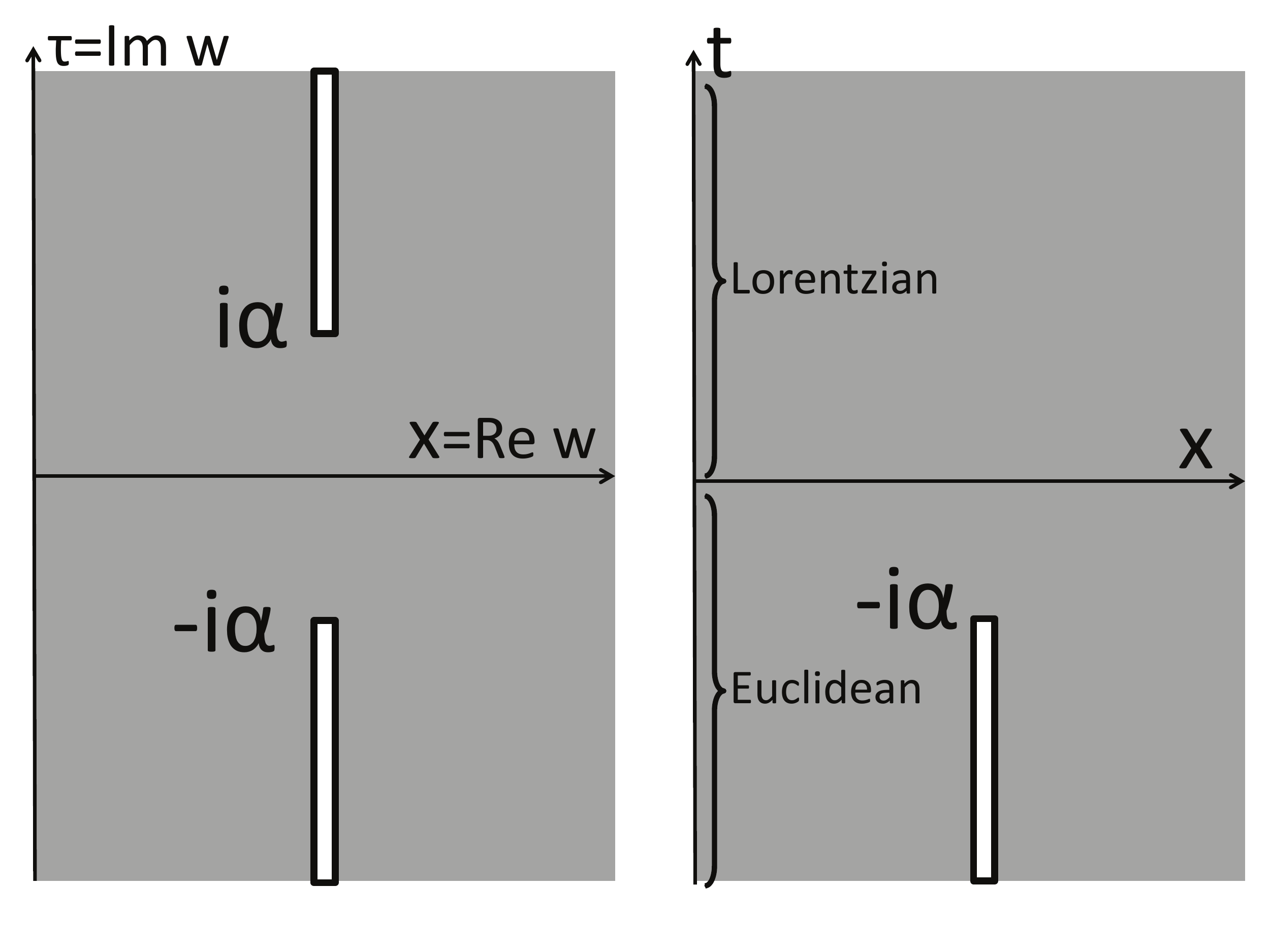}
 \caption{The geometries which realize the joining local quenches. The left figure describes the space on which we perform the Euclidean path-integral. This is mapped into an upper half plane by the map
 (\ref{pastef}). The right picture describes the path-integral realization of the time evolution after
 the joining process happened at $t=0$, where the Euclidean path-integral for $\tau<0$ creates the state just after this process.}
\label{fig:PSw}
\end{figure}

Consider a 2d CFT on two semi-infinite lines $x>0$ and $x<0$. We join each endpoint $x=0$ at $t=0$
as in the right picture of Fig.\ref{fig:SPPS}. This setup is described by the path-integral depicted in
Fig.\ref{fig:PSw}. We can map the Euclidean space (the left picture) into an upper half plane by the conformal map:
\be
\xi=i\s{\f{i\ap-w}{i\ap+w}}\equiv f(w).  \label{pastef}
\ee
The parameter $\ap$ again corresponds to the regularization of local quench.
We choose the subsystem $A$ to be $a\leq x\leq b$ at time $t$ as before. This corresponds to
\ba
w_a=a-t,\ \ \ \bar{w}_a=a+t,\ \ \  w_b=b-t,\ \ \ \bar{w}_b=b+t,\ \ \
\ea
where we performed the analytical continuation of the Euclidean time $\tau=it$.

\subsection{Joining Local Quench in Dirac Free Fermion CFT}

Consider a massless Dirac fermion CFT in this joining quench.
 The time evolution of EE can be computed from the formula (\ref{saf}) and (\ref{diracf}) as in the previous analysis of splitting quench.

The analytical expressions of $S_A$ at the time $t$ when $\ap$ is infinitesimally small, are given as follows (we choose $A$ to be the interval $a\leq x \leq b$ and we can again assume (\ref{asu}) without losing generality).
In the early time period, $0<t<|a|$, we have
\ba
S_A=\frac{1}{6}\log \frac{4|a|b(b-a)^2}{(b+|a|)^2\ep^2}, \label{psqa}
\ea
When $|a|<t<b$, we have
\ba
S_A=\frac{1}{6}\log \frac{4(b-a)b(b-t)(t^2-a^2)}{\ap(a+b)(b+t)\ep^2},\ \ \ \label{psqb}
\ea
At late time, $t>b$, we have
\ba
S_A=\frac{1}{3}\log (b-a)/\ep.  \label{psqc}
\ea
It is useful to note that when $b=-a>0$, $S_A$ takes a constant value for any $t$:
$S_A=(1/3)\log \left(2b/\ep\right)=S^{(0)}_A$, as we found also for the splitting quench.

\begin{figure}
  \centering
  \includegraphics[width=5cm]{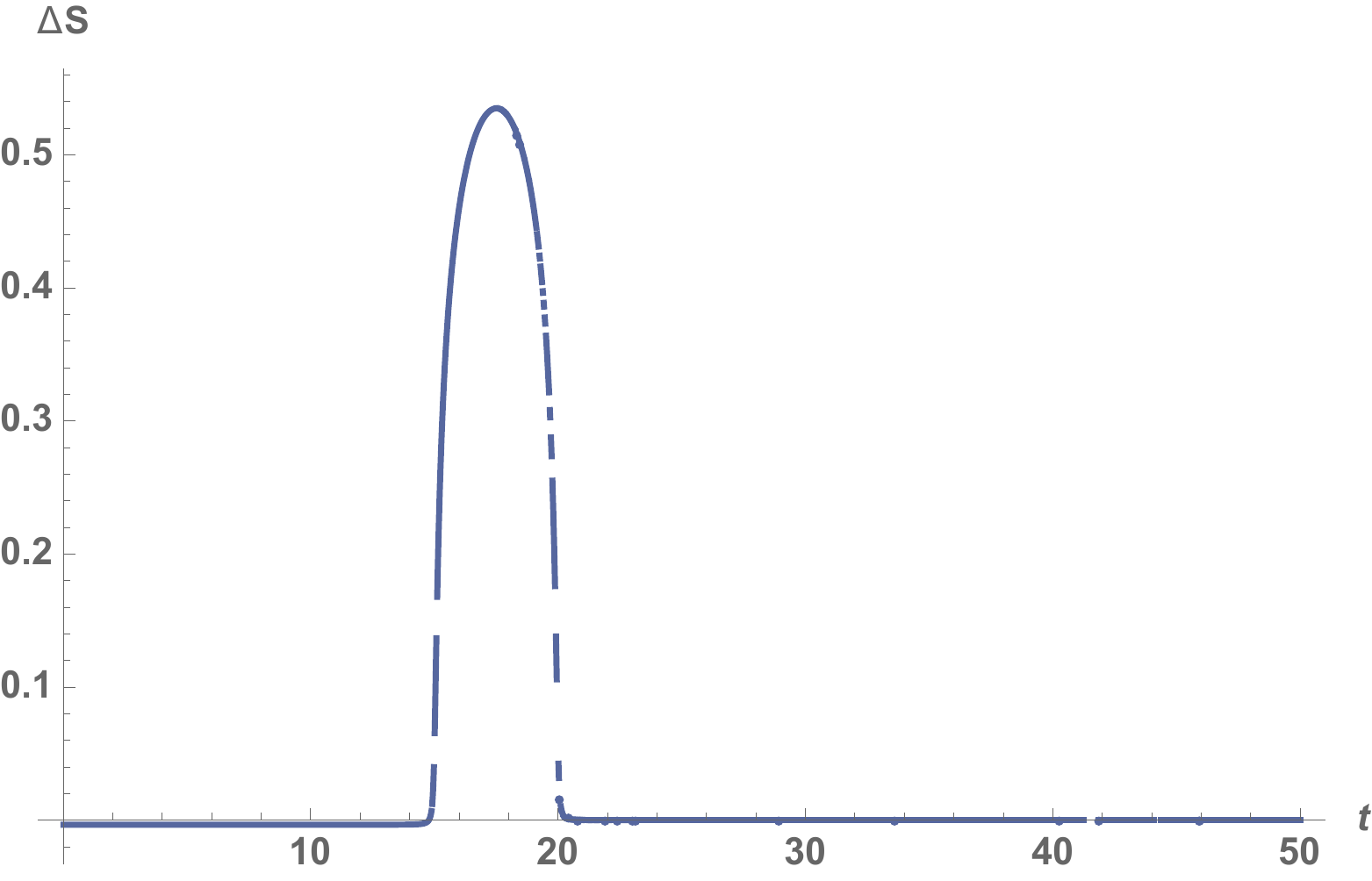}
  \includegraphics[width=5cm]{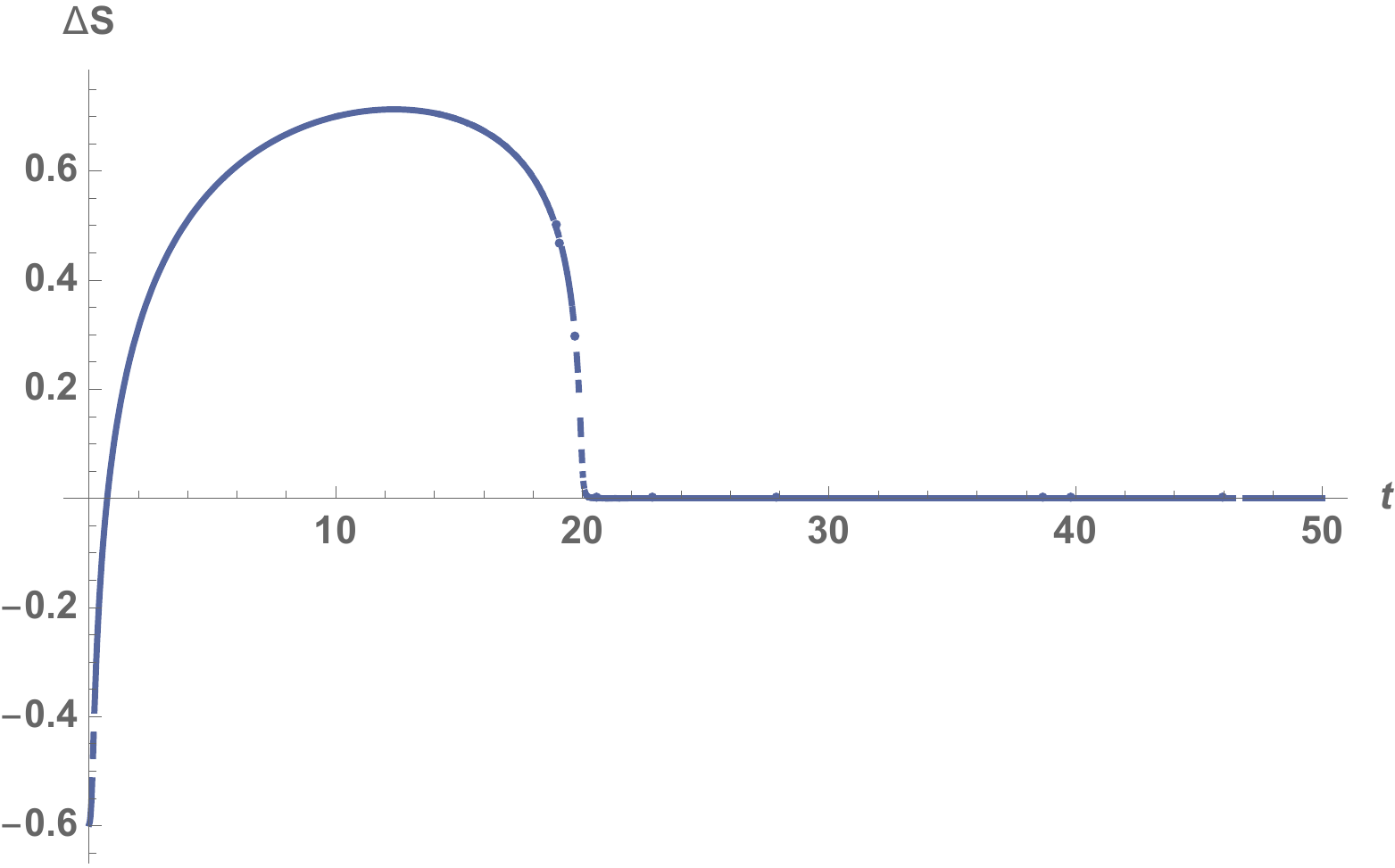}
  \includegraphics[width=5cm]{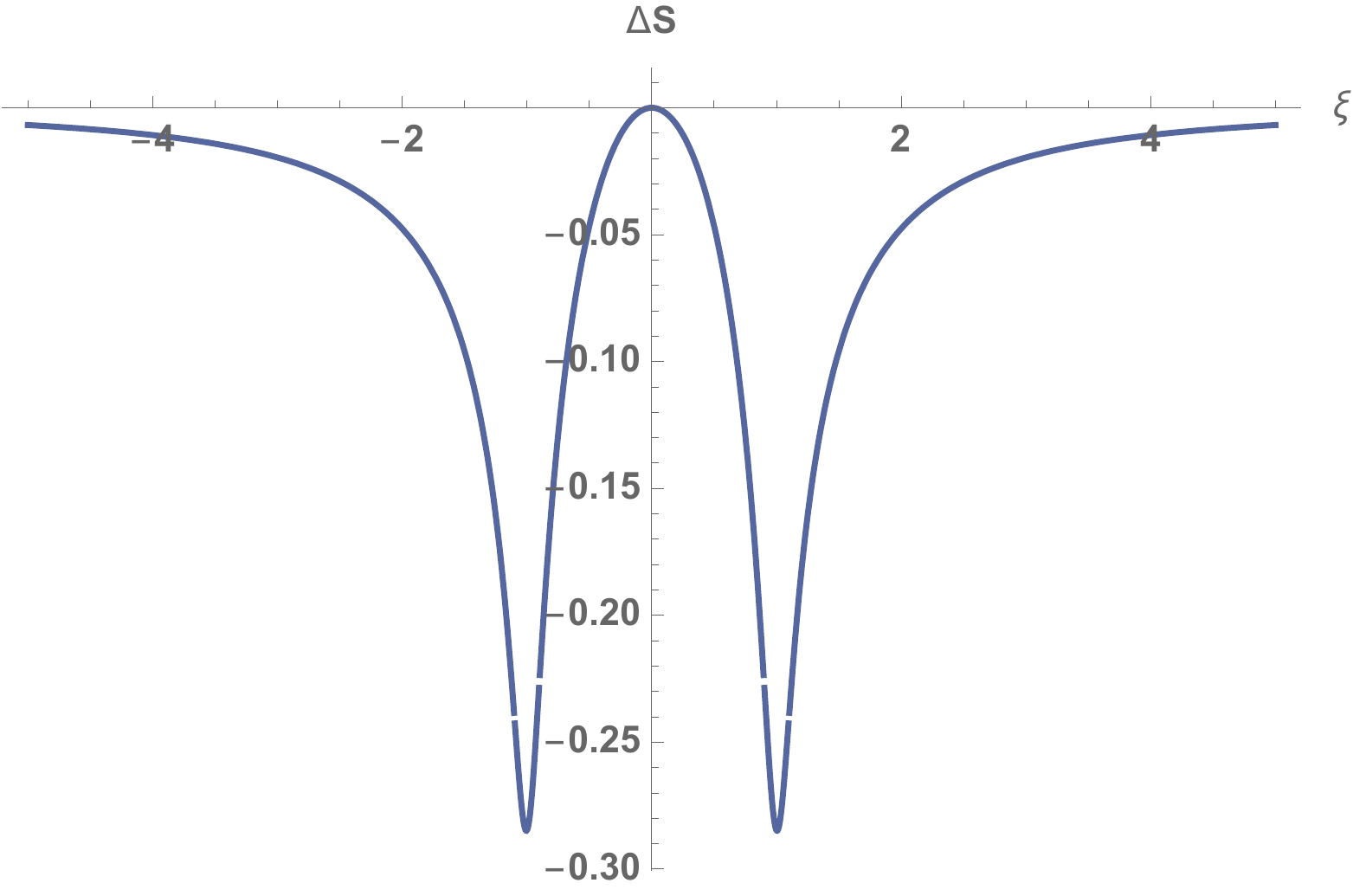}\\
 \caption{The plots of the entanglement entropy growth $S_A-S^{(0)}_A$ for $\ap=0.1$ for the joining local quench in the Dirac Fermion CFT. The left graph describes the time evolution when we take $(a,b)=(15,20)$. The middle one is the time evolution for  $(a,b)=(0.1,20)$. The right one
 is the plot for various $\xi$ when we fix $l=2$ and $t=0$.}
\label{fig:DPL}
\end{figure}

The behavior of entanglement entropy is also numerically plotted in Fig.\ref{fig:DPL}. The left
graph shows that the EE gets larger during only the time period where one of the entangled
pair is included in the subsystem A, while the EE is the same as its vacuum value for the
other time period. This fact is also true in the above analytical results as well as in all of
our numerical plots. This property is consistent with the relativistic particle propagation
picture for the massless Dirac fermion CFT.

The middle graph shows a logarithmic growth at an early time. Indeed, from (\ref{psqb}),
we find that when $|a|\ll t\ll b$, the EE shows the logarithmic behavior:
\be
S_A\simeq \frac{1}{3}\log\frac{2t}{\ep}+\frac{1}{6}\log\frac{l}{\ap}, \label{pslogg}
\ee
which agrees with the known local quench behavior \cite{CCL}.

\begin{figure}
  \centering
   \includegraphics[width=6cm]{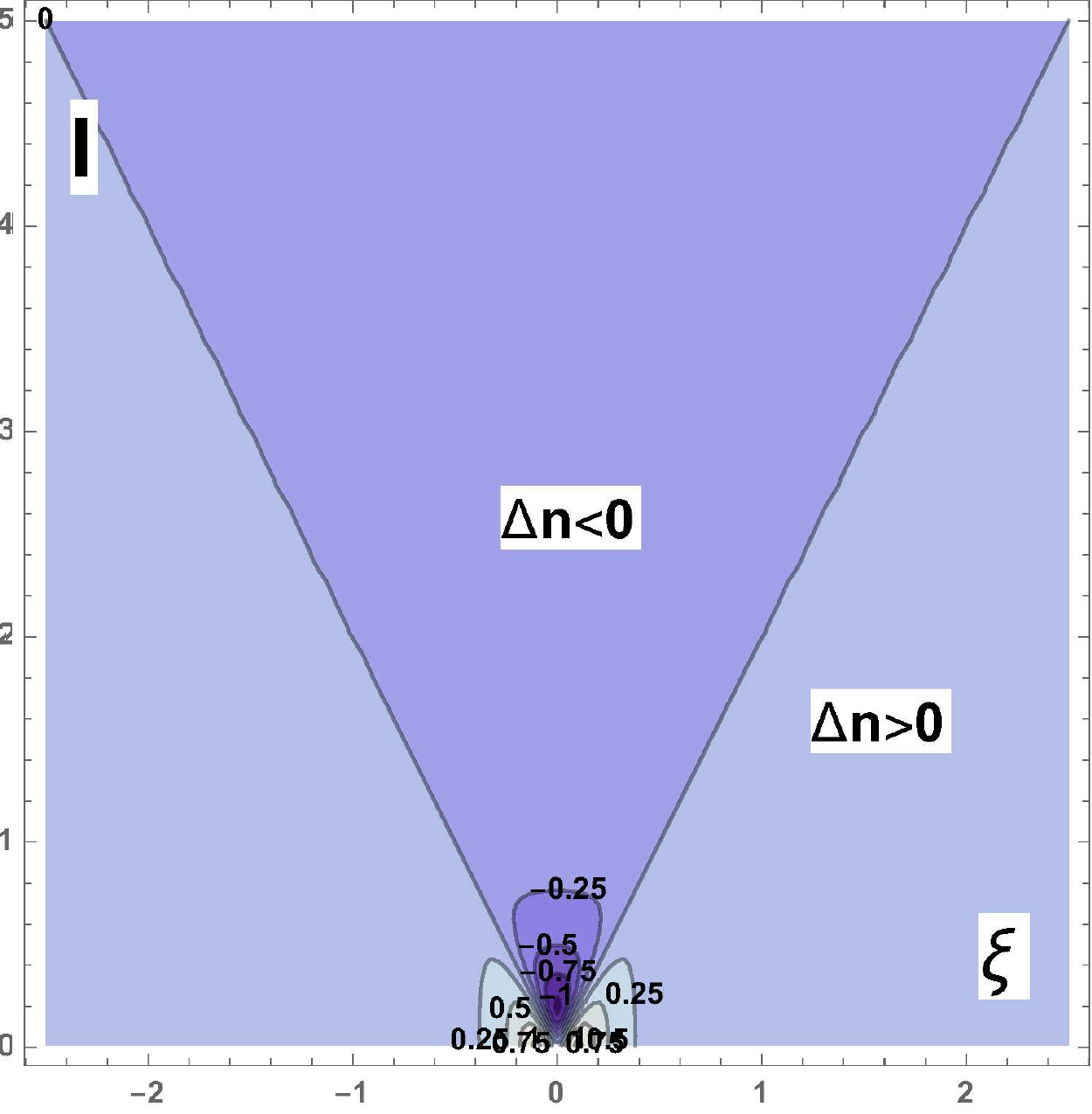}
  \includegraphics[width=6cm]{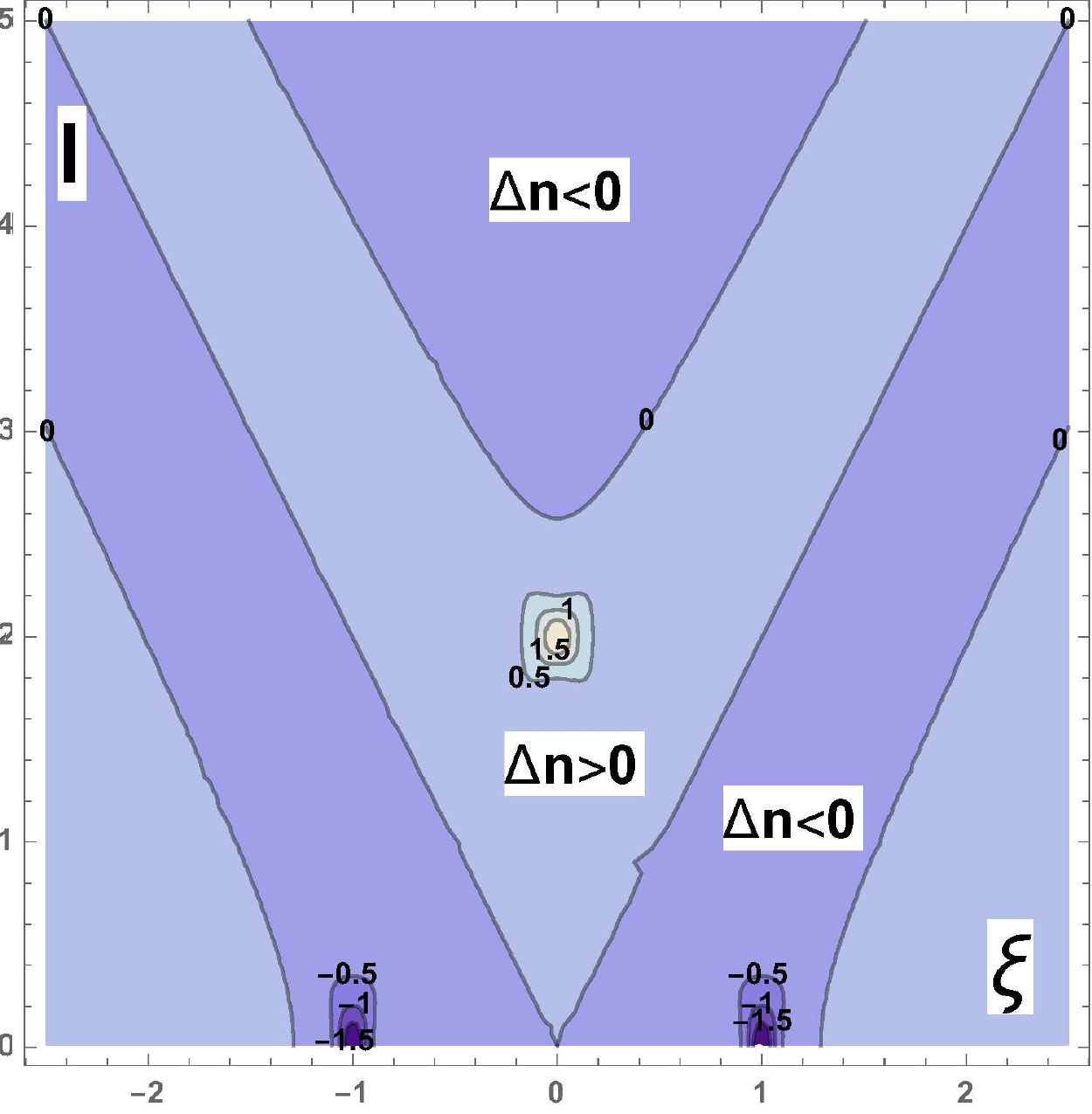}
\includegraphics[width=6cm]{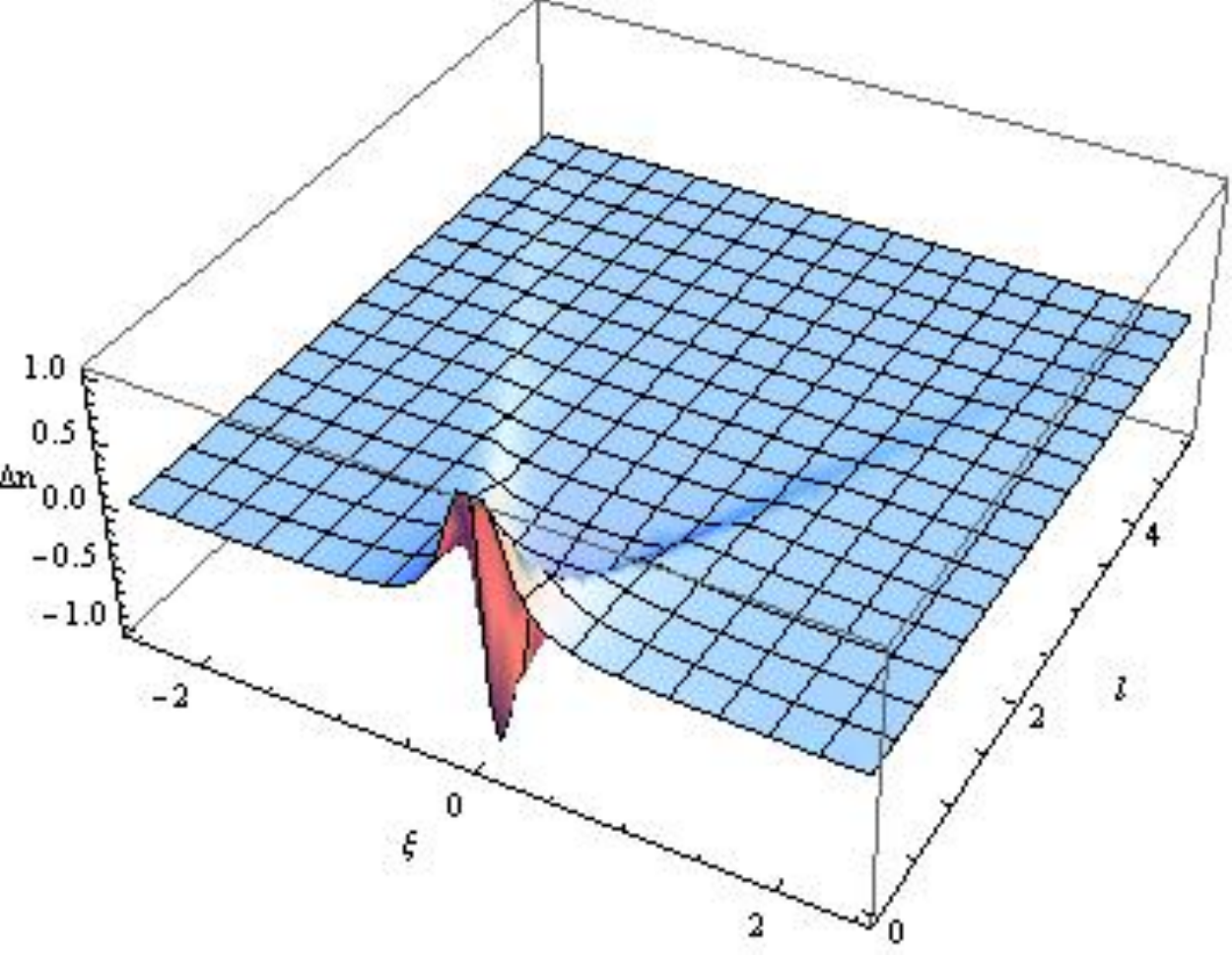}
\includegraphics[width=6cm]{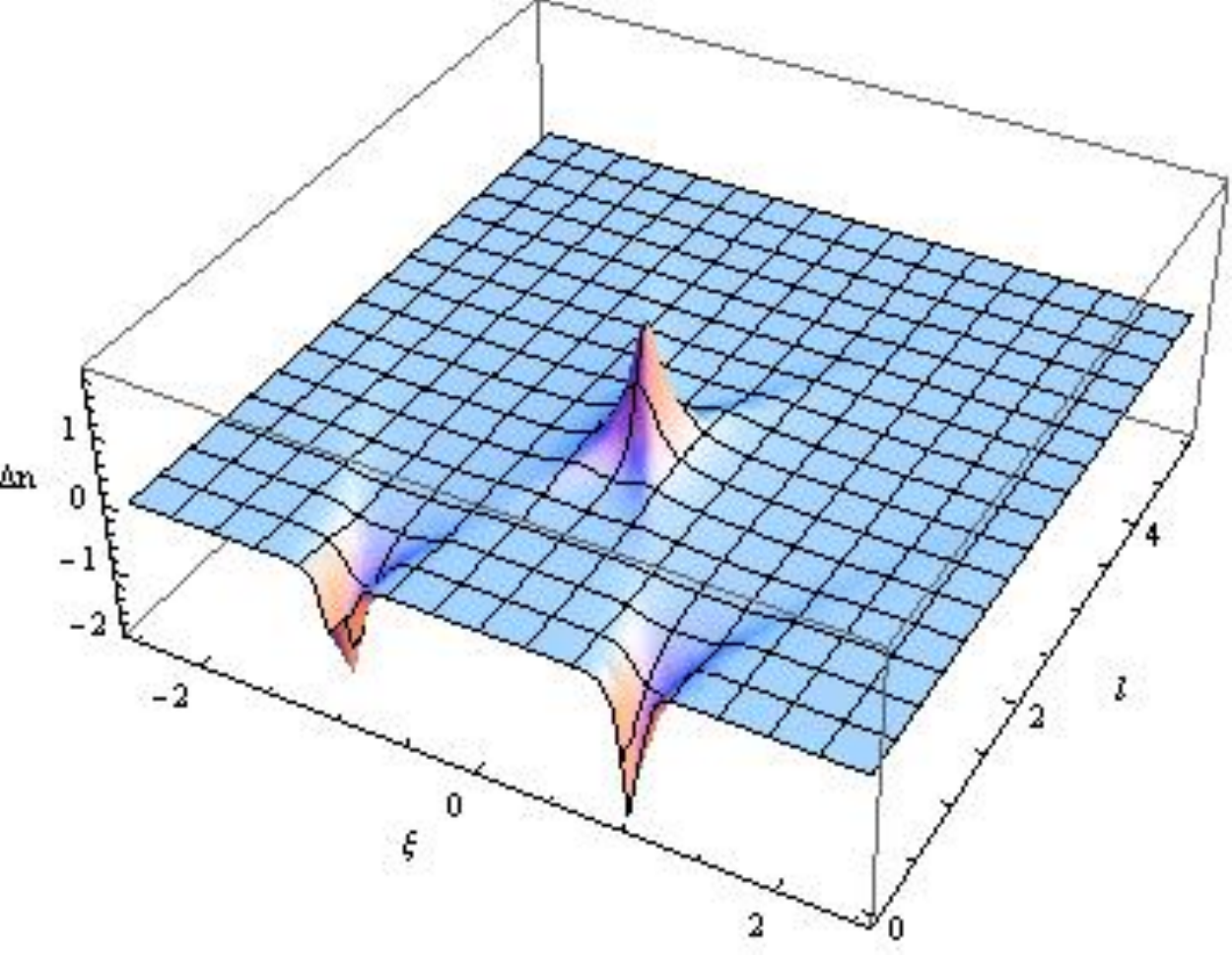}
 \caption{The plots of the entanglement density $\Delta n(\xi,l,t)$ for $\ap=0.1$
 at $t=0$ (left) and $t=1$ (right).
 The horizontal and depth coordinate corresponds to $\xi$ and $l$, respectively.}
\label{fig:DLPQ}
\end{figure}

The entanglement density (ED) can also be computed by taking differentiations of this entropy.
The results are plotted in Fig.\ref{fig:DLPQ}.  We can clearly see the positive peak at $(l,\xi)=(2t,0)$ as well as the negative peaks at $(l,\xi)=(0,\pm t)$. The former agrees with the relativistic particle picture. The latter coincides the peaks of the energy density. Even though in the previous splitting case of the Dirac fermion CFT, the negative region $\Delta n<0$ around $(l,\xi)=(t,0)$ was expanding, in the present joining case, the expanding region at the same place is positive $\Delta n>0$. This indeed violates the naive causality constraint (refer to the right picture in Fig.\ref{fig:Integ}) and we can again interpret this as the presence of non-local entanglement in the initial state just after the quench. This leads to the logarithmic growth at late time in the joining case (\ref{pslogg}) as opposed to the splitting case.

\subsection{Joining Local Quench in Holographic CFT}

For the holographic CFT, we can substitute the coordinate transformation (\ref{pastef}) to (\ref{corads}) in order to calculate the EE. The dual geometry is given by Im$[\xi]\geq 0$ in the Poincar\'{e} AdS coordinate (\ref{pol}).
The HEE can be computed as (\ref{conhol}) and (\ref{dishol}) using the AdS/BCFT, as already analyzed in the earlier paper \cite{Ugajin:2013xxa}.

The analytical expressions of $S^{con}_A$ and $S^{dis}_A$ at the time $t$ when $\ap$ is infinitesimally small, are given as follows (we choose $A$ to be the interval $a\leq x \leq b$ with (\ref{asu})).
In the early time period, $0<t<|a|$, we have
\ba
 S^{con}_A&=&\left\{\begin{aligned}\frac{c}{3}\log (b-a)/\ep, \quad\qquad\qquad (a>0),\\ \frac{c}{6}\log \frac{4(a^2-t^2)(b^2-t^2)}{\ap^2\ep^2}, \ \ \ (a<0)\end{aligned}\right.\no 
 S^{dis}_A&=&\frac{c}{6}\log \left(4b|a|/\ep^2\right)+2S_{bdy}. \label{hppa}
\ea

When $|a|<t<b$, we have
\ba
S^{con}_A=\frac{c}{6}\log \frac{2(b-a)(t-a)(b-t)}{\ap\ep^2},\ \ \
 S^{dis}_A=\frac{c}{6}\log \frac{4b(t^2-a^2)}{\ap\ep^2}+2S_{bdy}. \label{hppb}
\ea
At late time, $t>b$, we have
\ba
S^{con}_A=\frac{c}{3}\log (b-a)/\ep, \ \ \
 S^{dis}_A=\frac{c}{6}\log \frac{4(t^2-b^2)(t^2-a^2)}{\ap^2\ep^2}+2S_{bdy}.  \label{hppc}
\ea
In the late time limit $t \gg b$, we always find $S^{con}_A=S^{(0)}_A$ dominates as we expect.

\begin{figure}
  \centering
  \includegraphics[width=5cm]{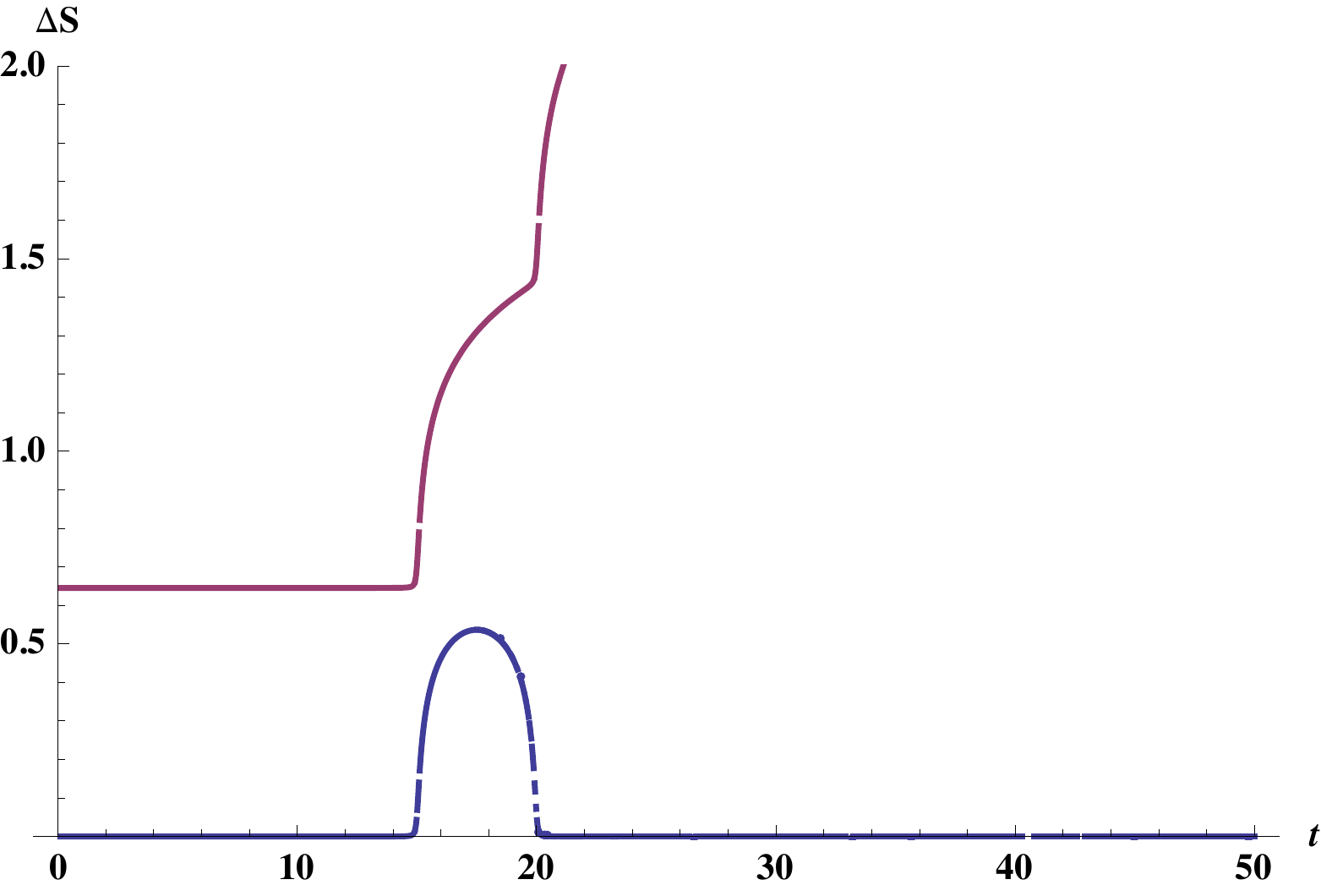}
  \includegraphics[width=5cm]{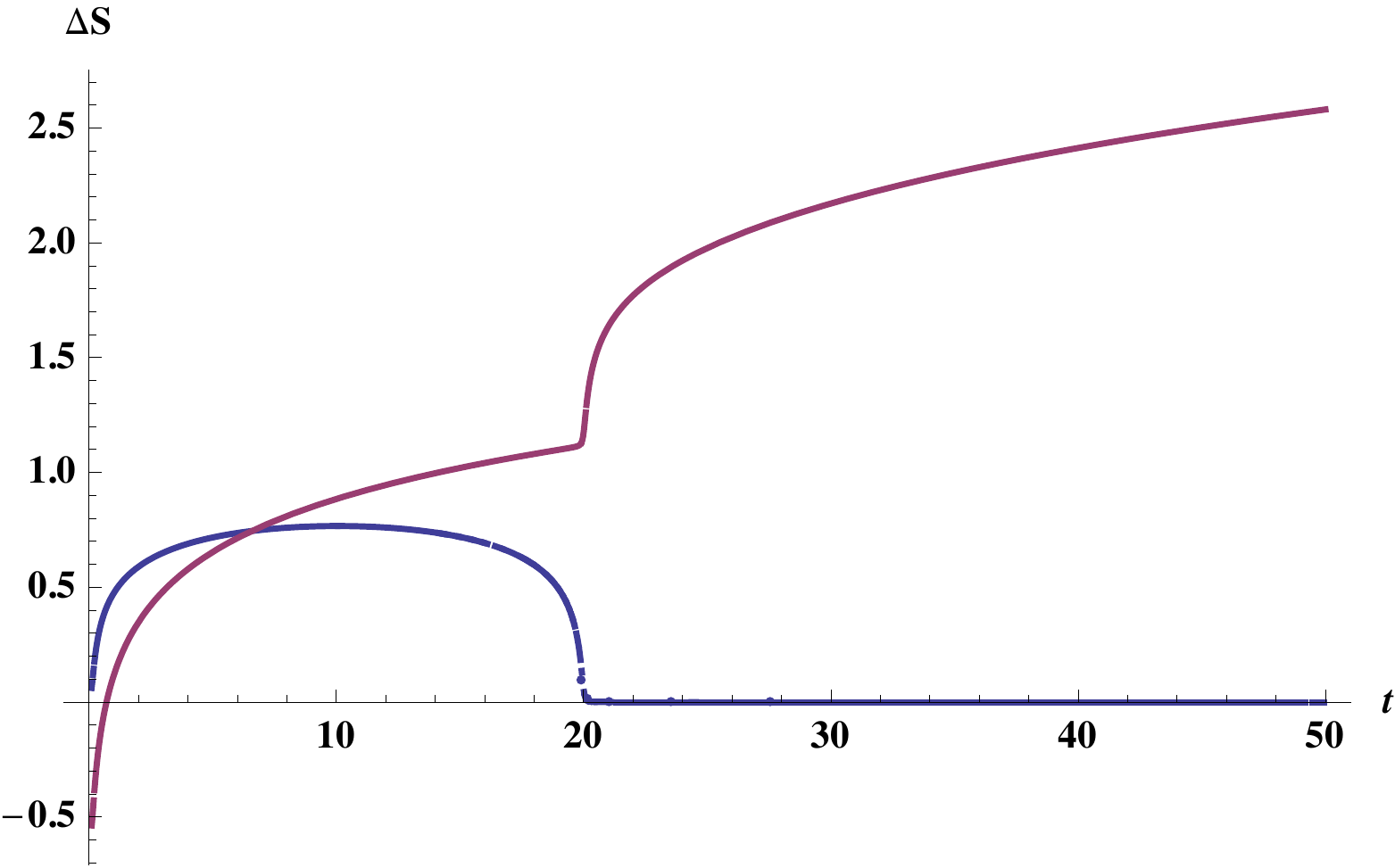}
  \includegraphics[width=5cm]{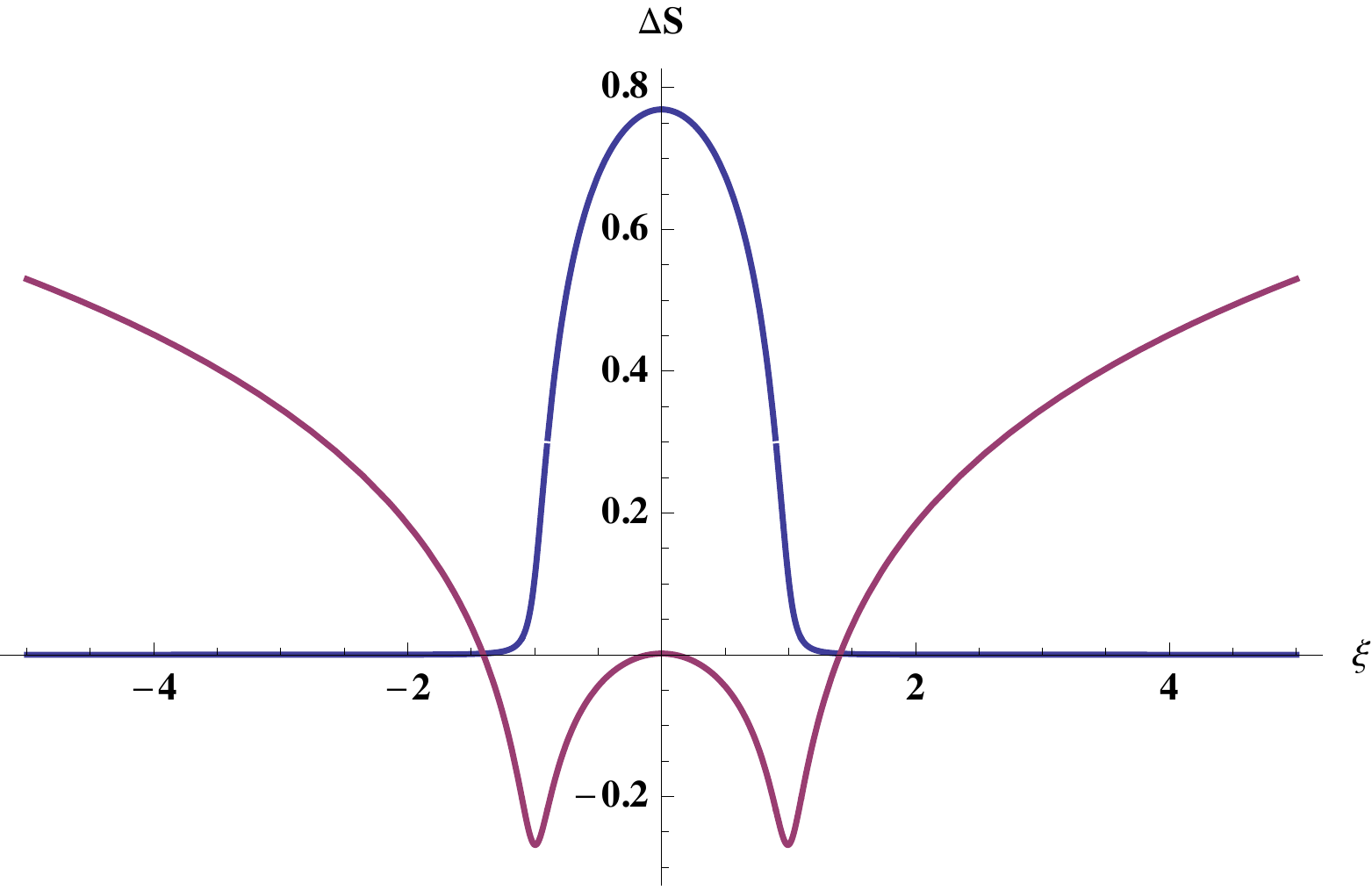}\\
 \caption{The plots of the holographic entanglement entropy growth $S_A-S^{(0)}_A$ for $\ap=0.1$ for the holographic joining local quench. We set the tension vanishing $T_{BCFT}=0$. The blue and red curve describes
 $S^{con}_A$ and $S^{dis}_A$, respectively. The left graph describes the time evolution when we take $(a,b)=(15,20)$. The middle one is the time evolution for  $(a,b)=(0.1,20)$. The right one
 is the plot for various $\xi$ when we fix $l=2$ and $t=0$.}
\label{fig:HPL}
\end{figure}

The behavior of entanglement entropy is also numerically plotted in Fig.\ref{fig:HPL}. By comparing
the holographic results (Fig.\ref{fig:HPL}) for vanishing boundary entropy $S_{bdy} = 0$ with
those for the Dirac fermion's one (Fig.\ref{fig:DPL}), we find they agree with each other qualitatively,
if we ignore the phase transition behavior in the former. If we take $|a| \ll t \ll b$,
then we find the logarithmic growth:
\be
S_A=S^{dis}_A\simeq \frac{c}{3}\log\frac{2t}{\ep}+\frac{c}{6}\log\frac{l}{\ap}, \label{ddd}
\ee
which indeed agrees with the Dirac fermion result (\ref{pslogg}).

However if the boundary entropy is large enough for $S^{con}_A$ to be favored, then we find
a slower logarithmic growth for $|a| \ll t \ll b$:
\be
S_A=S^{con}_A\simeq \frac{c}{6}\log\frac{2t}{\ap}+\frac{c}{3}\log\frac{l}{\ep}. \label{kkh}
\ee

\begin{figure}
  \centering
 \includegraphics[width=6cm]{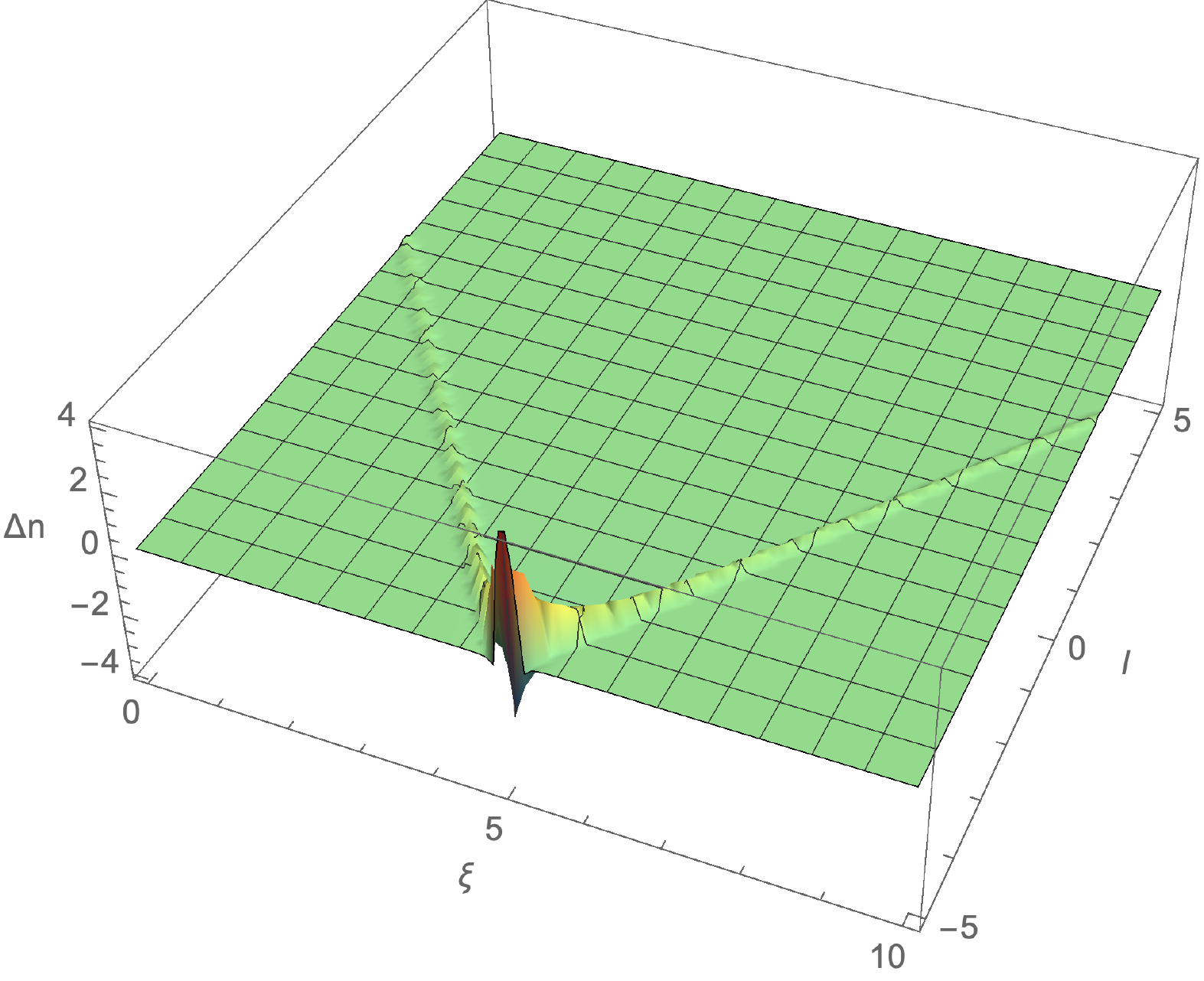}
   \includegraphics[width=6cm]{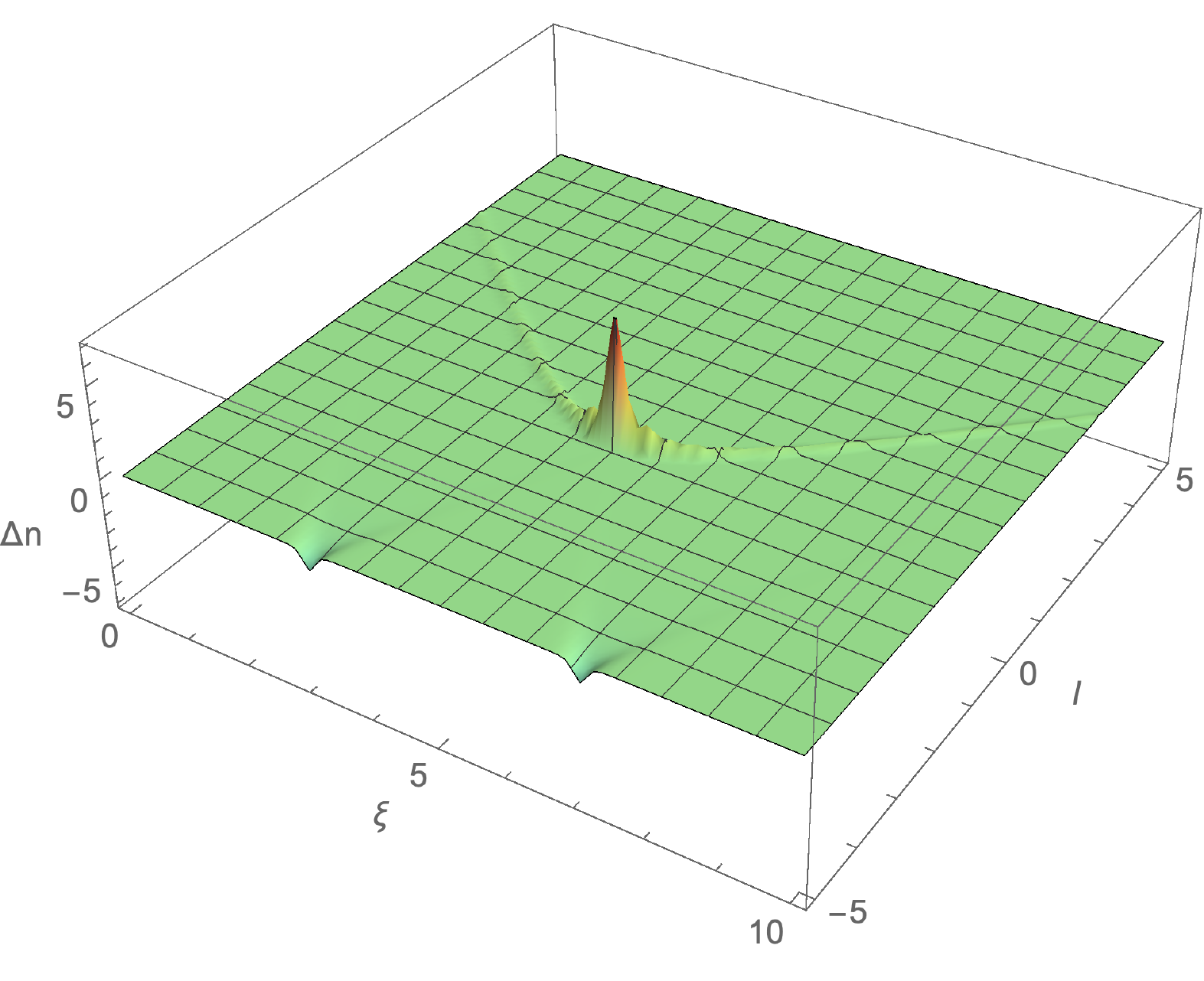}
    \includegraphics[width=6cm]{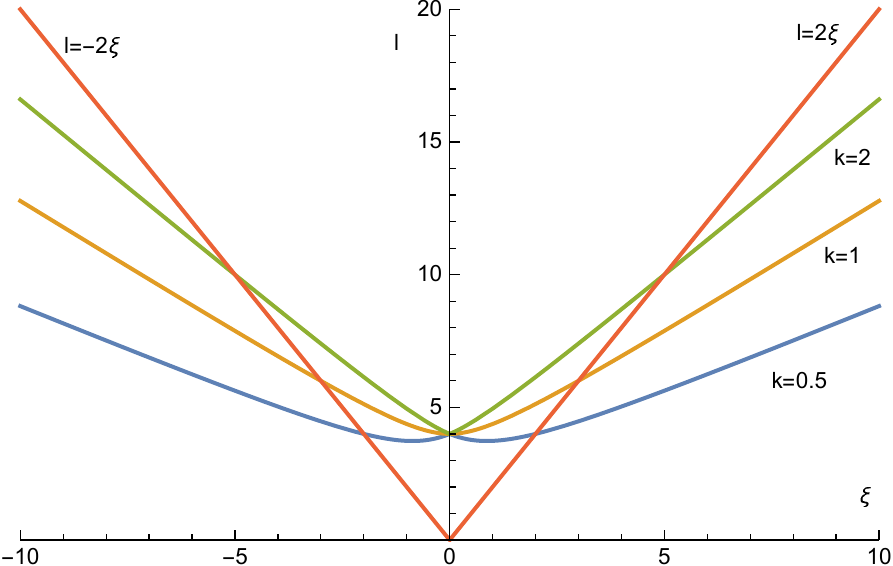}
       \includegraphics[width=6cm]{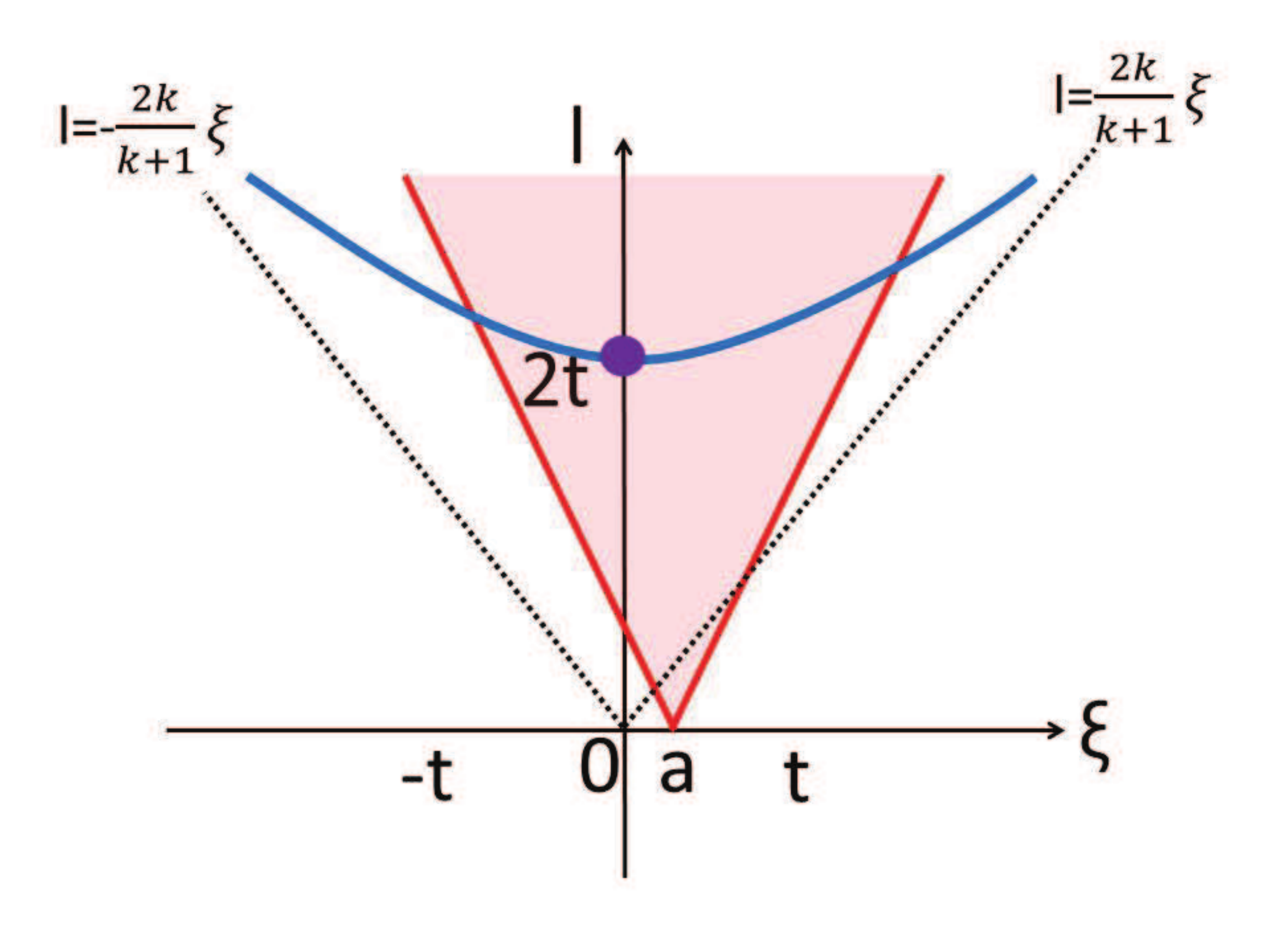}
 \caption{The behaviors of a smeared entanglement density for the holographic joining local quench. The
 upper left and right graph are the plots of $\Delta n(\xi,l,t)$ at $t=0$ and $t=2$, respectively, with $\ap=0.1$ (again we set $T_{BCFT}=0$). The horizontal and depth coordinate corresponds to $\xi$ and $l$, respectively. Here we smeared the originally delta functional behavior of the entanglement density by replacing the second derivatives w.r.t $l$ and $\xi$ with finite-differences $\delta=0.1$.
 We observe a delta functional behavior on the curve (\ref{curvpdf}) at $k=1$.
 The lower left picture describes the curve  (\ref{curvpdf})  for $k=0.5$ (blue), $k=1$ (orange) and
 $k=2$ (green) at $t=2$ as well as the line $l=2|\xi|$ (red). The lower right picture shows the computation of $S_A$ for $a\ll t\ll b$ by integrating the ED over the red region. The blue curve represents the peaks on (\ref{curveSP}). As the time evolves, the blue curve enters into the red region from both the left and right side. This explains the logarithmic growth of EE with the doubled coefficient.}
\label{fig:HPSED}
\end{figure}

The entanglement density is plotted in Fig.\ref{fig:HPSED}. In the upper two graphs, we
again smeared the derivatives in (\ref{delk}) by replacing it with the finite difference.
We observe that in addition to the expected positive peak at $(\xi,l)=(0,2t)$  and the two
negative ones at $(\xi,l)=(\pm t,0)$, we find a continuous peak along the following curve (we again defined $k$ as in (\ref{kdef})):
\ba
l=t-\frac{|\xi|}{1+k}+\s{\left(t-\frac{|\xi|}{1+k}\right)^2+\frac{4k}{k+1}(t+|\xi|)|\xi|}. \label{curvpdf}
\ea
The profile of this curve is plotted in the lower left picture of Fig.\ref{fig:HPSED}.
 This continuous peak on (\ref{curveSP})  arises since there is a phase transition from $S^{dis}$ to $S^{con}$  as we can see from (\ref{hppb}). At $t=0$, this curve is reduced to the line
 $l=\frac{2k}{k+1}|\xi|$. In the limit $t\to\infty$ and $|\xi|\to\infty$ we have
\ba
l\simeq 2t+\frac{2(k-1)}{k+1}|\xi|,\ \  (t\to\infty),\ \ \ \
l\simeq \frac{2k}{k+1}|\xi|+\frac{4k}{2k+1}t, \ \ (|\xi|\to\infty).
\ea

The presence of the codimension one peak is again peculiar to holographic CFTs and is missing in the free Dirac fermion CFTs or more generally RCFTs. Its presence shows that initial entanglement gets modified in highly non-local way. Moreover, this time, the curve (\ref{curvpdf}) extends to the region outside of the region $l>2|\xi|$ (i.e. the red region). Therefore the blue curve enters into the red region from both the left and right side as in the lower right picture of Fig.\ref{fig:HPSED}.
 This explains the logarithmic growth of $S_A$ (\ref{ddd}) with the doubled coefficient compared with that for the local operator quench (\ref{oplog}). Moreover, if $k$ gets very large, the curve (\ref{curvpdf}) approaches to $l\simeq 2|\xi|$ in the limit $|\xi|\to\infty$. Therefore the situation gets very similar to the local operator quench and therefore this explains the behavior (\ref{kkh}). Also we will provide the geometric explanation of the $\log t$ behaviors in section \ref{sec:log}.

\subsection{Holographic Geometry of Joining Local Quench}

\begin{figure}
  \centering
  \includegraphics[width=8cm]{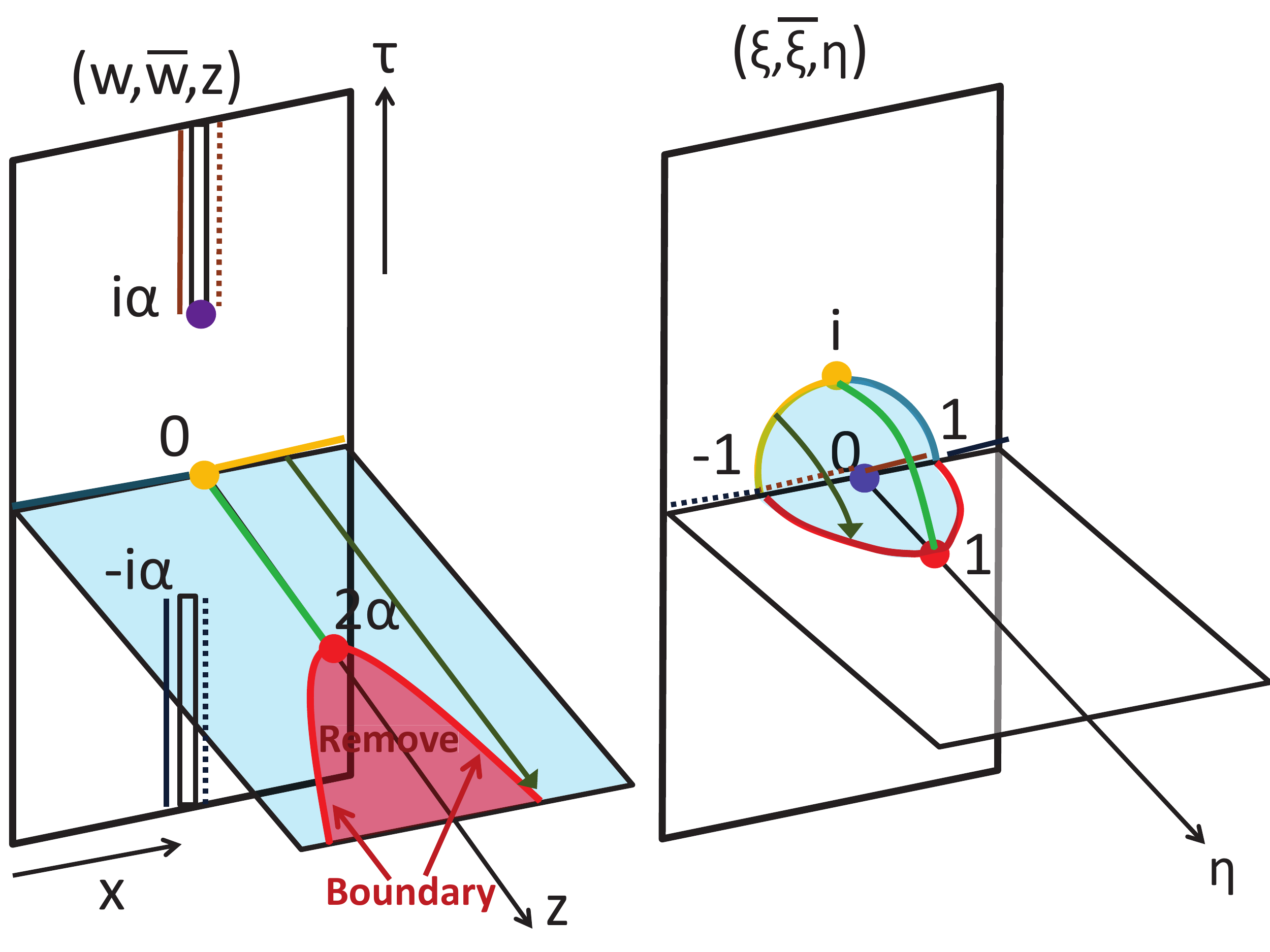}
\caption{The map (\ref{corads}) between the gravity dual of joining local quench (left) and
the upper half of the Poincar\'{e} AdS$_3$ (right). The red curve in the left is given by $z=2(x^2+\ap^2)/\ap$
and this corresponds to the boundary surface $Q$. Thus the region inside this curve should be
removed.}
\label{fig:PSmapt0}
\end{figure}

Now let us study the geometry of the gravity dual for the joining local quench. Since this has
many similarities with that for the splitting local quench, our explanation will be brief.
Indeed since the energy stress tensor (\ref{sttw}) remains the same, the metric of gravity dual
for the joining quench is the same as that for the splitting one. The difference comes
from the location of the boundary surface $Q$. The time slice $t = 0$ in the $(w,\bar{w}, z)$
coordinate is again mapped into the quarter sphere $|\xi|^2+\eta^2 = 1$ in the half Poincar\'{e} AdS:
Im$[\xi] > 0$. However the details of the mapping
is different from the previous one as sketched in Fig.\ref{fig:PSmapt0}.
In the present case, the region
(\ref{identif}) corresponds to Im$[\xi] < 0$ and thus should be removed. Its boundary (the red curve)
given by $z = 2(x^2+\ap^2)/\ap$ now represents the boundary surface $Q$.
The region $-i\ap < \tau < i\ap$ and $x = 0$ is now mapped into a region Re$[\xi] = 0$ in the half
Poincar\'{e} AdS$_3$. Since this map is identical to that for the splitting quench i.e. Fig.\ref{fig:SPmapt0a}, we will omit its detail.

\begin{figure}
  \centering
  \includegraphics[width=8cm]{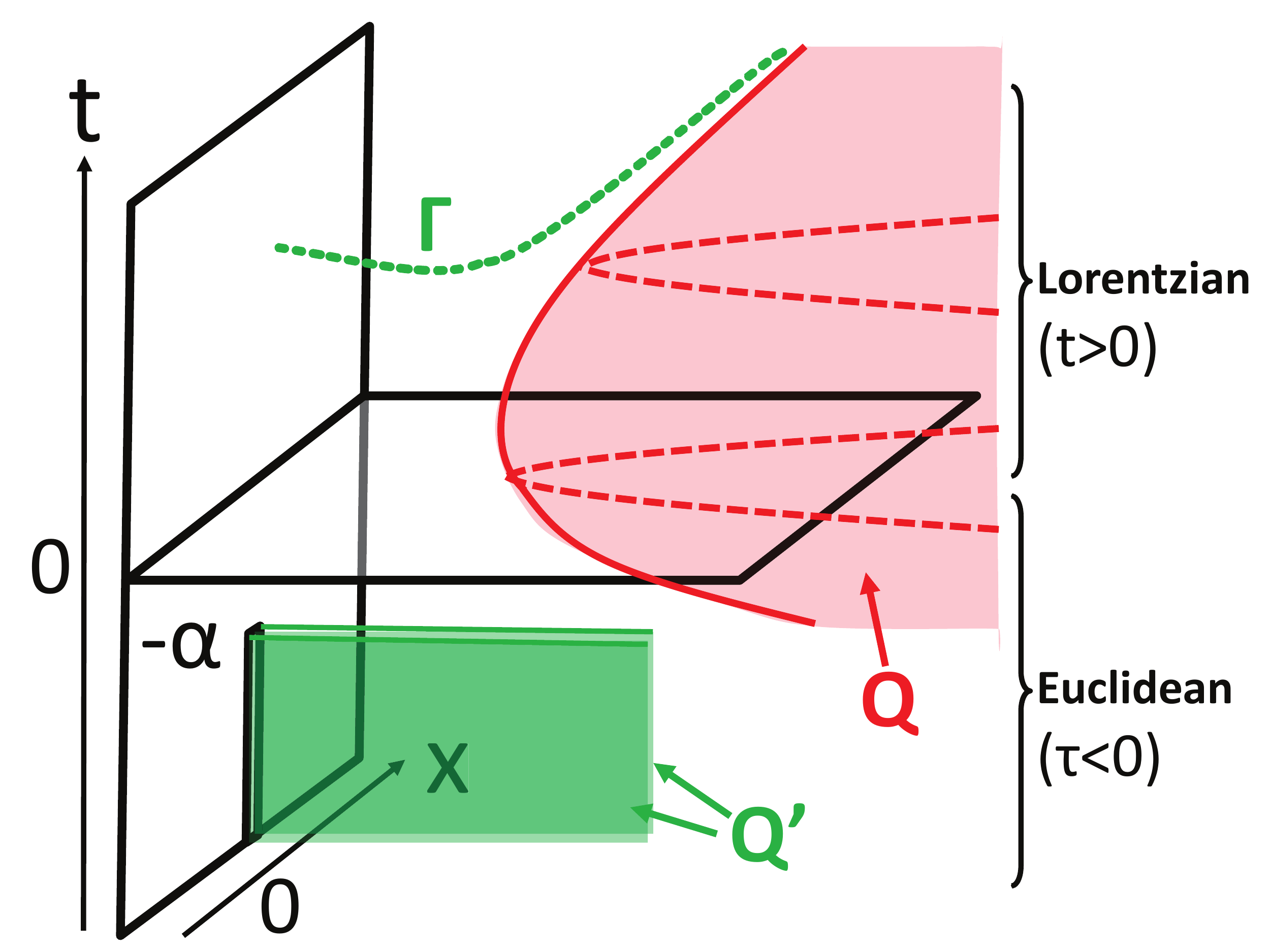}
\caption{The spacetime geometry of gravity dual of the joining local quench. The red
region defined by $z>G_L(x,t)$
should be removed. The boundary of this region, which we call the red surface,
corresponds to the boundary surface $Q$. Though the green
surface $Q'$ is also boundary surface, it should be identified with the red one.}
\label{fig:PSHOLgeo}
\end{figure}

Finally we obtain the spacetime geometry sketched in Fig.\ref{fig:PSHOLgeo}. The boundary surface
$Q$, expressed by the red surface, is given by the same expression (\ref{idencc}) and (\ref{idenccl})
in the Euclidean and Lorentzian case, respectively. In the Euclidean geometry, there is the green surface $Q'$ which extends from the CFT boundary and this is actually identified with the red surface $Q$, whose details we will omit as they are involved with complicated numerics. In some sense, in this geometry, the location of the boundary surface $Q$ is opposite to that for the splitting local quenches.

In summary, the time evolution after the joining local quench is described
by the spacetime with the boundary surface $Q$ which moves toward the horizon at the
speed of light.\footnote{This motion of boundary surface is analogous to a string world-sheet
description of holographic local quenches proposed in \cite{AM}, though the choice of geodesics responsible for the HEE is different from ours.} This intuitively
agrees with our expectation that after the joining, the
CFT state gradually approaches to the vacuum state on a connected line. It is also intriguing to note that in the Lorentzian description, the boundary surface $Q$ does not end on the AdS boundary but is localized in the bulk.

So far we implicitly assumed $T_{BCFT}=0$. For more general cases $T_{BCFT}\neq 0$,
the AdS/BCFT setup in $(\xi,\bar{\xi},\eta)$ coordinate is given by the boundary surface
(\ref{tensionb}). Then we can map $Q$ into the original coordinate $(w,\bar{w},z)$
as in Fig.\ref{fig:PSmapt0}. When $T_{BCFT}>0$ $(T_{BCFT}<0)$, the boundary surfaces $Q$ is moved in the larger (smaller) $z$ direction. In other words, the physical region of the gravity dual, surrounded by the boundary surface $Q$ and the AdS boundary $z=0$, expands as $T_{BCFT}$ gets larger as expected.

In this subsection we studied the Lorentzian motion of the boundary surface $Q$.
For the holographic entanglement entropy, we actually need the profile of geodesic in Lorentzian 
geometry, though in the previous subsection we employed the formal Wick rotation arguments to avoid this
analysis. We will explicitly study how the Lorentzian geodesic looks like in section \ref{sec:log}.

\section{Log $t$ Geodesic Length in Holographic Lorentzian Geometry}\label{sec:log}

In our previous calculations of holographic local quenches we mainly worked in 
Euclidean setups and we took the analytic continuation to Lorentzian 
time evolutions only for the final results of entanglement entropy. 
This procedure went straightforwardly or mechanically due to the simplicity of Euclidean gravity duals, while
this raises the question whether we really have sensible Lorentzian gravity duals whose 
geodesic lengths give the correct holographic entanglement entropy. We will confirm this by explicitly 
finding out the Lorentzian geodesics in this section. Note that in the true Lorentzian gravity dual,
coordinates $(t,x,z)$ should take real values, while the coordinates in Poincare coordinate  $(\xi,\bar{\xi},\eta)$ take complex values in general due to the coordinate transformation (\ref{corads}) gets complex
valued under the Wick rotation. 
 
The results of the time evolutions of HEE under (i) local operator quench $S^{(O)}_A$ (\ref{oplog}),
(ii) splitting local quench $S^{(S)}_A$ (\ref{hpsa})-(\ref{hpsc}), and (iii) joining local quench
$S^{(J)}_A$ (\ref{hppa})-(\ref{hppc}) are characterized by two logarithmic behavior $\log t/\ep$ and
 $\log t/\ap$. Moreover, we can find the same expression of $S_A$ for these different types of local quenches. Our analysis of Lorentzian geodesics in this section, will also explain such logarithmic behaviors in a systematical way.

\subsection{Late Time Behavior}

 First note that in the late time limit $t_0\gg b>a$, we find
 \be
 S^{con(S)}_A(t_0)\simeq S^{dis(J)}_A(t_0)\simeq \frac{c}{3}\log\frac{t_0}{\ep}+\frac{c}{3}\log\frac{t_0}{\ap}.  \label{lateq}
 \ee
 First we would like to point out that both are identified with the twice of the
geodesic length between the boundary point $(t,x,z)=(t,0,0)$ and the tip
of the surface $Q$, given by
\be
z=\frac{2(\ap^2+t^2)}{\s{\ap^2+4t^2}}\equiv g(t).  \label{ztq}
\ee
When $t\gg \ap$, we have
\be
g(t)\simeq t+\frac{7\ap^2}{8t}+\ddd.  \label{gtas}
\ee

These geodesics are sketch as $\Gamma_{\pm}$  in Fig.\ref{fig:SPHOLgeo} and $\Gamma$
 in Fig.\ref{fig:PSHOLgeo}. In the former (splitting quench) case, we consider a connected geodesic and therefore should avoid touching on the boundary surface $Q$. On the other hand, in the latter (joining) case, the disconnected geodesic should end on the boundary surface $Q$. Therefore these two should approximately coincide in the late time limit, by setting
 $a,b\to 0$.

Next we would like to explain why the geodesic length looks like (\ref{lateq}).
Instead of finding a geodesic directly in $(t,x,z)$ coordinate, it is easier to start with
the transformed coordinate $({\rm Re}[\xi],{\rm Im}[\xi],\eta)$. However, we have to note that since we are considering an analytical continuation into the Lorentzian geometry, the coordinates $({\rm Re}[\xi],{\rm Im}[\xi],\eta)$ take complex values in general, though $(t,x,z)$ should take real values.
By remembering the conformal map (\ref{pastef}) for the joining quench and the holographic transformation (\ref{corads}), it is straightforward to find that the correct geodesic is given by\footnote{
We can find this form by noting that the solution to the geodesic equation takes the form 
$\xi\bar{\xi}+\eta^2=$const. and by plugging the coordinate values at the AdS boundary.}
\be
\xi\bar{\xi}+\eta^2=e^{-i\beta}, \label{geoxe}
\ee
where the real parameter $\beta(>0)$ is related to the boundary time $t_0$ at $z=0$ via
\be
t_0=\ap\cdot \tan\frac{\beta}{2}.
\ee
Therefore for $t_0\gg \ap$ we have
\be
\beta\simeq \pi-\frac{2\ap}{t_0}.
\ee

\begin{figure}
  \centering
  \includegraphics[width=5cm]{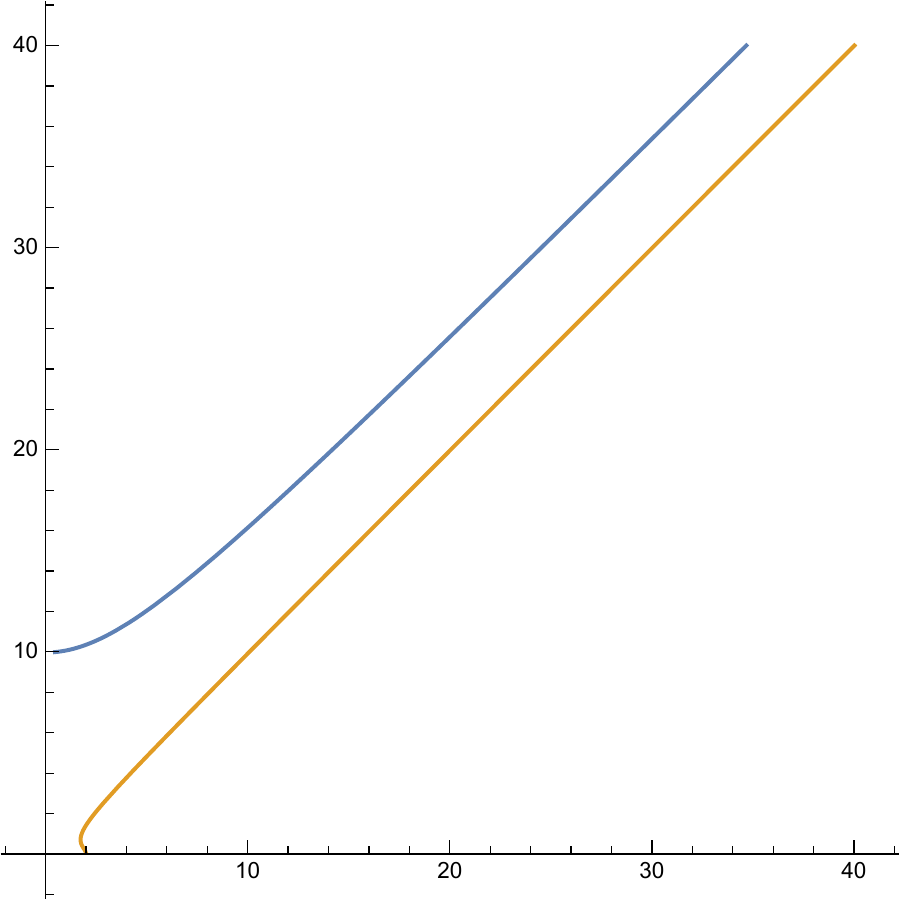}
  \includegraphics[width=7cm]{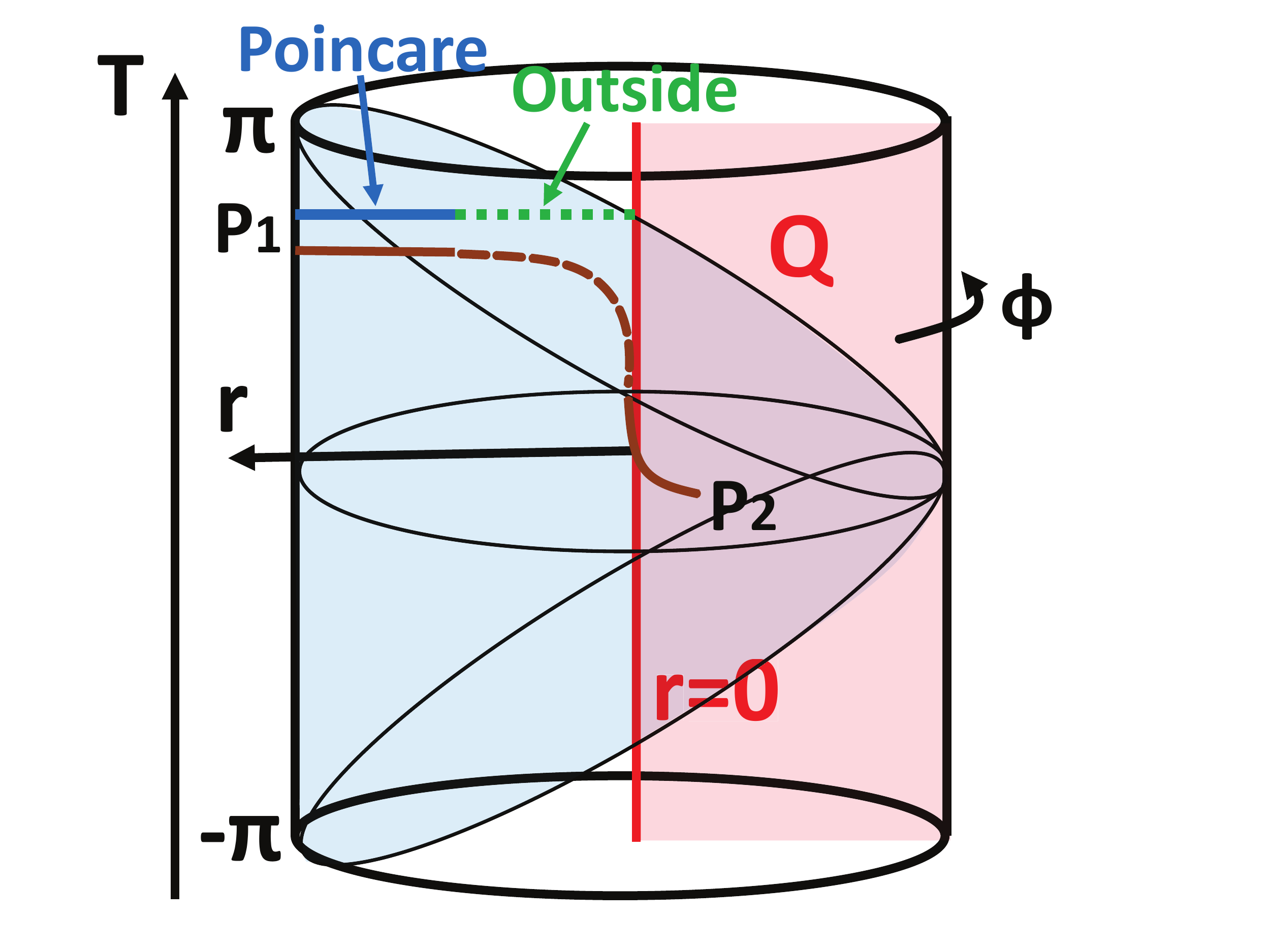}
\caption{Profiles of the geodesic. In the left graph, we show the geodesic (blue) and
the boundary surface $Q$ (orange) in the Poincar\'{e} coordinate $(z,t)$ with $t_0=10$ and $\ap=1$.
The right picture shows the outside part of the disconnected geodesic $L_{outside}$
(green dotted line)  as well as the one in the Poincar\'{e} patch $L_{poincar\acute{e}}$
(blue line) in the global AdS coordinate. The point $P_1$ is situated at $0<x\ll t$, while the point $P_2$ is at $0<t\ll x$. The blue region corresponds to the Poincar\'{e} patch. The red vertical line is the tip (\ref{ztq}) of the boundary surface $Q$. The brown curve describes the connected geodesic that corresponds to (\ref{thcoin}).}
\label{fig:geomap}
\end{figure}

In terms of the original coordinate, (\ref{geoxe}) is equivalent to
\be
z=\frac{2(t^2+1)\left(((1+\cos\beta)t-\ap\sin\beta)^2+(t\sin\beta-\ap(1-\cos\beta))^2\right)^{1/4}}
{\left(((1+\cos\beta)(3t+4t^3)+\ap\sin\beta)^2+(\sin\beta(3t+4t^3)+\ap(1-\cos\beta))^2\right)^{1/4}},
\label{geozp}
\ee
which is plotted in Fig.\ref{fig:geomap}.

 At late time, we have $z(t)\simeq t-\frac{t_0}{2}+O(1/t)$ and thus the geodesic extends to infinity almost in a light-like way. In the limit $t\to t_0\gg \ap$ we have
 \be
 z(t)\simeq \s{t_0(t-t_0)}+\frac{(t-t_0)^{3/2}}{2\s{t_0}}+\ddd.
 \ee
Indeed we can confirm that this is a solution to the geodesic equation in the holographic spacetime, obtained from (\ref{sttw}) and (\ref{metads}), given by (on $x=0$):
\be
ds^2=\frac{1}{z^2}\left[dz^2-(1+f(t,z)z^2)dt^2\right],  \label{fxx}
\ee
\be
f(t,z)=\frac{9\ap^4 z^2-24\ap^2(t^2+\ap^2)^2}{16(t^2+\ap^2)^4}.
\ee
We can estimate the geodesic length of (\ref{geozp}) as
\be
\int^{\infty}_{t_0} dt\frac{\s{(z')^2-1-fz^2}}{z(t)}\simeq \frac{c}{6}\log \frac{t}{\ep}.
\label{pogf}
\ee
The twice of this length can explain only one of the logarithmic term in (\ref{lateq}).

The missing contribution actually comes from another part of the geodesic which is not covered by the
coordinate (\ref{fxx}). Indeed, if we parameterize the geodesic (\ref{geoxe}) as
\be
\xi=e^{-i\beta/2}\cos\theta, \ \ \ \eta=e^{-i\beta/2}\sin\theta, \ \ (0\leq \theta<\frac
{\pi}{2}). \label{geoful}
\ee
Note that $\theta=0$ corresponds to the AdS boundary $\eta=0$ and $\theta=\pi/2$ corresponds to the
boundary surface $Q$ on Im$[\xi]=0$. Actually from the map (\ref{corads}), we find that the trajectory
(\ref{geozp}) for $t_0\leq t \leq \infty$ only covers a part of the full geodesic
(\ref{geoful}) given by $0\leq \theta\leq \beta/2$. The geodesic length for this part is computed as
\ba
\int^{\cos(\beta/2)e^{-i\beta/2}}_{\hat{\ep}} \frac{e^{-i\beta/2} d\eta}{\eta\s{e^{-i\beta}-\eta^2}}
=\log\left(\frac{2e^{-i\beta/2}}{\hat{\ep}}\right)
+\log\left(\frac{\cos(\beta/2)}{1+\sin(\beta/2)}\right),
\ea
where $\eta=\hat{\ep}$ is the cut off corresponding to the original one $z=\ep$.
The first term in the right hand side is the full
contribution
\be
\int^{e^{-i\beta/2}}_{\hat{\ep}}\frac{e^{-i\beta/2} d\eta}{\eta\s{e^{-i\beta}-\eta^2}}=\log\left(\frac{2e^{-i\beta/2}}{\hat{\ep}}\right)\simeq \log\frac{t_0}{\ep}+\log\frac{t_0}{\ap}, \label{fullg}
\ee
which agrees with a half of (\ref{lateq}). The second term is estimated when $t_0\gg \ap$ as
\be
\log\left(\frac{\cos(\beta/2)}{1+\sin(\beta/2)}\right)\simeq -\log\frac{t_0}{\ap}.
\ee
This shows that the part of geodesic length in the Poincar\'{e}-like coordinate patch (\ref{fxx}) gives (identical to (\ref{pogf}))
\be
L_{poincar\acute{e}}=\log\frac{t_0}{\ep}, \label{pogeo}
\ee
and the one outside of the Poincar\'{e} patch does
\be
L_{outside}=\log\frac{t_0}{\ap}. \label{outgeo}
\ee

To understand this outside contribution well\footnote{Such a contribution is also familiar in
holographic global quenches \cite{AAL,Ba,HaMa} and local operator quenches \cite{JaTa}.}, let us ignore the back reaction and focus on the Poincar\'{e} AdS (restricted on $x=0$) by setting $f(t,z)=0$ in (\ref{fxx}).
 This is embedded in the global AdS$_3$ (refer to  Fig.\ref{fig:geomap})
\be
ds^2=-(r^2+1)dT^2+\frac{dr^2}{r^2+1}+r^2d\phi^2,
\ee
at $\theta=0$, via the map (refer to e.g.\cite{NNT})
\ba
&& \s{r^2+1}\cos T=\frac{\ap+(z^2-t^2)/\ap}{2z},\no
&& \s{r^2+1}\sin T=\frac{t}{z}.
\ea
Here the parameter $\ap$ is introduced such that the static trajectory at $r=0$ is mapped
into that for an accelerated particle $z=\s{t^2+\ap^2}$ in the Poincar\'{e} AdS, which imitates
the behavior (\ref{gtas}).

At the AdS boundary, the late time limit $t_0\gg \ap$ in the Poincar\'{e} AdS corresponds to
the time $T\simeq \pi-\ap/t_0$ and the cut off $r_{\infty}=t_0^2/(\ap\ep)$ in the global AdS. We are considering a geodesic which starts from this boundary point and ends on the tip of the surface $Q$, namely the straight line $r=0$ in the global AdS$_3$. It is clear that this geodesic is given by $T=$constant and the radial coordinate $r$ runs from $r=r_\infty$ to $r=0$.
Along this geodesic, the part outside of Poincar\'{e} patch is $0<r<r_*$, where $r_*$ is the solution to $\s{r^2+1}\cos T+r=0$, from which we find $r_*\simeq {t_0}/{\ap}\gg 1$.

Thus, in this simplified model, the geodesic length in the Poincar\'{e} patch and that for the outside
read (refer to  Fig.\ref{fig:geomap}):
\ba
&& L_{poincar\acute{e}}=\int^{r_\infty}_{r_*} \frac{dr}{\s{r^2+1}}\simeq \log\frac{t_0}{\ep},\no
&& L_{outside}=\int^{r_*}_{0} \frac{dr}{\s{r^2+1}}\simeq\log\frac{t_0}{\ap}.
\ea
These reproduce the previous results from the exact geodesic (\ref{pogeo}) and (\ref{outgeo}).

\subsection{Middle Time Period}

Now we move on to the middle time period and we focus on the region $a\ll t\ll b$ to see the
logarithmic behavior clearly. Then we find the following coincidences between the three
different types of local quenches\footnote{Note that for the holographic local operator quench, we
need only connected geodesics as there is no boundary surface $Q$.} :
\ba
S^{con(O)}_A\simeq S^{con(S)}_A\simeq S^{con(J)}_A\simeq
\frac{c}{6}\log\frac{t_0}{\ap}+\frac{c}{3}\log \frac{b}{\ep}. \label{thcoin}
\ea
On the other hand, the disconnected geodesic length in joining quench behaves differently:
\ba
S^{dis(J)}_A\simeq \frac{c}{6}\log\frac{t_0}{\ep}
+\frac{c}{6}\log\frac{t_0}{\ap}+\frac{c}{6}\log \frac{b}{\ep}.  \label{djd}
\ea

The coincidence (\ref{thcoin}) can be easily understood because they are described by a geodesic which connects the two end points, where the existence of slits is not a crucial obstruction, and because the background metric is the same (refer to the brown curve in Fig.\ref{fig:geomap}). We can regard the first term
$({c}/{6})\log({t}/{\ap})$ as the geodesic length outside of the Poincar\'{e} patch as in
(\ref{outgeo}).
The second term $({c}/{3})\log ({b}/{\ep})$ is simply interpreted as the
familiar geodesic length in the Poincar\'{e} AdS$_3$.

On the other hand, the geodesic length (\ref{djd}) is divided into two contributions:
the one which connects $x=a$ and $Q$ and the other one which connects $x=b$ and $Q$. The former one
is clearly equal to $({c}/{6})(L_{poincar\acute{e}}+L_{outside})$ in (\ref{pogeo}) and (\ref{outgeo}).
The latter one is simply given by the standard result of AdS/BCFT: $({c}/{6})\log ({b}/{\ep})$.
These explain the behavior (\ref{djd}).

\section{Tensor Network Interpretation of Local Quenches}\label{sec:tnlq}

 Here we would like to present a sketch of tensor network description of time evolution of our 
local quench states and their entanglement entropies so that they matches with our gravity duals. 
Our argument in this section will be heuristic and qualitative.

\begin{figure}
  \begin{center}
    \begin{tabular}{c}

      \begin{minipage}{0.5\hsize}
        \begin{center}
          \includegraphics[clip, width=8cm]{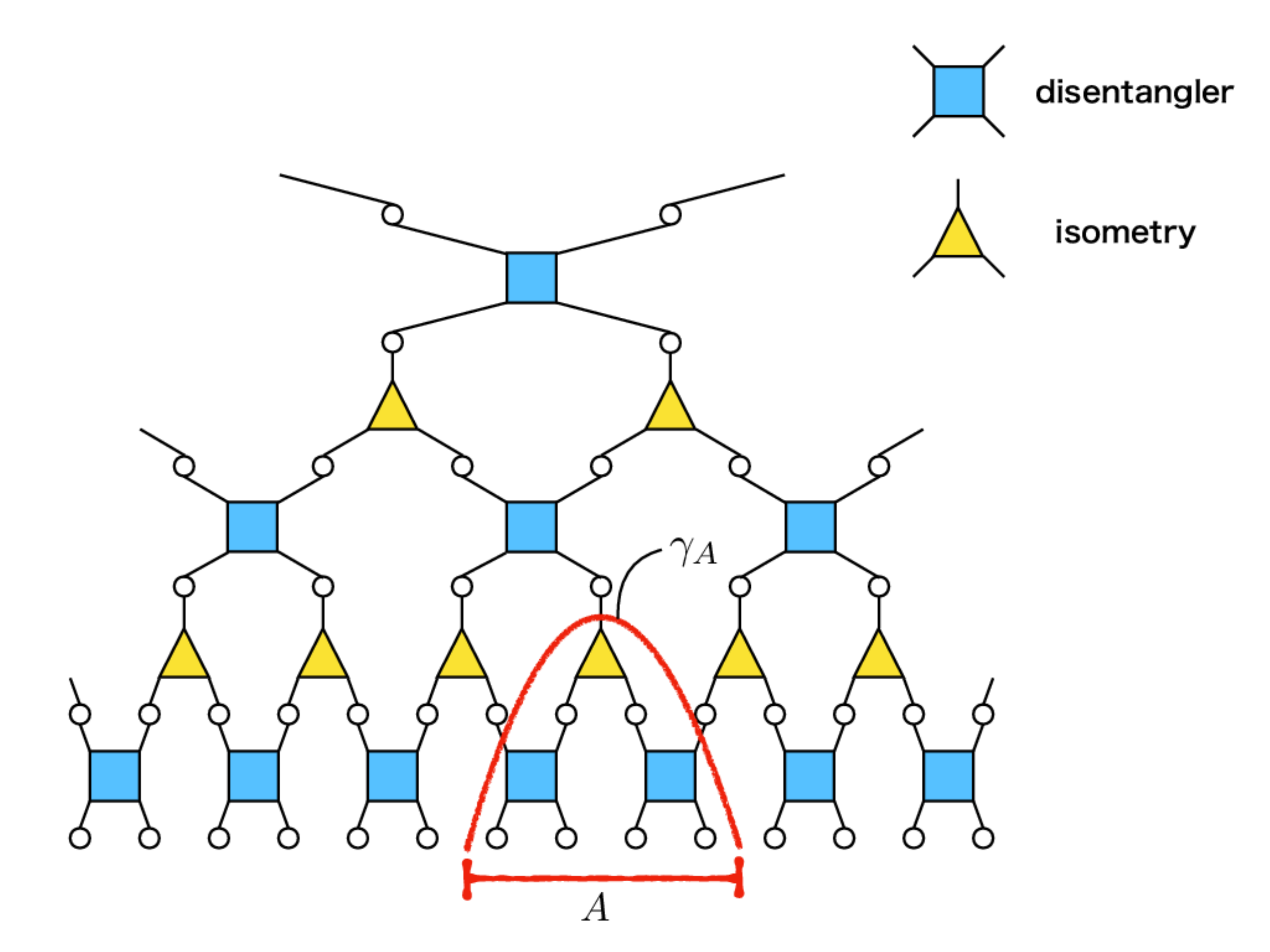}
          \hspace{1.6cm}
        \end{center}
      \end{minipage}

      \begin{minipage}{0.5\hsize}
        \begin{center}
          \includegraphics[clip, width=8cm]{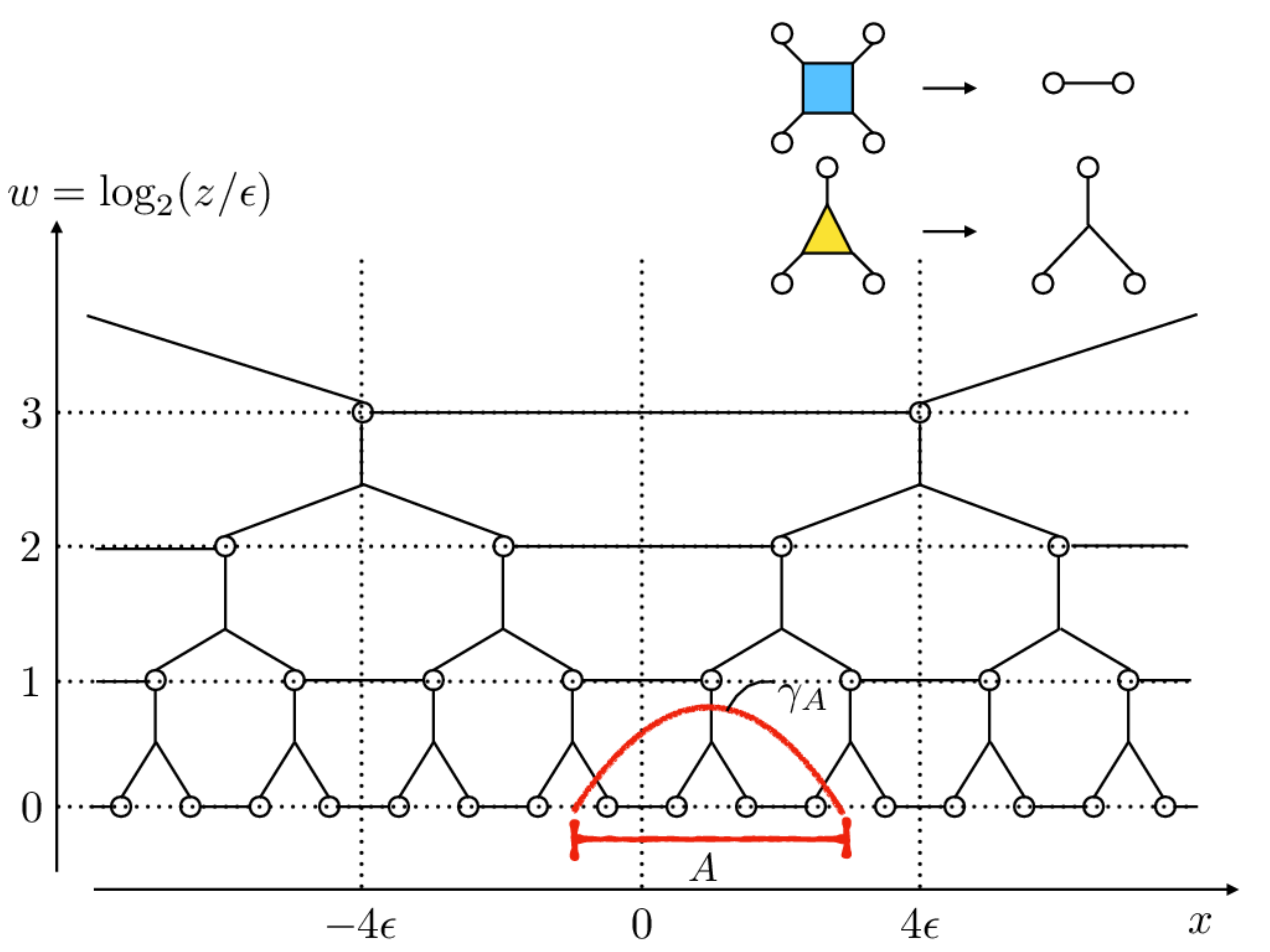}
          \hspace{1.6cm}
        \end{center}
      \end{minipage}
      \end{tabular}

    \caption{(Left) A standard description of MERA tensor network. (Right) A simplified representation of MERA. This representation can help us capture the relationship between MERA and AdS. (Both) $A$ is a subsystem of the boundary of MERA, and $\gamma_A$ is a curve in the bulk who share the boundary with $A$. $(z,x)$ forms a Poincar\'{e} coordinate, while $w = \log_2 (z/\ep)$ corresponds to the number of layer.}\label{fig:aMERA}
  \end{center}
\end{figure}

We employ the MERA (Multiscale Entanglement Renormalization Ansatz) \cite{MERA} for our qualitative interpretation, though our arguments can equally hold for other holographic tensor networks.
 MERA is a class of tensor networks which are constructed by two kinds of tensors: isometry and disentangler  (Fig.\ref{fig:aMERA}). This is manifestly scale invariant and is known to describe  ground states of critical quantum systems with high accuracy. In tensor networks, by contracting all interior tensors, a desired quantum state is realized at the boundary. We regard the physical system as the boundary, and the tensor network structure as the bulk in AdS/CFT \cite{TNa}.

 Consider a curve $\gamma$ in the bulk, whose length can be defined by the number of tensors it crosses\footnote{It would be more precise to define the length of a curve in a tensor network by the number of tensor legs it crosses. In our case, however, since we only care about the shortest curve in MERA, this definition is also allow for convenience.}. For a subsystem $A$ on the boundary, its EE $S_A$ is upper bounded by
 \be
 S_A \leq C\times\pa{{\rm length~of~}\gamma_A}
 \ee
where $C$ is a constant and $\gamma_A$ is an arbitrary curve who shares the boundary with $A$ (i.e.
$\de \gamma_A=\de A$).

Particularly, in typical examples of MERA, such as that of a critical transverse Ising model, $S_A$ is roughly equal to
\be
S_A \simeq C\times\pa{{\rm length~of~}\gamma_A^{min}}
\ee
where $\gamma_A^{min}$ is the minimal curve, whose length is proportional to $\log L_A$, with $L_A$ the size (i.e. the number of spins) of subsystem $A$.

A qualitative correspondence between MERA and AdS is as follows \cite{TNa}.
 If we set the coordinate $x$ parallel to the layers, and the coordinate $w=\log_2 (z/\ep)$ vertical to them (the right picture in Fig.\ref{fig:aMERA}), then the geometry of MERA is interpreted as a discrete version of the time slice of AdS (i.e. hyperbolic space), whose metric is
 \be
 ds^2 =\PA{\frac{dz^2+dx^2}{z^2}} = R^2\ep^{-2}{\frac{dw^2}{\pa{\log_2 e}^2} + 2^{-2w}dx^2},
 \ee
where $\ep$ is introduced as an UV cutoff. Refer to \cite{Tak} for a possible interpretations of various slices in AdS spacetime as tensor networks.

Now we would like to consider an tensor network interpretation of our three types of local quenches 
(local operator/splitting/joining). 
We will take a heuristic strategy: first we assume specific structures of tensor networks under the 
time evolutions and then confirm if they reproduce the correct time evolutions of entanglement entropy.
In our construction of tensor networks corresponding to local quenches, we assume that an appropriate time slice $t= {\rm const}$. of a gravity dual spacetime after a local quench, is approximately a tensor network at the time $t$. We note that the information of tensor network at each time can be separated in to two ingredients: one is the geometric structure of tensor network itself, and another one is the parameters of each tensor.\footnote{In a MERA corresponding to the ground state of a critical system, for example, network structure means the way tensors linked with each other. Besides, all the isometries (disentanglers) carry the same parameters due to the scale invariance of the state. Network structure and parameters together gives a corresponding quantum state.}  One essential observation we will make below is that we can 
divide the growth of EE into a ``shock wave contribution" and ``splitting/joining contribution", which can 
also be suggested from the holographic analysis in the previous section. 
In the discussion below, we will regard the ``shock wave contribution" as modifications of the parameters and ``splitting/joining contribution" as modifications of the network structure. We will see that under this assumption, by choosing appropriate tensor network structure and neglecting modifications on the parameters, we can realize qualitative features  of the entanglement structure in the AdS/CFT setup.

 \begin{figure}
  \centering
  \includegraphics[width=12cm]{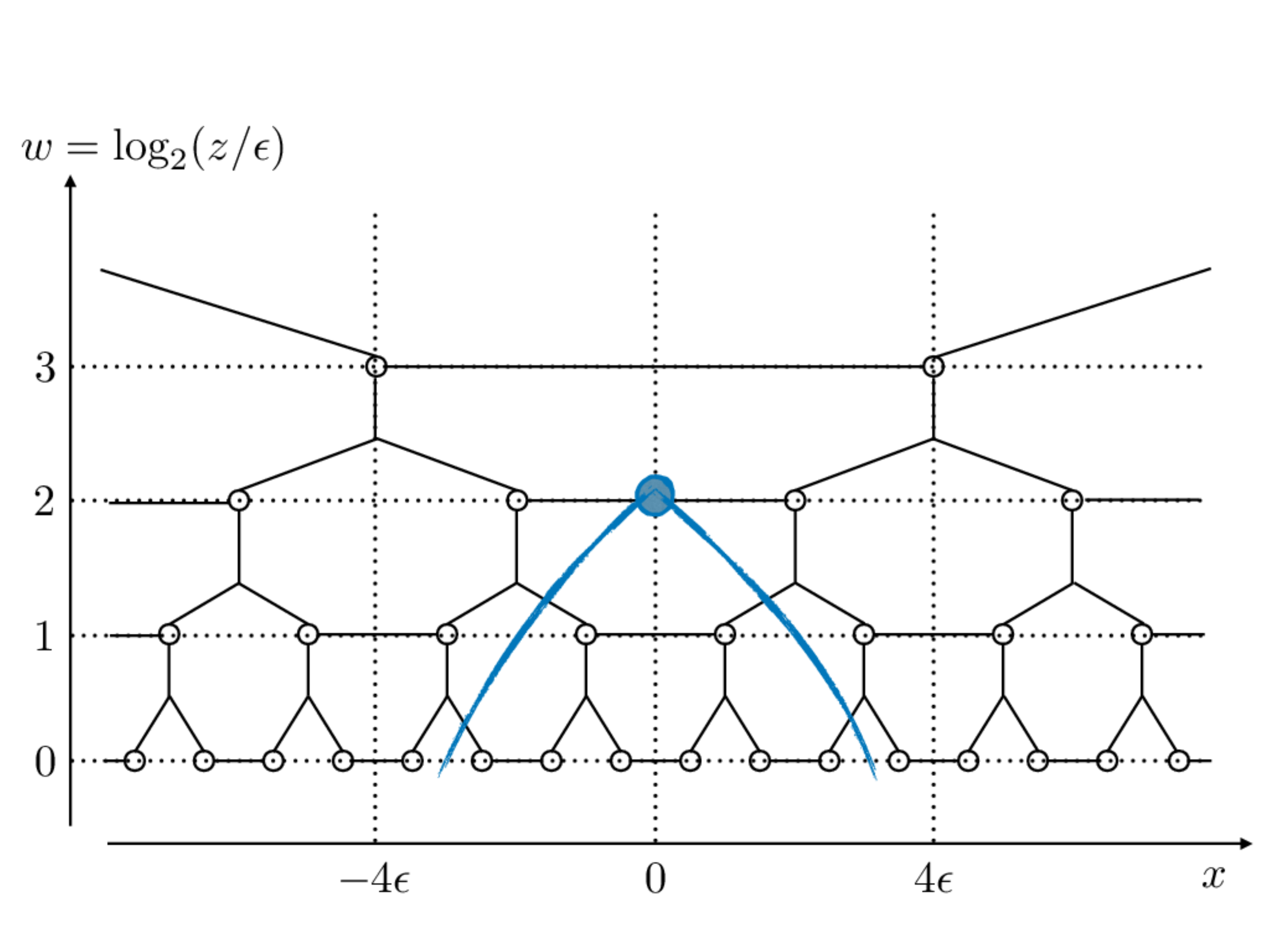}
 \caption{A tensor network description of local operator quench at $t=4\ep$. The blue dot is the
 large modification of tensor dual to the falling particle. The  blue curves describe the image of shock wave deformation of tensors. The parameters of tensors are expected to be modified, and tensors close to the blue curve are expected to accept the strongest modification.}
\label{fig:MERAAdS}
\end{figure}

 \subsection{Local Operator Quench}

A local operator quench in a holographic CFT can be regarded as a particle falling towards the AdS black hole horizon on the gravity side \cite{HLQ}. This particle trajectory is analogous to the tip of the boundary surface $Q$ in the splitting/joining local quench.
In this picture, its back-reaction on spacetime spreads as a shock wave \cite{HoIz}, and the EE of a subregion on the boundary will change when the shock wave crosses its minimal surface.

Accordingly in the tensor network, we expect the local operator quench causes a modification of tensor at $x=0$ and $w=0$ due to the initial operator insertion, though the structure of tensor network does not
change. As the time evolves, this modification propagates into the interior 
as $w\sim \log t$. At the same time, tensors are modified along the shock wave. These modifications increases quantum entanglement locally. These are depicted in Fig.\ref{fig:MERAAdS}.

The behavior of the EE is consistent with the causal behavior that the EE gets non-trivial only for the time interval $a<t<b$, as we found both for the RCFT and holographic CFT.
In the holographic case (\ref{oplog}), however, we observed a logarithmic growth
\be
S_A \simeq \underbrace{\frac{c}{6}\log \frac{t}{\ap}}_{\rm shock~wave}+..., \label{locoptn}
\ee
if $a\ll t\ll b$. Our entanglement density analysis showed such a logarithmic contribution is due to
the presence of non-local entanglement at the initial state. Our geometric analysis in the gravity dual
shows this contribution comes from outside of the Poincar\'{e} patch (\ref{outgeo}). In tensor network, this hidden contribution is expected to come from modifications of tensor parameters, though we cannot figure it out precisely. Since this is owing to the huge back-reactions, we would like to still call this a ``shock wave contribution''. To find out the details of this contribution, we need to make the conjectured correspondence 
between the AdS/CFT and tensor networks more precisely beyond qualitative arguments, which is not available at present.

 \subsection{Splitting Quench}

Now we moved on to the splitting quench.
We split the disentangler in the MERA at $x = 0, z = \ep~\pa{w = 0}$, initially (refer to \cite{EvVi} for introducing boundaries in MERA). This is motivated by the gravity dual where the boundary surface extends to the bulk as in Fig.13. We identify the boundary in the gravity dual as the termination of tensor network 
(i.e. cutting the disentanglers or unitaries). This elimination of disentanglers propagates under the time evolutions as in the gravity dual. Therefore at the time $t=2^n\ep$, the disentangler at $x = 0, w = n$ is eliminated.  At the same time, from the gravity dual, we expect there are shock wave propagations as in the local operator quench case. Thus we also expect that in our tensor network description 
a shock wave spreads from $w=n$  at the same time $t=2^n\ep$.
Both of them will contribute to the change of EE. We call them ``shock wave contribution" and ``splitting contribution" respectively.

From the CFT viewpoint, we can heuristically understand this prescription as follows. 
When we triggered the splitting quench, we eliminate the nearest neighbor interactions between two 
adjacent lattice sites. This is equivalent to cutting the disentangler in the very UV. After the time evolution
this cutting of entanglement propagates  towards the IR. 

\begin{figure}[h]
  \begin{center}
    \begin{tabular}{c}

      \begin{minipage}{0.5\hsize}
        \begin{center}
          \includegraphics[clip, width=8cm]{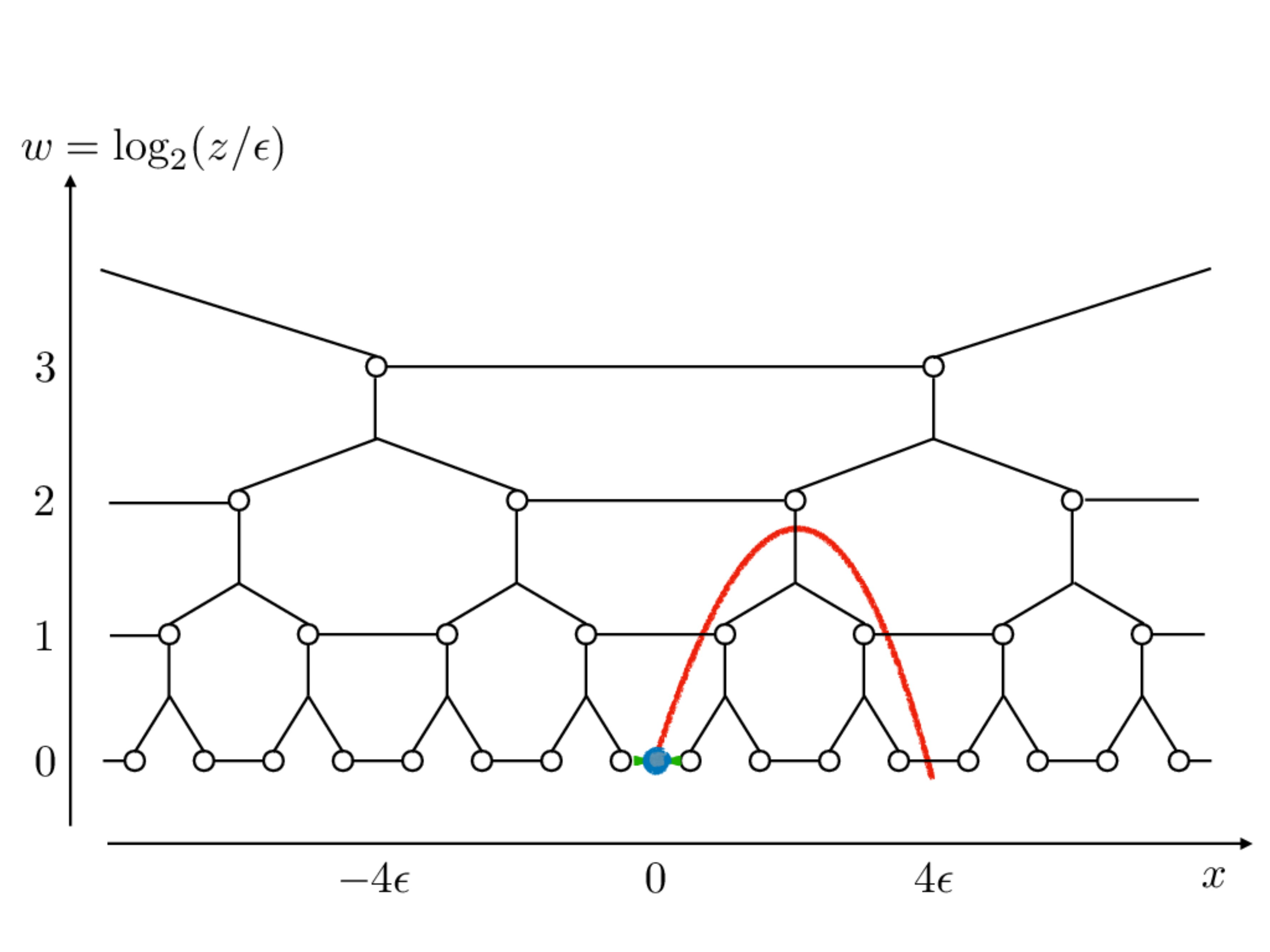}
          \hspace{1.6cm} $t = \ep$
        \end{center}
      \end{minipage}

      \begin{minipage}{0.5\hsize}
        \begin{center}
          \includegraphics[clip, width=8cm]{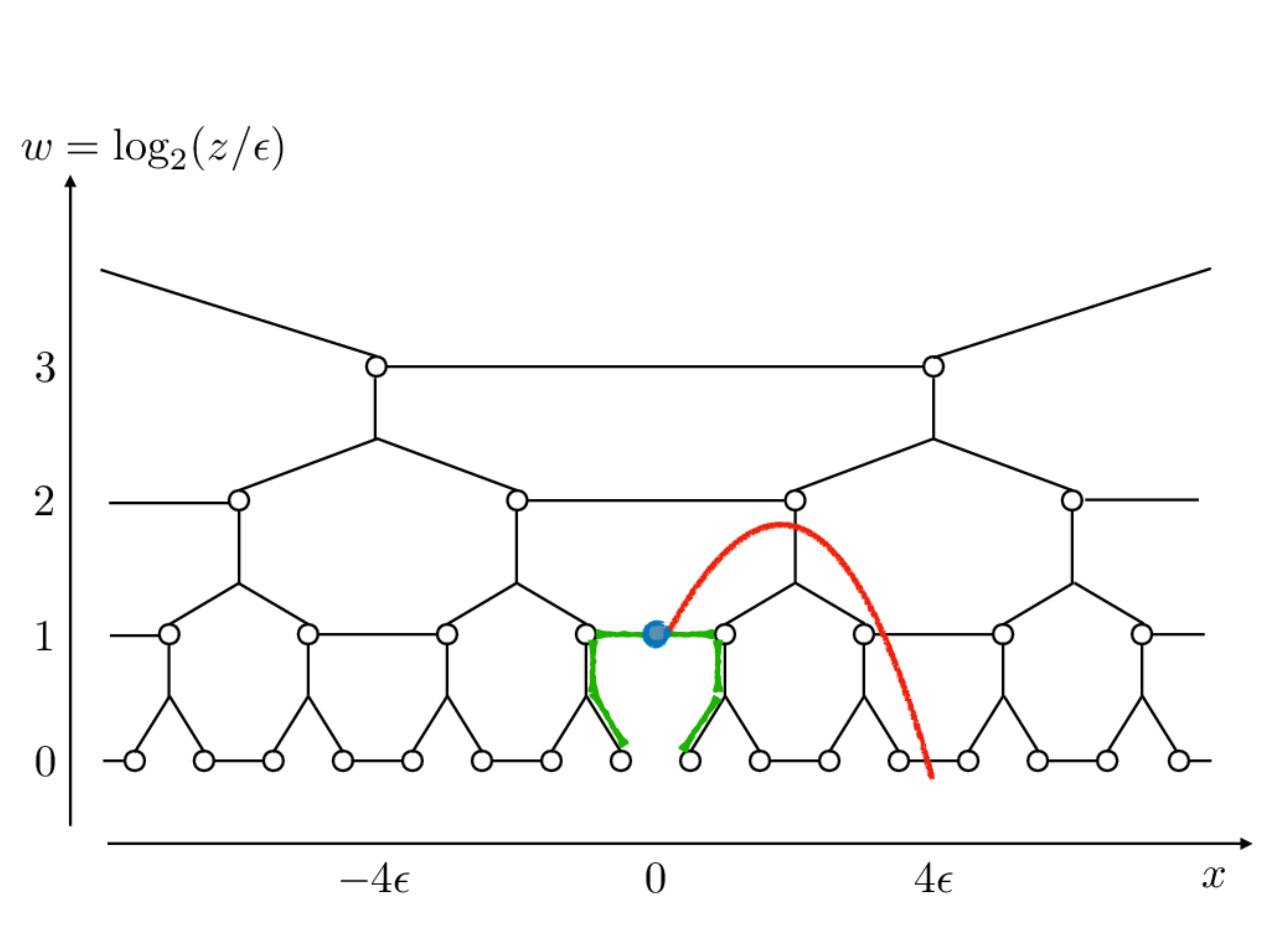}
          \hspace{1.6cm} $t = 2\ep$
        \end{center}
      \end{minipage}
      \end{tabular}

    \caption{Time evolution of the tensor network after splitting quench. The red curve shows the minimum curve in the bulk. The blue dot and green curve are dual to the falling particle and
    boundary surface $Q$.  Here, $A = \{ x|x\in\pa{0, 4\ep}\}$.\label{fig:Msplit}}
  \end{center}
\end{figure}

 We show in Fig.\ref{fig:Msplit} how the network changes after splitting quench. If we focus on the subsystem $A = \{ x|x\in\pa{0, l\ep}\}$  and neglect modifications of  tensor parameters (equally neglect ``shock wave contribution"), we will find $S_A$ is given by the length of minimal disconnected curve, which decreases logarithmically in time at first, and then stays as a constant. In other words, splitting contribution gives a negative logarithmic growth $\propto -\log t/\ep$ to $S_A^{dis}$. On the other hand, as we have mentioned in the case of local operator quench (\ref{locoptn}), the shock wave contribution should give a positive logarithmic growth $\propto \log t/\ap$ to EE. In this way, we reach the following estimation:
\be
S^{dis}_A \simeq \underbrace{\frac{c}{6}\log \frac{t}{\ap}}_{\rm shock~wave} \underbrace{-~ \frac{c}{6} \log \frac{t}{\ep}}_{\rm splitting} + ...,
\ee
where we determined the coefficient of the second logarithmic term so that we do not totally have
any logarithmic growth as in our previous calculation of HEE (\ref{hpsb}).

 \subsection{Joining Quench}

 For the joining local quench, we start with two copies of semi infinite MERA, which are disconnected
 initially. The quench is triggered by adding a new disentangler on $x = 0, z = \ep~\pa{w = 0}$, at $t = \ep$. Then, besides a shock wave spreading at $w=n$ in the bulk, a new disentangler on $x = 0$ comes out to connect the $n$th layer at $t = 2^n\ep$.  Both of them will contributes to the EE evolution. We call them ``shock wave contribution" and ``joining contribution" respectively.

\begin{figure}[h]
     \begin{center}
          \includegraphics[clip, width=5cm]{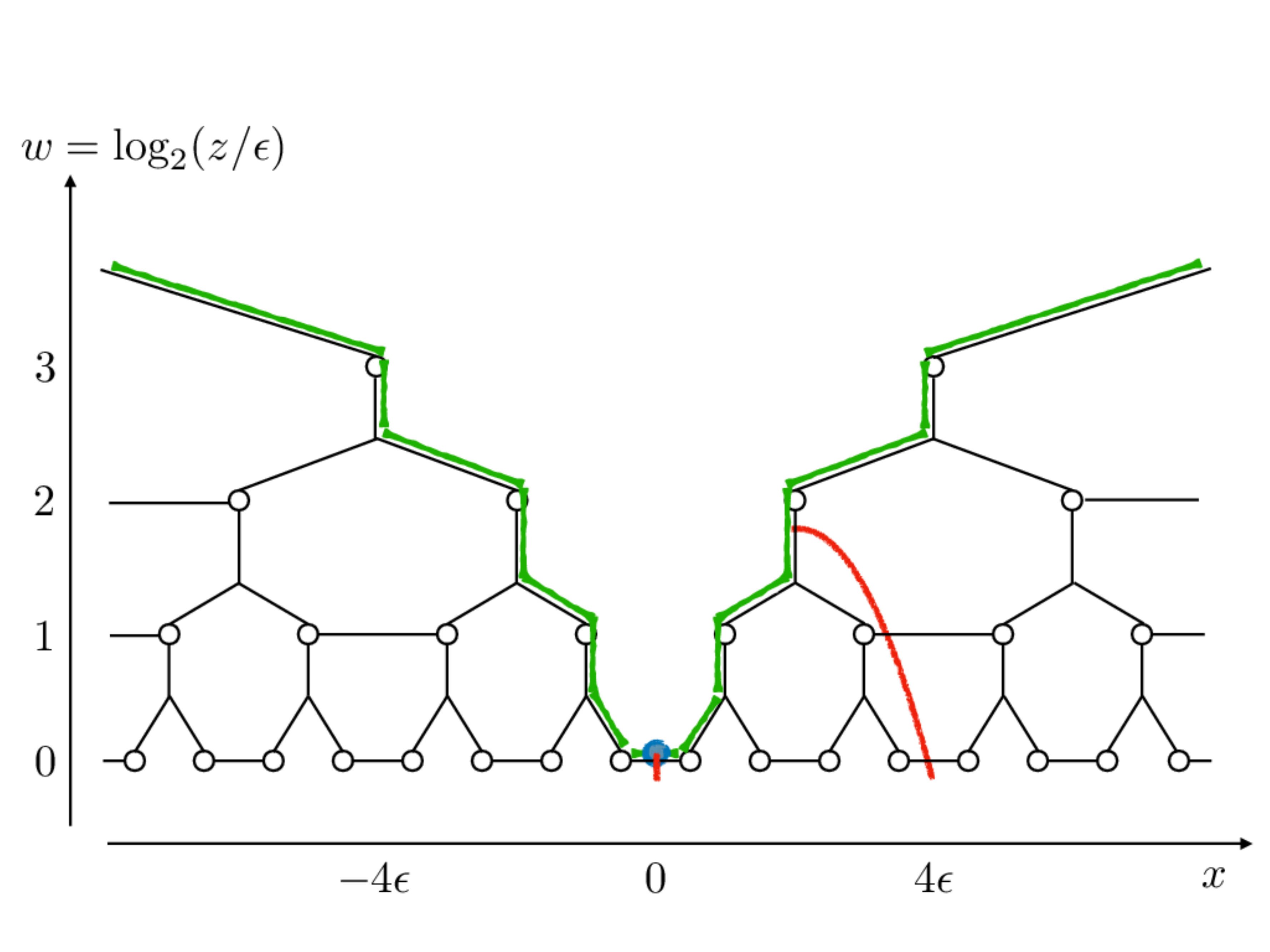}
          \includegraphics[clip, width=5cm]{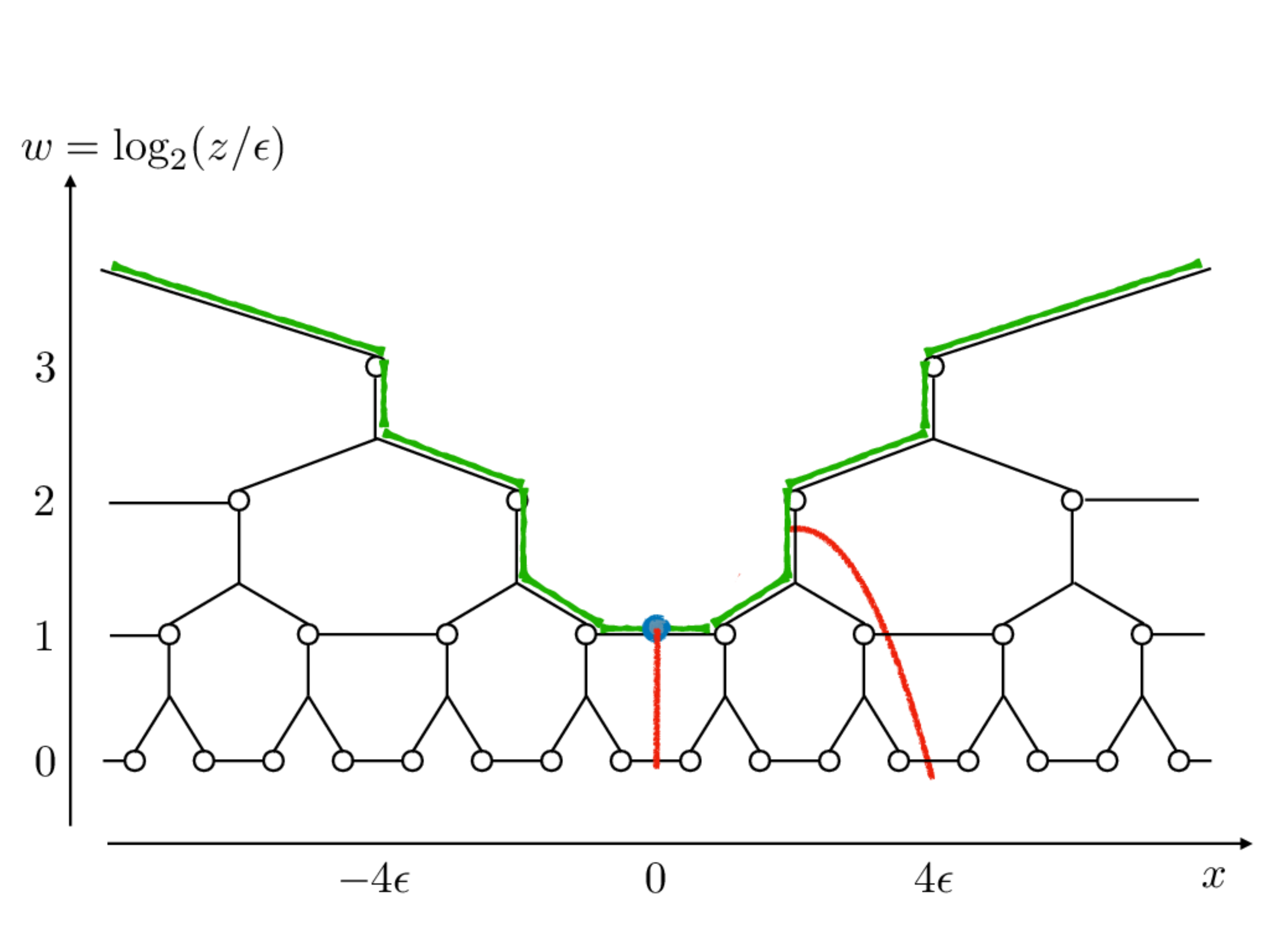}
           \includegraphics[clip, width=5cm]{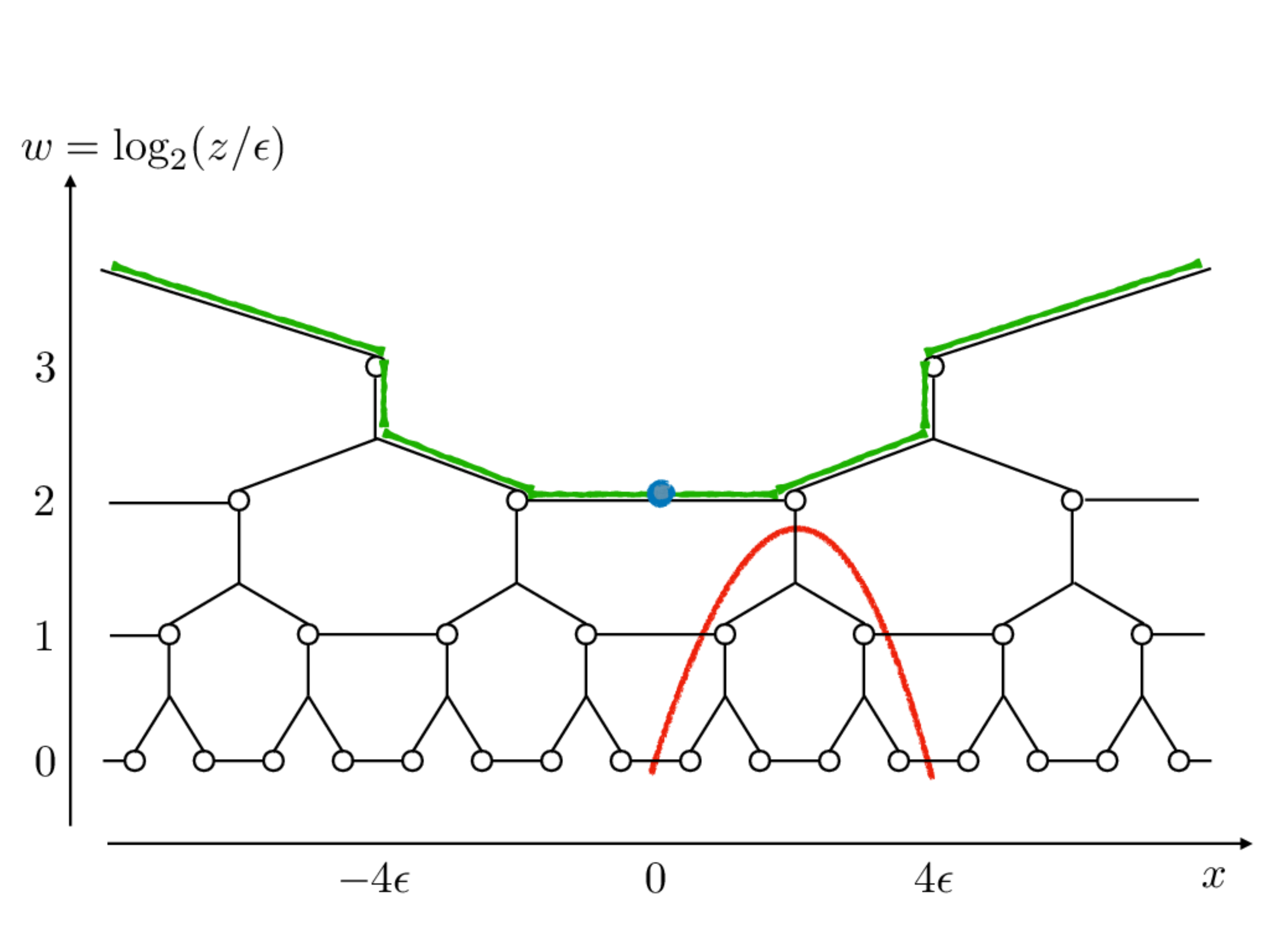}
    \caption{Time evolution of the tensor network after joining quench.
    The left, middle and right picture correspond to $t=\ep$, $2\ep$ and $4\ep$, respectively.
    The red curve shows the minimum curve in the bulk. The blue dot and green curve are dual to the falling particle and boundary surface $Q$.
    Here, $A = \{ x|x\in\pa{0, 4\ep}\}$.  \label{fig:Mjoin}}
  \end{center}
\end{figure}

 We show in Fig.\ref{fig:Mjoin} how the network will change after joining quench. If we focus on block $A = \{ x|x\in\pa{0, l\ep}\} $ and neglect modifications of  tensor parameters (equally neglect ``shock wave contribution"), we will find $S_A$ is firstly given by the length of minimal disconnected curve, which increases logarithmically in time. In other words, the joining contribution gives a positive logarithmic growth to $S_A^{dis}$, which is expected to be equal to the absolute value of splitting contribution, since they should cancel with each other if they occur at the same time. On the other hand, since the shock wave contribution should give a positive logarithmic growth that cancels with split contribution, it is proper to think that the two contributions together give a double logarithmic growth, which is exactly consistent with the results in Section 5.2. That is,
 \be
S^{dis}_A \simeq \underbrace{\frac{c}{6}\log \frac{t}{\ap}}_{\rm shock~wave} +\underbrace{~ \frac{c}{6} \log \frac{t}{\ep}}_{\rm joining} + ...
\ee
In this way, we can explain the logarithmic growth we saw in (\ref{hppb})

\section{Holographic Quantum Circuits from AdS/BCFT}
\label{sec:circuits}

\begin{figure}
  \centering
  \includegraphics[width=6cm]{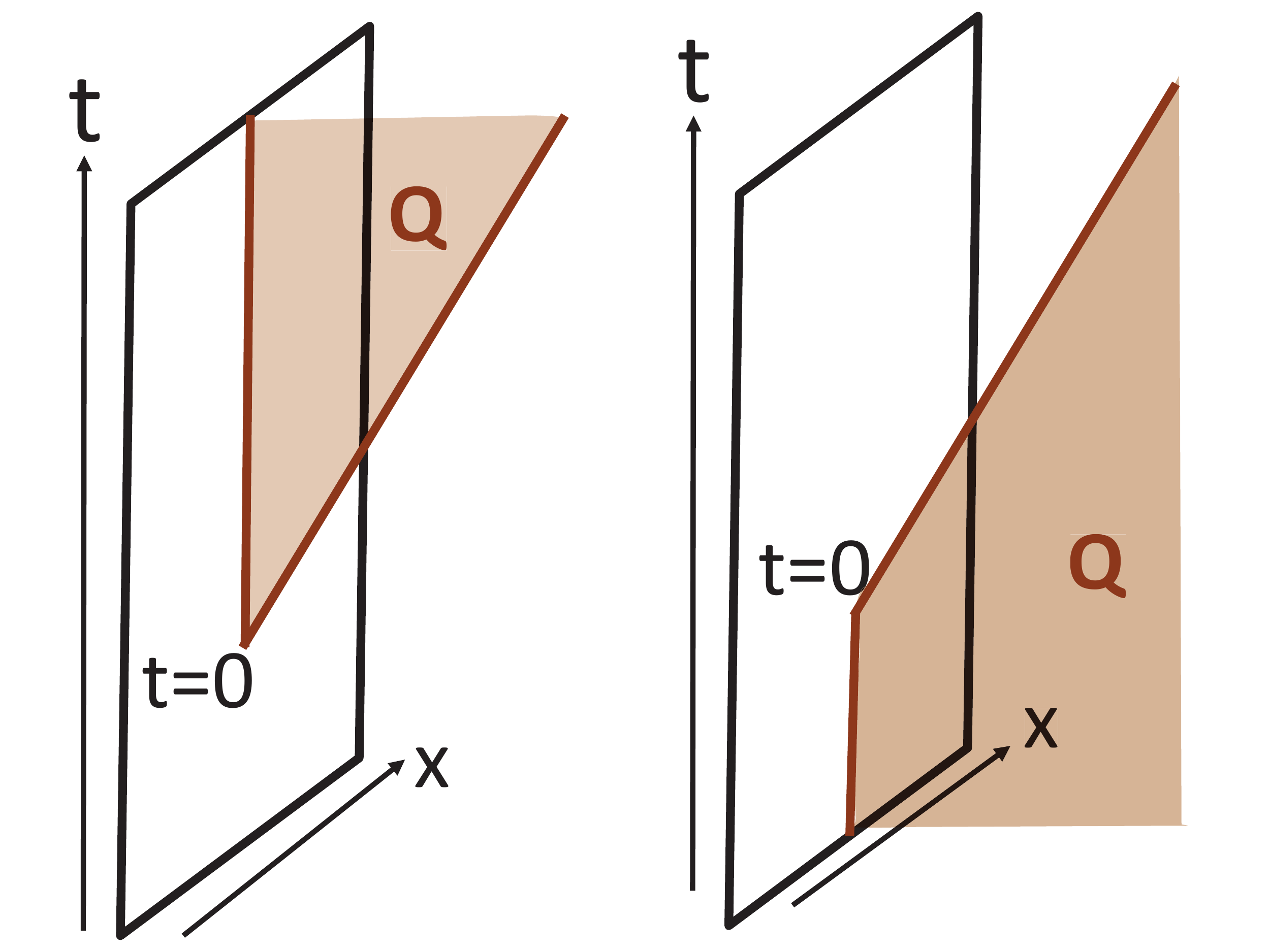}
  \includegraphics[width=6cm]{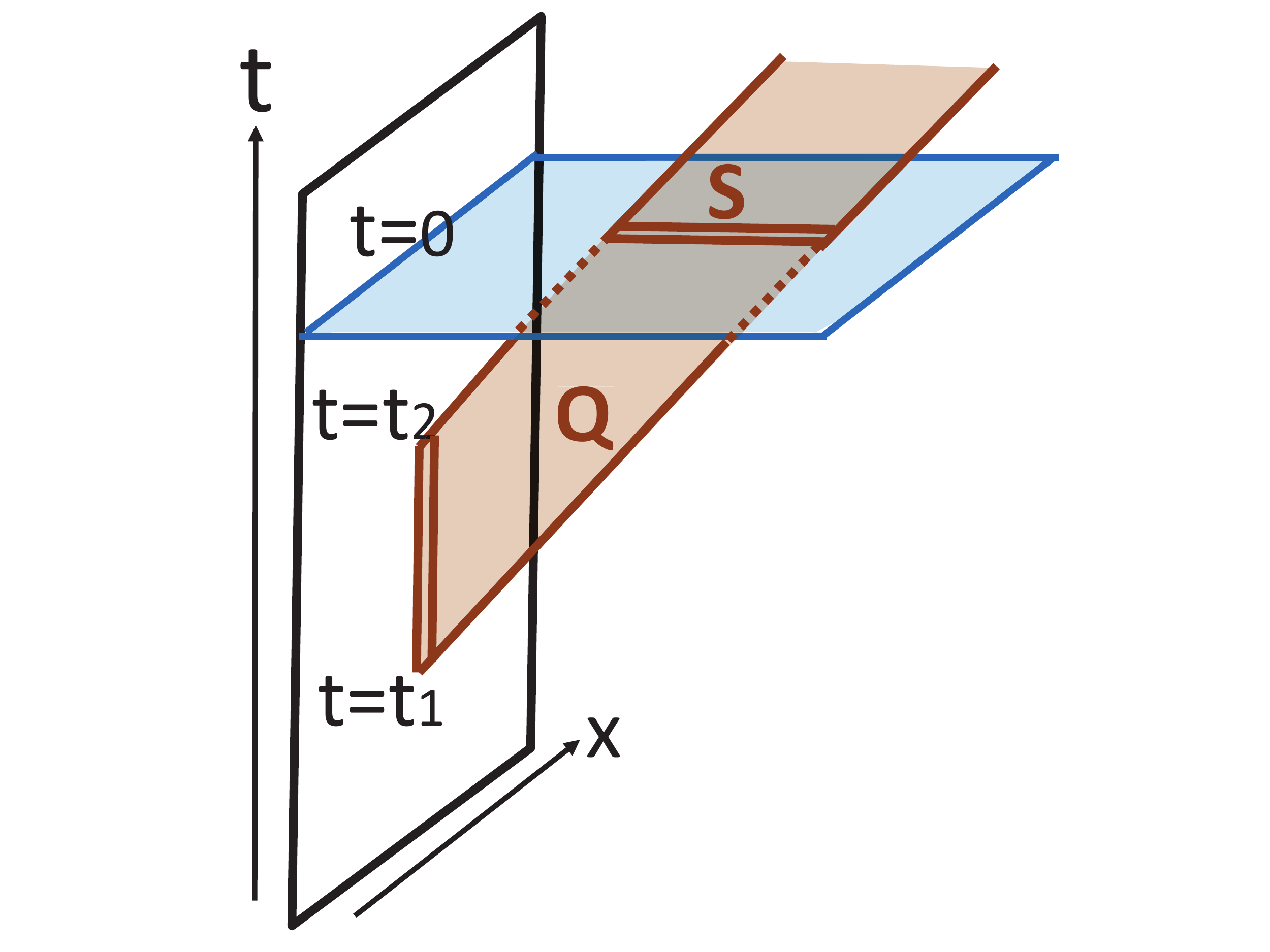}
\caption{A simplified sketch of the gravity dual spacetimes for the splitting (left),
the joining (middle) local quench and their combination (right). The brown surfaces describe the boundary surfaces $Q$. In the right picture, the boundary surface $Q$ intersects with the blue horizontal plane which describes the time slice $t=0$. Their intersection $S$ makes a slit on the time slice.}
\label{fig:SPPSHOLgeo}
\end{figure}

We can simply summarize gravity dual spacetimes of the splitting and joining local quench as in the left and middle picture of Fig.\ref{fig:SPPSHOLgeo} in the limit $\ap\to 0$. If we combine them such that we join two CFTs at $t=t_1$ and split them again at $t=t_2$, then the expected holographic dual is given by the right picture in Fig.\ref{fig:SPPSHOLgeo}. The important lesson from this example is that we can create a slit ($S$ in that picture) which looks ``floating" in the bulk on a time slice of gravity dual. Of course, the location of such a floating slit moves under the time evolution following the condition (\ref{KT}).

\begin{figure}
  \centering
  \includegraphics[width=6cm]{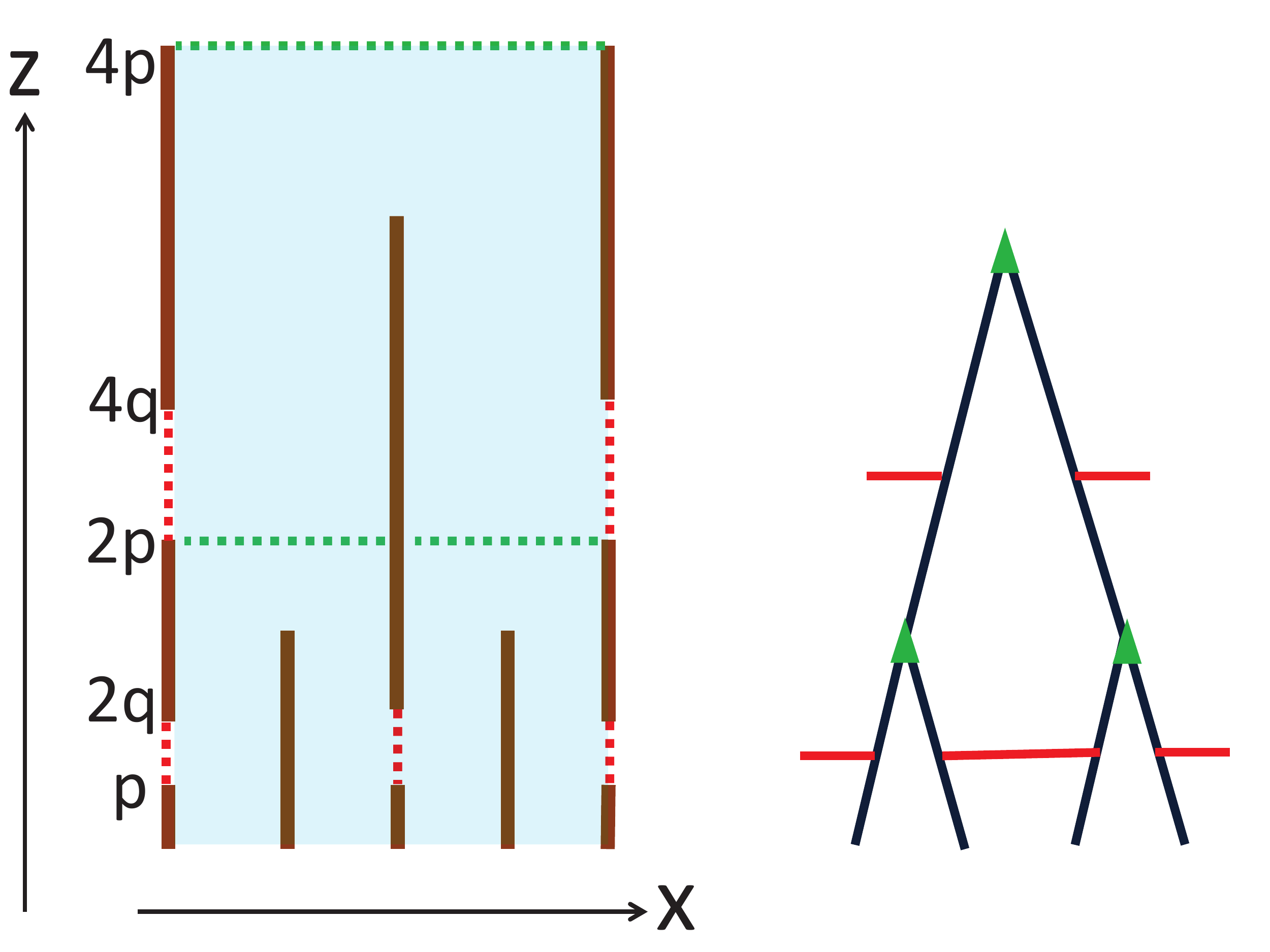}
  \hspace{1cm}
  \includegraphics[width=6cm]{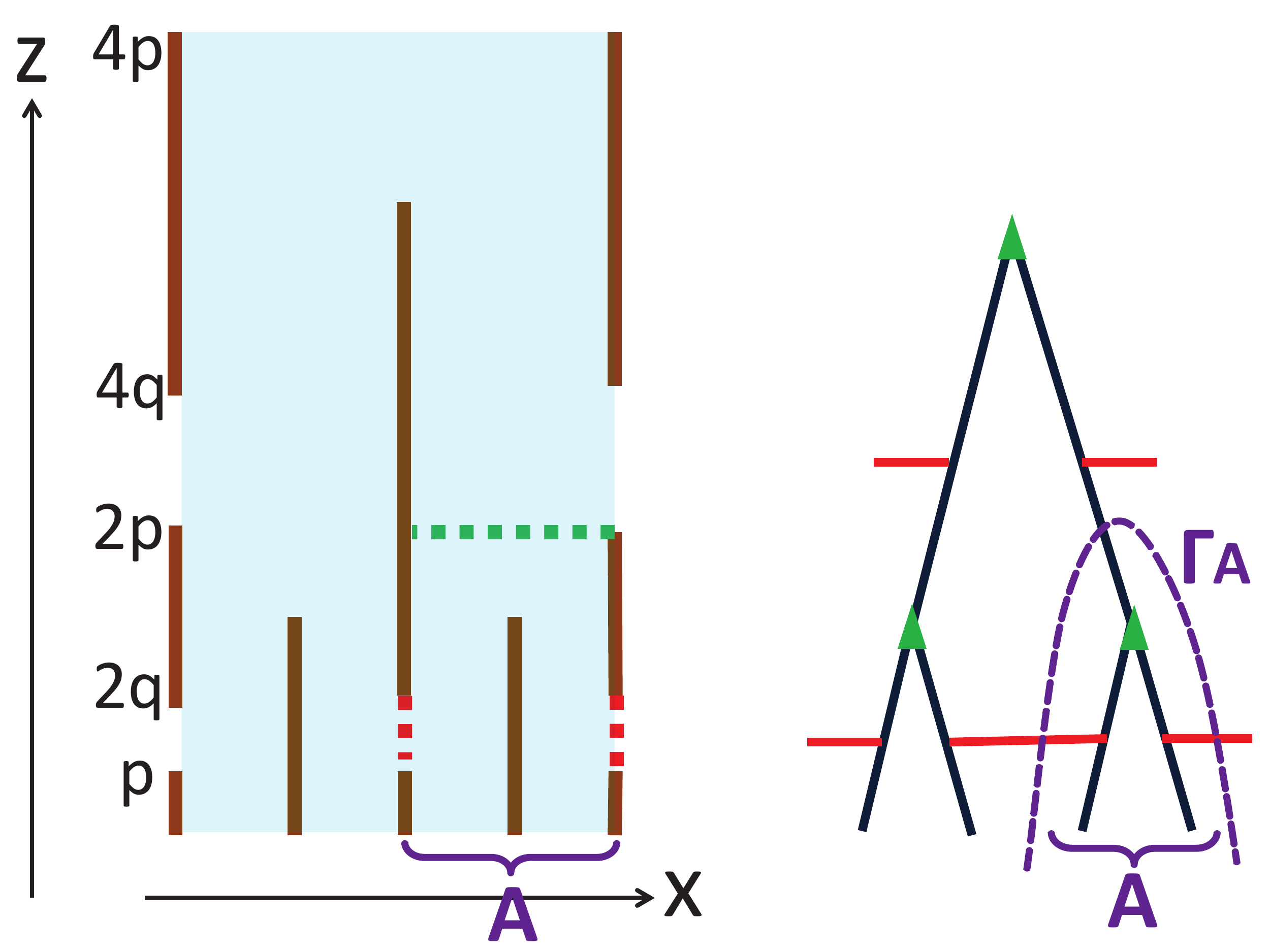}
\caption{A sketch of the time slice of our proposed AdS$_3/$BCFT$_2$ setup
dual to a MERA-like tensor network (left two pictures), and the estimation of EE $S_A$ in the AdS/BCFT and tensor network (right two pictures). The blue region in each picture is the bulk time slice and we added its tensor network description in its next right. The brown thick lines are the slits created by the boundary surface $Q$ in the AdS/BCFT. The red lines and green triangles in the tensor network describe the disentanglers and the coarse-graining tensors. The red and green dotted lines are the geodesics whose length compute the holographic entanglement entropy, corresponding to the entanglement removed by the disentanglers or coarse-graining tensors. We ignore the back-reactions just for simplicity and choose the locations of the slits such that the values of the EE for each red/green dotted lines are the same, which leads to the scale invariance. By comparing the right two pictures, we find that the calculation of HEE in our AdS/BCFT setup is equivalent to that for the tensor network.}
\label{fig:SPPSHOLgeoo}
\end{figure}

As an ambitious attempt which finalizes this paper, we would like to propose a holographic counterpart of tensor network states in the AdS/BCFT setup by repeating such procedures.\footnote{
For another connection between BCFTs and tensor networks refer to \cite{RBC}.}
 Consider the MERA tensor network \cite{MERA} and try to realize an analogous geometry on a time slice of a gravity dual. As we explained in section \ref{sec:tnlq}, the MERA network consists of the disentangling and coarse-graining operations, which cut the entanglement in the ground state of a critical spin chain.
We depicted a simple setup to achieve this in Fig.\ref{fig:SPPSHOLgeoo}. In this analogy, each ``spin'' in MERA corresponds to the (blue) region between two (brown) slits. A coarse-graining and disentangler correspond to terminating and creating a hole in a (brown) slit, respectively. The standard estimation of EE $S_A$ in the MERA nicely agrees with the HEE calculation in the AdS/BCFT as we show in  Fig.\ref{fig:SPPSHOLgeoo}, where we simply ignored the back-reactions. The entanglement entropy reduced by the coarse-graining/disentangler operation correspond to the green/red dotted line in the pictures.

Notice that in this section we consider a different interpretation of the AdS spacetime in terms of tensor networks, as compared with the arguments in section 7.
Here we approximately regard the narrow strip in AdS space as a link in a tensor network and
fill the gap between the discretized lattice network (such as MERA) between the continuous AdS spacetime  by cutting the latter spacetime into lattices by inserting various boundary surfaces of AdS/BCFT. 
This idea opens up a new approach to the conjectured connection between the AdS/CFT and tensor networks.

Next we discuss how to realize a gravity dual whose time slice is given by the geometry in Fig.\ref{fig:SPPSHOLgeoo}. One way is to perform an Euclidean path-integration on the manifold
 given by the left picture in Fig.\ref{fig:SPPSHOLgeoo} with $z$ interpreted as the Euclidean time
  $-\tau$ and consider its gravity dual. Note that the initial state $z=-\tau=\infty$ is expected to be a completely disentangled state. Therefore
  we choose the state at $\tau=-\infty$ to be the (regularized) boundary state $|B\lb$ \cite{MRTW,Numa}.
Indeed, we expect that the time slice $\tau=0$ of such a Euclidean gravity dual is given by what we want, according to what we learn from the holographic local quenches summarized in Fig.\ref{fig:SPPSHOLgeo}.

Another idea is based on a Lorentzian path-integration in the dual CFT.
Now we regard $z$ as the real time $-t$ in
 Fig.\ref{fig:SPPSHOLgeoo} and again choose the initial state at $t=-\infty$ to be the boundary state
 $|B\lb$. Since each end points of the slits moves toward the horizon at the speed of light, we again expect that the gravity dual of this time-dependent setup is approximately given by the one in  Fig.\ref{fig:SPPSHOLgeoo}.

\section{Conclusions}\label{sec:conclusion}

In this paper, we studied the time evolutions of entanglement entropy (EE) under three types of local quenches (local operator, splitting and joining) in two dimensional CFTs. Our main examples of CFTs are the free massless Dirac fermion CFT and holographic CFTs in two dimensions. The typical behavior of EE for each of the three types of local quenches in holographic CFTs is the logarithmic growth for a large subsystem. On the other hand, for operator local quenches in free or RCFTs, we observe simple step functional behaviors. However, for splitting and joining local quenches, we still have logarithmic time evolutions even for free or RCFTs. We also noted that there are two types of logarithmic growth: $(c/6)\log(t/\ap)$ and $(c/6)\log(t/\ep)$, where $\ap$ is a regularization parameter of the local quench and $\ep$ is the lattice spacing of the CFT.
In this paper, we got systematic understandings of these differences from various points of view, including entanglement density (ED), holographic geometry, and tensor network descriptions. We also calculated evolution of EE in spin systems for splitting quenches.

In the holographic CFT case, the EE is computed as the length of geodesics in the dual AdS$_3$ spacetime based on the holographic entanglement entropy (HEE). Furthermore, in the splitting/joining quench case, these geodesics can end on the boundary surfaces $Q$, following the AdS/BCFT prescription. Therefore there are two kinds of geodesics: connected and disconnected. Thus the HEE is given by ${\rm min}\{S_A^{con}, S_A^{dis}\}$.
One of them, $S_A^{dis}$, depends on the tension parameter $T_{BCFT}$ in AdS/BCFT.

In the splitting and joining quench, both gravity duals share the same metric but have different boundary surfaces $Q$. In order to see the logarithmic growth of HEE, 
we have to consider a region which is not covered by the Poincar\'{e} patch because the 
geodesic penetrates the Poincar\'{e} horizon. We found that the growth $(c/6)\log(t/\epsilon)$ comes from the geodesic length in the Poincar\'{e} patch, while the other one $(c/6)\log(t/\alpha)$ comes from that near the surface $Q$, which is hidden inside the 
 Poincar\'{e} horizon. 
 
 In the splitting quench, the boundary surface $Q$ expands from the AdS boundary toward the bulk at the speed of light. We can regard the two semi-infinite lines at each time are connected through an expanding horizon or
 equally (non-traversal) wormhole.
 Note that since $Q$ is time-like, there is no causal influence between them.
In the joining quench, the boundary surface $Q$ can be regarded as a ``falling string" towards the horizon, which is analogous to the falling particle in the holographic local operator quench.
  In both examples, the precise location of $Q$ depends on the tension $T_{BCFT}$ in a way
that when $T_{BCFT}>0$ (or $T_{BCFT}<0$), the  region of gravity dual expands (or shrinks).

Entanglement density (ED) turns out to be a good tool to systematically probe entanglement structures between two space points. The analysis with the ED helped us to figure out that the initial entanglements created by our ``local" quenches is highly non-local, except for the local operator quench in RCFTs. In general, the ``local'' quenches are not described as local unitary transformations. In holographic case, the ED can straightforwardly explain logarithmic time evolutions under different local quenches. However, the two evolutions $(c/6)\log(t/\alpha)$ and $(c/6)\log(t/\epsilon)$ cannot be easily distinguished from this perspective.

The gravity duals of local quenches also have a qualitative interpretation in terms of MERA-like tensor networks, where $\sim\log(t/\epsilon)$ behavior can be understood as splitting/joining disentanglers. However, we could not find a clear realization of the $\sim\log(t/\alpha)$ behavior, called  ``shock wave" contribution. One possibility is to explain this by inserting extra tensors into the networks and we would like to leave more details for a future problem.

In the final part of this paper, we presented an analogue of quantum circuits using both of splitting and joining quenches and gave their gravity duals based on the AdS/BCFT. This offers a method of discretizing gravity dual spacetimes. It is expected that this can further lead to a deeper understanding of the conjectured AdS/tensor network correspondence.

\section*{ Acknowledgements}

We are grateful to Pawel Caputa, Masahiro Nozaki, Tokiro Numasawa and Tomonori Ugajin for useful comments.
TT is supported by the Simons Foundation through the ``It from Qubit'' collaboration and by World Premier International Research Center Initiative (WPI Initiative) from the Japan Ministry of Education, Culture, Sports, Science and Technology (MEXT). TT is supported by JSPS Grant-in-Aid for Scientific Research (A) No.16H02182 and by JSPS Grant-in-Aid for Challenging Research (Exploratory) 18K18766.

\appendix

\section{Vacuum Entanglement Density in Massless Dirac Fermion CFT} \label{dfee}

Here we would like to show that the simple profile of entanglement density (we set $c=1$ in (\ref{ved}))
\be
n_0(x,y)=\frac{1}{6(x-y)^2}, \label{EDDi}
\ee
reproduces arbitrary entanglement entropy in the massless Dirac fermion CFT, including cases where
the subsystem $A$ consists of multiple disconnected intervals:
\be
A=I_1\cup I_2\cup\ddd\cup I_n.
\ee

We choose the $n$ intervals are parameterized by $I_i=[a_i,b_i]$ for $i=1,2,\ddd,n$.
For convenience, we introduce $b_0=-\infty$ and $a_{n+1}=\infty$.
Then, by summing all bipartite entanglement, using the entanglement density (\ref{EDDi}),
$S_A$ is estimated as follows:
\ba
S_A&=&\sum_{k=0}^n\sum_{l=1}^n \int^{a_{k+1}}_{b_k}dx\int^{b_l}_{a_l}dy \frac{1}{6(x-y)^2},\no
&=&\frac{1}{6}\sum_{k=0}^n\sum_{l=1}^n \left(\log\left(\frac{a_{k+1}-b_l}{a_{k+1}-a_l}\right)
+\log\left(\frac{b_k-a_l}{b_k-b_l}\right)\right), \no
&=&\frac{1}{3}\sum_{i,j=1}^{n}\log|b_i-a_j|-\frac{1}{6}\sum_{i,j=1}^{n}\log|a_i-a_j|
-\frac{1}{6}\sum_{i,j=1}^{n}\log|b_i-b_j|.
\ea
We interpret the diagonal terms in the final expression by introducing the UV cut off $\ep$ as $a_i-a_i=b_i-b_i=\ep$. Thus finally we obtain
\ba
S_A=\frac{1}{3}\sum_{i,j=1}^{n}\log\frac{|b_i-a_j|}{\ep}
-\frac{1}{3}\sum_{i>j=1}^n\log\frac{|a_i-a_j|}{\ep}
-\frac{1}{3}\sum_{i>j=1}^n\log\frac{|b_i-b_j|}{\ep}.
\ea
This indeed agrees with the known expression of entanglement entropy in the massless Dirac fermion CFT
\cite{CH}. A closely related property is that the multi partite mutual information is vanishing
\cite{Blanco:2011np}.

\section{Detailed Computations of Evolutions of Holographic Entanglement Entropy}\label{sec:HLQ}

Here we present the details of calculations of HEE under splitting/joining local quenches.
We define the Euclidean coordinate $(\tau,x)$ and complex coordinate $(w,\bar{w})$ as
\begin{equation}
w = x+i\tau,\quad \bar{w}=x-i\tau.
\end{equation}
Both splitting/joining quench setup are described by almost the same conformal maps:
\begin{eqnarray}
\bigskip
(\textnormal{splitting}) \quad f_{+}(w) &=& i\sqrt{\zeta(w)}\quad \arg(\zeta ) \in (-\pi,\pi]\\
(\textnormal{joining}) \quad f_{-}(w) &=& i\sqrt{-\zeta(w)} \quad \arg(-\zeta ) \in (-\pi,\pi].
\end{eqnarray}
Here we define $\zeta(w)$ by
\begin{equation}
\zeta (w) = \frac{w+i\alpha}{w-i\alpha}.
\end{equation}

The HEE for the connected/disconnected geodesics $S_{A,\pm}^{con},S_{A,\pm}^{dis}$ for the subsystem $A(=[w_a,w_b])$ is computed by (when the tension $T_{BCFT}$ is vanishing):
\begin{eqnarray}
S_{A,\pm}^{con} &=& \frac{c}{12}\log \left(\frac{|f_{\pm}(w_a)-f_{\pm}(w_b)|^4}
{\epsilon^4|f_{\pm}^{\prime}(w_a)|^2|f_{\pm}^{\prime}(w_b)|^2}\right),\\
S_{A,\pm}^{dis} &=& \frac{c}{12}\log \left(\frac{16(\textnormal{Im}f_{\pm}(w_a))^2(\textnormal{Im}f_{\pm}(w_b))^2}{\epsilon^4|f_{\pm}^{\prime}(w_a)|^2|f_{\pm}^{\prime}(w_b)|^2}\right).
\end{eqnarray}
The plus and minus subscript in $S_{A,\pm}$ corresponds to splitting and joining quench, respectively.

When computing these quantities, it will be helpful to use the results below.
\begin{eqnarray}
\zeta = \frac{w+i\alpha}{w-i\alpha} &=& \frac{x+i(\tau+\alpha)}{x+i(\tau-\alpha)} = \frac{x^2+\tau^2-\alpha^2+i(2x\alpha)}{x^2+(\tau-\alpha)^2}\\
|\zeta| &=& \frac{\sqrt{(x^2+\tau^2-\alpha^2)^2+(2x\alpha)^2}}{x^2+(\tau-\alpha)^2}=\sqrt{\frac{x^2+(\tau+\alpha)^2}{x^2+(\tau-\alpha)^2}}\\
\frac{d \sqrt{\pm \zeta}}{d w} &=& -\frac{i\alpha}{w^2+\alpha^2}\sqrt{\pm \zeta}\\
|w^2+\alpha^2| &=& \sqrt{(x^2+(\tau+\alpha)^2)(x^2+(\tau-\alpha)^2)}\\
 &=&\sqrt{(x^2+\tau^2-\alpha^2)^2+(2x\alpha)^2}\\
\left|\frac{d \sqrt{\pm \zeta}}{d w}\right|^2 &=& \frac{\alpha^2}{(x^2+(\tau-\alpha)^2)\sqrt{(x^2+\tau^2-\alpha^2)^2+(2x\alpha)^2}}
\end{eqnarray}
For simplicity we define $R,A_{\pm}$ as
\begin{eqnarray}
R &=& \sqrt{(x^2+\tau^2-\alpha^2)^2+(2x\alpha)^2}\\
A_\pm &=& \sqrt{(x^2+\tau^2-\alpha^2)^2+(2x\alpha)^2}\pm(x^2+\tau^2-\alpha^2).
\end{eqnarray}
And we get more results,
\begin{eqnarray}
\sqrt{\zeta} &=& \left\{
\begin{array}{l}
\bigskip
\displaystyle\frac{\sqrt{A_{+}}+i\sqrt{A_{-}}}{\sqrt{2\left(x^2+(\tau-\alpha)^2\right)}}\quad (x>0)\\
\bigskip
\displaystyle\frac{\sqrt{A_{+}}-i\sqrt{A_{-}}}{\sqrt{2\left(x^2+(\tau-\alpha)^2\right)}}\quad (x<0).
\end{array}
\right. \\
\sqrt{-\zeta} &=& \left\{
\begin{array}{l}
\bigskip
-i\sqrt{\zeta}\quad (x>0)\\
i\sqrt{\zeta}\quad (x<0).
\end{array}
\right.\\
\textnormal{Im} (i\sqrt{\pm\zeta}) &=& \sqrt{\frac{A_{\pm}}{2(x^2+(\tau-\alpha)^2)}}\\
(\textnormal{Im} (i\sqrt{\pm\zeta}))^2 &=& \frac{\sqrt{(x^2+\tau^2-\alpha^2)^2+(2x\alpha)^2}\pm(x^2+\tau^2-\alpha^2)}{2(x^2+(\tau-\alpha)^2)}
\end{eqnarray}
Also we can compute $|f(w_a)-f(w_b)|^4$ in the connected EE. If $a,b>0$
\begin{equation}
|\sqrt{\pm\zeta_a}-\sqrt{\pm\zeta_b}|^4 = \frac{\left| (\sqrt{A_{+a}}+i\sqrt{A_{-a}})\sqrt{b^2+(\tau-\alpha)^2}-(\sqrt{A_{+b}}+i\sqrt{A_{-b}})\sqrt{a^2+(\tau-\alpha)^2} \right|^4}{4(a^2+(\tau-\alpha)^2)^2(b^2+(\tau-\alpha)^2)^2}
\end{equation}
If $a<0<b$,
\begin{equation}
|\sqrt{\pm\zeta_a}-\sqrt{\pm\zeta_b}|^4 = \frac{\left| (\sqrt{A_{+a}}-i\sqrt{A_{-a}})\sqrt{b^2+(\tau-\alpha)^2}\mp(\sqrt{A_{+b}}+i\sqrt{A_{-b}})\sqrt{a^2+(\tau-\alpha)^2} \right|^4}{4(a^2+(\tau-\alpha)^2)^2(b^2+(\tau-\alpha)^2)^2}.
\end{equation}
Here we write $\zeta(x,\tau)=\zeta_x,\ A_{\pm}(x,\tau)=A_{\pm x}$.

Now we can derive Euclidean HEE  for the connected/disconnected geodesics, with $R(x,\tau)=R_x$,
\begin{eqnarray}
S_{A,\pm}^{con} &=& \frac{c}{12}\log\left\{
\begin{array}{l}
\bigskip\displaystyle\frac{R_a R_b( Q- (\sqrt{A_{+a}A_{+b}}+\sqrt{A_{-a}A_{-b}}) )^2}{\alpha^4\epsilon^4}\quad (a,b>0)\\
\bigskip\displaystyle\frac{R_a R_b( Q\mp (\sqrt{A_{+a}A_{+b}}-\sqrt{A_{-a}A_{-b}}) )^2}{\alpha^4\epsilon^4}\quad (a<0<b)
\end{array}
\right. \\
Q &=& \sqrt{(b^2+(\tau-\alpha)^2)(a^2+(\tau+\alpha)^2)}+(a\leftrightarrow b)\\
&=&  \sqrt{2}\sqrt{(a^2+\tau^2+\alpha^2)(b^2+\tau^2+\alpha^2)-(2\tau\alpha)^2+R_aR_b}\\
S_{A,\pm}^{dis} &=& \frac{c}{12}\log\frac{4R_a R_b A_{\pm a}A_{\pm b}}{\alpha^4\epsilon^4}
\end{eqnarray}

We do analytic continuation to the real time $\tau\to it$, and then we approximate $x,t,|x^2-t^2|\gg\alpha$. Under this approximation we get
\begin{eqnarray}
R_x &\sim& |x^2-t^2|+\frac{x^2+t^2}{|x^2-t^2|}\alpha^2\\
A_{\pm x} &\sim& \left(|x^2-t^2|\pm(x^2-t^2)\right)+\left(\frac{x^2+t^2}{|x^2-t^2|}\mp 1 \right)\alpha^2
\end{eqnarray}

Consequently we derive the Lorentzian HEE formula. Here we write the results only in splitting quench case (i.e. $S_{A,+}$ case),
\begin{equation}
S_{A,+}^{dis} \sim \frac{c}{6}\log\left\{
\begin{array}{l}
\bigskip
\displaystyle\frac{4(a^2-t^2)(b^2-t^2)}{\alpha^2\epsilon^2}\quad (t<|a|<|b|)\\
\bigskip
\displaystyle\frac{4|a|(b^2-t^2)}{\alpha\epsilon^2}\quad (|a|<t<|b|) \\
\bigskip
\displaystyle\frac{4|a||b|}{\epsilon^2}\quad (|a|<|b|<t)
\end{array}
\right.
\end{equation}
And for connected HEE we have to consider the sign of $a$. If $a,b>0$
\begin{equation}
S_{A,+}^{con} \sim \frac{c}{6}\log\left\{
\begin{array}{l}
\bigskip
\displaystyle\frac{(b-a)^2}{\epsilon^2}\quad (0<t<a,b<t)\\
\bigskip
\displaystyle\frac{2(b-a)(t-a)(b-t)}{\alpha\epsilon^2}\quad (a<t<b)
\end{array}
\right.
\end{equation}
If $a<0<b,\ -a<b$,
\begin{equation}
S_{A,+}^{con} \sim \frac{c}{6}\log\left\{
\begin{array}{l}
\bigskip
\displaystyle\frac{(b-a)^2}{\epsilon^2}\quad (0<t<-a)\\
\bigskip
\displaystyle\frac{2(b-a)(t+a)(b+t)}{\alpha\epsilon^2}\quad (-a<t<b)\\
\bigskip
\displaystyle\frac{4(t^2-a^2)(t^2-b^2)}{\alpha^2\epsilon^2}\quad (b<t)\\
\end{array}
\right.
\end{equation}
These give the results (\ref{hpsa}), (\ref{hpsb}) and (\ref{hpsc}). Similarly we can get the results for the holographic joining quenches (\ref{hppa}), (\ref{hppb}) and (\ref{hppc}).

\end{document}